# Imaging Nanoscale Carrier, Thermal, and Structural Dynamics with Time-Resolved and Ultrafast Electron Energy-Loss Spectroscopy

Wonseok Lee,[1,†] Levi D. Palmer, [1,†] Thomas E. Gage,[2,*] Scott K. Cushing[1,*]

[1]*Division of Chemistry and Chemical Engineering, California Institute of Technology, Pasadena, CA 91125, USA*
[2]*Center for Nanoscale Materials, Argonne National Laboratory, Lemont, IL 60439, USA*
† These authors contributed equally to this work.
*Authors to whom correspondence should be addressed: tgage@anl.gov and scushing@caltech.edu

## ABSTRACT

Time-resolved and ultrafast electron energy-loss spectroscopy (EELS) is an emerging technique for measuring photoexcited carriers, lattice dynamics, and near-fields across femtosecond to microsecond timescales. When performed in either a specialized scanning transmission electron microscope or ultrafast electron microscope (UEM), time-resolved and ultrafast EELS can directly image charge carriers, lattice vibrations, and heat dissipation following photoexcitation or applied bias. Yet recent advances in theoretical calculations and electron optics are often required to realize the full potential of ultrafast EEL spectrum imaging. In this review, we present a comprehensive overview of the recent progress in the theory and instrumentation of time-resolved and ultrafast EELS. We begin with an introduction to the technique, followed by a physical description of the loss function. We outline approaches for calculating and interpreting ground-state and transient EEL spectra spanning low-loss plasmons to core-level excitations analogous to X-ray absorption. We then survey the current state of time-resolved and ultrafast EELS techniques beyond photon-induced near-field electron microscopy, highlighting abilities to image carrier and thermal dynamics. Finally, we examine future directions enabled by emerging technologies, including electron beam monochromation, *in situ* and operando cells, laser-free UEM, and high-speed direct electron detectors. These advances position time-resolved and ultrafast EELS as a critical tool for uncovering nanoscale dynamic processes in quantum materials and solar energy conversion devices.

# HIGHLIGHT IMAGE

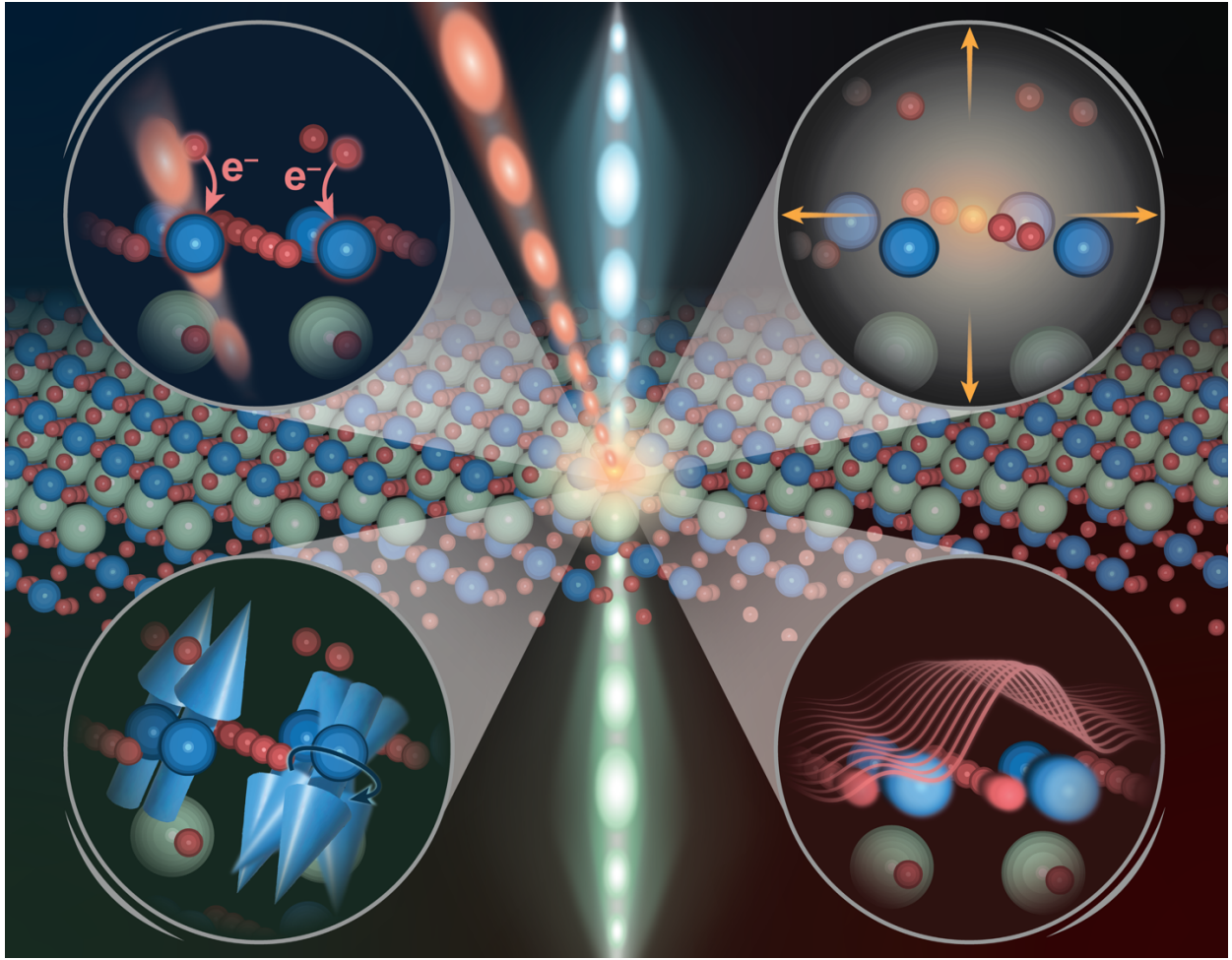

# Article Contents

# NOMENCLATURE

| | |
|---:|---|
| 4D-STEM | Four-Dimensional Scanning Transmission Electron Microscopy |
| AuNT | Gold Nanotriangle |
| BSE | Bethe–Salpeter Equation |
| CCD | Charge-Coupled Device |
| CDEM | Charge Dynamics Electron Microscopy |
| cDFPT | Constrained Density Functional Perturbation Theory |
| CL | Cathodoluminescence |
| CNT | Carbon Nanotubes |
| CTM4XAS | Charge Transfer Multiplet Program for X-ray Absorption Spectroscopy |
| Cryo | Cryogenic |
| DC | Direct Current |
| DFPT | Density Functional Perturbation Theory |
| DFT | Density Functional Theory |
| DOS | Density of States |
| e-Beam | Electron Beam |
| EEG | Electron Energy-Gain |
| EELS | Electron Energy-Loss Spectroscopy |
| EFTEM | Energy-Filtered Transmission Electron Microscopy |
| ELNES | Energy-Loss Near-Edge Structure |
| EM | Electron Microscopy |
| ETEM | Environmental TEM |
| EXELFS | Extended Energy-Loss Fine Structure |
| FDTD | Finite-Difference Time-Domain |
| FEFF | Full Extended Fine Field |
| FEM | Finite Element Method |
| fs | Femtosecond |
| FTIR | Fourier Transform Infrared |
| FWHM | Full Width at Half-Maximum |
| LL | Liouville–Lanczos |
| LSP | Localized Surface Plasmon |
| MD | Molecular Dynamics |
| MNPBEM | Simulation of Metallic Nanoparticles using a Boundary Element Method |
| NA | Numerical Aperture |
| NR | Not Reported |
| ns | Nanosecond |
| OCEAN | Obtaining Core Excitations from the Ab initio electronic structure and the NIST BSE solver |
| PEET | Plasmon Energy Expansion Thermometry |
| PINEM | Photon-Induced Near-Field Electron Microscopy |
| ps | Picosecond |
| q-EELS | Momentum-Resolved EELS |
| Quantum ESPRESSO | Quantum Open-Source Package for Research in Electronic Structure, Simulation, and Optimization |
| RF | Radio-frequency |
| RMS | Root-Mean Squared |
| SBR | Signal-to-Background Ratio |
| SNR | Signal-to-Noise Ratio |
| STEM | Scanning Transmission Electron Microscopy |
| TDC | Time-to-Digital Converter |
| TDDFPT | Time-Dependent Density Functional Perturbation Theory |
| TDDFT | Time-Dependent Density Functional Theory |
| TEM | Transmission Electron Microscopy |
| THz | Terahertz |
| ToA | Time-of-Arrival |
| Tr-q-EELS | Time- and Momentum-Resolved Electron Energy-Loss Spectroscopy |
| UED | Ultrafast Electron Diffraction |
| UEM | Ultrafast Electron Microscopy |
| UV-VIS | Ultraviolet-Visible |
| VASP | Vienna Ab initio Simulation Package |
| XAS | X-ray Absorption Spectroscopy |
| XLD | X-ray Linear Dichroism |
| XTA | X-ray Transient Absorption |
| XUV | Extreme Ultraviolet |
| ZLP | Zero-Loss Peak |

# I. INTRODUCTION

Studying ultrafast and time-resolved dynamics is crucial for understanding complex transient processes in a material. These studies are particularly significant at the nanoscale, where a material's behavior can be drastically different under non-equilibrium conditions. On timescales ranging from femtoseconds to milliseconds, the interactions and energy transfer between electrons, holes, and phonons in a material determine its electronic, optical, and structural properties. These processes are fundamental to the operation of energy conversion and optoelectronic devices. For example, understanding the excitation, thermalization, transport, and recombination dynamics of charge carriers can lead to the design of solar cells with improved photovoltaic efficiencies,[1–4] while monitoring the trajectories of hot electrons in plasmonic materials can guide effective photocatalytic platforms.[5–8] In addition, nanoscale systems often exhibit size- and shape-dependent properties,[9–11] and photoexcited dynamics at interfaces, grain boundaries, and defects deviate significantly from their bulk counterparts.[12,13] This highlights the need for imaging techniques with nanometer spatial and femtosecond temporal resolutions.

The spatial resolution limit of imaging techniques is determined by the Abbe diffraction limit, given by $d = \lambda/2\text{NA}$ where $\lambda$ is the wavelength and NA is the numerical aperture.[14] Since electrons can have shorter wavelengths than photons, transmission electron microscopy (TEM) provides superior spatial resolution compared to optical microscopy. For example, an electron with an energy of 10 keV and a NA of 1.0 has an Abbe limit of 6.1 pm—shorter than the smallest atomic radius (37 pm). Technical advances in aberration-corrected electron optics,[15–18] scanning TEM (STEM),[19,20] and direct electron detectors[21–24] have enabled picometer-scale imaging with resolution limited primarily by inherent atomic vibrations.

For decades, TEM has proven to be a powerful and robust imaging tool; however, its temporal resolution was limited by the detection rate of charge-coupled device (CCD) cameras with scintillators, making it unsuitable for imaging dynamics that occur on timescales faster than milliseconds. To overcome this limitation, ultrafast electron microscopy (UEM) integrates a pump–probe technique within the TEM, enabling spatially resolved measurements at ultrafast timescales.[25–30] **Figure 1(a)** shows a standard UEM setup, where a femtosecond (fs) pump pulse photoexcites the specimen, while a photoelectron probe is generated from the photocathode via the photoelectric effect using a femtosecond ultraviolet (UV) pulse. The time delay between the pump and probe pulses can be controlled using a mechanical delay stage to measure a material's photoexcited response. The generated electron pulses are then shaped and focused using condenser lenses. After passing through the specimen, the image is magnified and projected onto the detector through post-specimen lenses. In STEM, scan coils deflect the electron beam to raster across the specimen with a focused probe, providing more localized data compared to conventional TEM. A q-slit selects specific momentum directions for momentum-resolved measurements. Real-space imaging in a UEM can be correlated with reciprocal and energy spaces via diffraction and spectroscopy, respectively.

We focus on using electron energy-loss spectroscopy (EELS) to image charge carrier, thermal, and structural dynamics at the nanoscale. Time-resolved and ultrafast EELS in a UEM can probe localized photoexcited dynamics, including charge and thermal transport within a specimen, by utilizing a wide energy range.[30] As the name suggests, EELS measures the energy distribution of incident electrons that have interacted with the specimen as a function of energy loss.[31,32] For a specimen with a thickness less than the electron's inelastic mean free path, most electrons are elastically scattered, which results in negligible energy loss. This is represented by the most intense peak at 0 eV energy loss, known as the zero-loss peak (ZLP), in the EEL spectrum. In contrast, inelastic scattering involves the loss of energy by the incident electrons. The low-loss region, extending from 0 to 50 eV, corresponds to quasiparticle excitations such as phonons, magnons, excitons, and surface and bulk plasmons, as well as excitations of valence electrons to unoccupied states (inter/intraband transitions). This region provides access to a range of a material's properties, such as atomic and lattice vibrations,[33,34] band gap,[35] specimen thickness,[36] dielectric function,[37] valence electron density,[38] surface or interface states,[39] and joint density of states (DOS).[40] The core-loss region, generally above 50 eV, corresponds to the excitation of electrons from atomic core levels, adding element specificity to the technique. This enables the measurement of elemental distributions,[41] local coordination geometry,[42] element-specific unoccupied DOS,[43] and radial distribution function.[44] Time-resolved and ultrafast EELS can therefore theoretically measure all these inelastic excitations along femtosecond to microsecond timescales.

Most ultrafast EELS studies have employed photon-induced near-field electron microscopy (PINEM), where free electrons interact with photons, resulting in an EEL spectrum characterized by multiple side peaks around the ZLP.[45] Under normal circumstances, free electrons do not interact with photons due to energy-momentum mismatch. However, this limit can be overcome at high pump fluences or in the presence of nanostructures,[46] and PINEM enables the visualization of the near-fields of plasmonic nanoparticles.[47–50] As this review mainly focuses on time-resolved and ultrafast EELS for studying carrier, structural, and thermal dynamics, readers seeking a more in-depth treatment of the PINEM technique are encouraged to consult Refs. 45 and 46.

In this review, we start by outlining the key features observed in transient EEL spectra. Then, we discuss how these spectra can be simulated using *ab initio* and numerical computational methods. The primary focus of this review is to highlight recent advances in time-resolved and ultrafast EELS techniques beyond PINEM in Sec. IV, highlighting their advantages and applications. Finally, we conclude by discussing potential improvements and future directions for time-resolved and ultrafast EELS in imaging dynamics in materials. Readers can further explore a guide to technical considerations for time-resolved and ultrafast EELS techniques in the Appendix.

## II. BACKGROUND

Photoexcitation alters the spectral features of the EEL spectrum by both local electromagnetic fields and changes in electron or phonon populations (**Figs. 1(b)–1(e)**). For example, a local electromagnetic field modifies characteristics of the electron probe, enabling ultrafast imaging of charge dynamics at nanometer spatial and femtosecond temporal resolutions (**Fig. 1(b)**). Photoexcitation also influences the energy, intensity, and linewidth of the bulk plasmon and core-loss peaks, which allows spatially and temporally resolved EELS measurements across the full accessible energy range. These changes to spectral features, called differential spectra, are central to time-resolved and ultrafast EELS techniques. Therefore, understanding the origin of these ground-state peaks and how photoexcitation induces differential features is essential for analyzing dynamics in such studies.

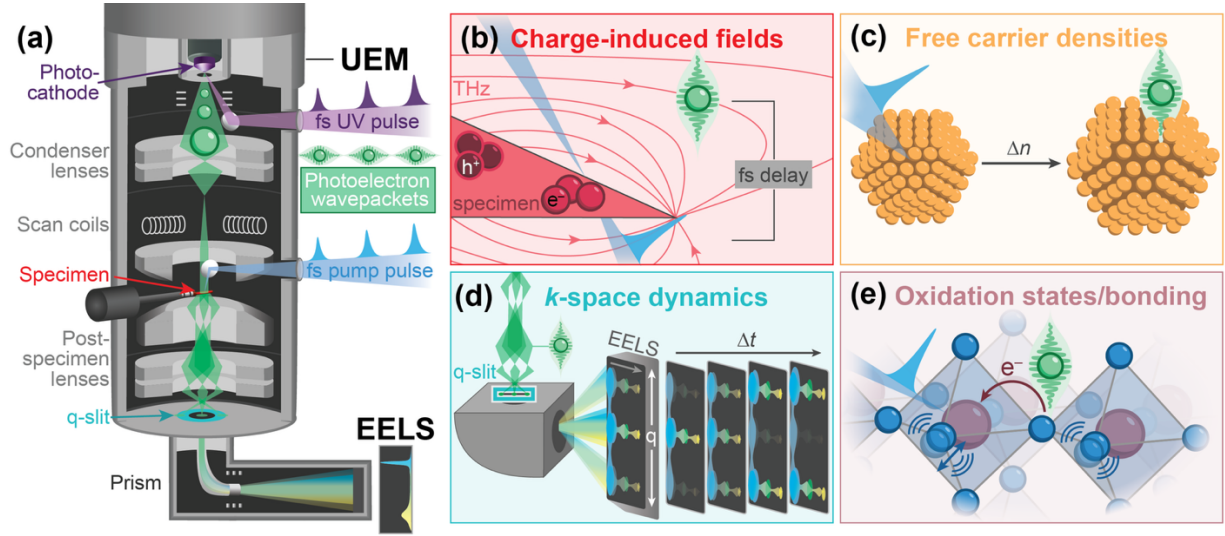

**Figure 1. Time-resolved and ultrafast electron energy-loss spectroscopy (EELS) in the ultrafast electron microscope (UEM).** (a) A standard UEM uses an ultraviolet (UV) pulse-triggered photocathode to generate ultrafast photoelectron wavepackets (green). Scan coils deflect the electron beam in scanning TEM (STEM). A femtosecond (fs) pump (blue) photoexcites the specimen (red) at fixed time delays relative to the photoelectron probe. A q-slit (teal) enables momentum-resolved EELS. Apertures are excluded for clarity. (b–e) Time-resolved and ultrafast EELS imaging methods. (b) Charge dynamics electron microscopy (CDEM) maps charge-induced terahertz (THz) fields around a specimen. (c) Ultrafast low-loss EELS quantifies a material's free carrier density change due to thermal expansion or strain. (d) Time- and momentum-resolved EELS measures carrier dynamics in reciprocal space with a q-slit. (e) Ultrafast core-loss EELS probes photoinduced changes in oxidation state and bonding due to charge transfer and phonons.

**Low-Loss EELS.** The bulk plasmon is a collective longitudinal oscillation of loosely bound electrons. In metals, it corresponds to the collective excitation of the free-electron gas, while in semiconductors and insulators, it involves the collective oscillation of valence electrons.[51] In EELS, the bulk plasmon excitation typically occurs between 5 and 30 eV and the peak is in the low-loss region of the spectrum, which is dependent on the free or valence carrier density of materials. Since plasmon scattering has the highest scattering cross-section, the plasmon loss feature appears as the most prominent feature after the ZLP.

According to the Drude model,[52–54] the energy-loss function for a bulk plasmon is given by:

$$\text{Im}\left[-\frac{1}{\epsilon(\omega)}\right] = \frac{\omega_p^2 \omega \Gamma}{\left(\omega^2 - \omega_p^2\right)^2 + \omega^2 \Gamma^2} \quad (1)$$

where $\omega_p = \sqrt{ne^2/m_e\epsilon_0}$ is the bulk plasmon frequency. Here, $n$ is the valence or free electron density, $e$ is the elementary charge of an electron, $m_e$ is the effective electron mass, and $\epsilon_0$ is the vacuum permittivity. The parameter $\Gamma$ is the inverse of relaxation time and depends on both electronic and lattice temperatures. Described further in Sec. III. A, the energy-loss function of a perfect conductor has a frequency peaking at $\omega_p$, full-width at half maximum (FWHM) of $\Gamma$, and peak height of $\omega_p/\Gamma$. Therefore, the bulk plasmon provides information about the dielectric function, carrier density, and electron scattering time.

Importantly, the bulk plasmon energy is sensitive to the electron density $n = N/V$, where $N$ is the number of electrons and $V$ is the lattice volume. In photoexcited semiconductors and insulators, the valence electron density is modified by 1) the promotion of electrons from the valence band to the conduction band and 2) thermal expansion or contraction of the lattice. As a result, ultrafast low-loss EELS can measure changes to a material's carrier density due to photoexcitation while capturing both the electronic and structural effects (**Fig. 1(c)**).

The bulk plasmon energy is affected not only by the carrier density but also exhibits dependence on momentum transfer. At small momentum transfer $q$, the dispersion relation of the bulk plasmon is approximately given by:

$$E_p \approx E_{p,0} + \left(\frac{\hbar^2}{m}\right)\alpha q^2 \qquad (2)$$

where $E_{p,0}$ is the plasmon energy at $q = 0$, $\hbar = \frac{h}{2\pi}$ is the reduced Planck constant, and $\alpha = \frac{3}{5}\frac{E_F}{\hbar\omega_p}$ is the dispersion coefficient with $E_F$ being the Fermi energy.[54] This relation indicates that the bulk plasmon energy increases quadratically with momentum transfer. It is also noted that electrons can carry a wide range of momentum transfer, unlike photons. This difference enables the extraction of detailed insights into photoexcitation, scattering, and relaxation processes in reciprocal space using momentum-resolved EELS (**Fig. 1(d)**).

**Core-Loss EELS.** In addition to bulk plasmon excitation, electrons can be excited from core levels to unoccupied states of a material, leading to peaks or "edges" in the core-loss region. The energy of the ionization edge is determined by the binding energy of the atomic subshell, making this spectral feature element-specific. According to Fermi's golden rule, the transition rate $W_{i \to f}$ for a system transitioning from an initial state $\Psi_i$ to a final state $\Psi_f$ upon interaction with an incident electron of energy $\hbar\omega$ is given by:[55]

$$W_{i \to f} = \frac{2\pi}{\hbar}\left|\langle\Psi_f|\hat{T}|\Psi_i\rangle\right|^2 \delta(E_f - E_i - \hbar\omega). \qquad (3)$$

Here, $\hat{T}$ is the transition operator for inelastic scattering, expressed as $\hat{T} = e^{i\mathbf{q}\cdot\mathbf{r}}$. The delta function accounts for energy conservation, indicating that the transition occurs when the energy of the final state matches the energy of the initial state plus the energy loss of probe electrons. The transition operator can be expressed as a Taylor series:

$$\hat{T} = e^{i\mathbf{q}\cdot\mathbf{r}} = 1 + i\mathbf{q}\cdot\mathbf{r} - \frac{1}{2}(\mathbf{q}\cdot\mathbf{r})^2 + \cdots. \tag{4}$$

For small momentum transfer, higher-order terms can be neglected. Under the dipole approximation and assuming the initial core-level and final state wavefunctions are atomic-like, Eq. (3) can be simplified as:

$$W_{i\to f} = \frac{2\pi}{\hbar}\left|\langle\Psi_f|\mathbf{q}\cdot\mathbf{r}|\Psi_i\rangle\right|^2 \delta(E_f - E_i - \hbar\omega). \tag{5}$$

The wavefunction overlap integral in Eq. (5) will only be nonzero if the symmetry of the wavefunction changes during the transition. This leads to the dipole selection rule $\Delta l = \pm 1$, where $\Delta l$ is the change in angular momentum quantum number during the transition. For instance, a K edge probes unoccupied states with *p*-character, while L$_{2,3}$ edges probe unoccupied *s*- and *d*-like states. However, if the momentum transfer is sufficiently large to consider the higher-order terms, EELS can detect dipole-forbidden transitions.[56,57]

An understanding of core-level electronic transitions is crucial for interpreting core-loss EELS, as they provide insight into the chemical bonding, local structure, and coordination geometry of materials. In solids, the unoccupied electronic states can be modified by chemical bonding and oxidation states, which appear in the first 30–40 eV above the edge threshold. This region, known as the electron energy-loss near edge structure (ELNES), reveals details about the local chemical structure and bonding.[58] At energy loss beyond the ELNES region, oscillations in the extended energy-loss fine structure (EXELFS) are used to determine interatomic distances and coordination numbers.[59] Ultrafast core-loss EELS measurements can, therefore, probe changes in the chemical bonding and coordination geometry of a material caused by photoexcitation, such as charge transfer and lattice changes (**Fig. 1(e)**).

However, ultrafast core-loss EELS presents several challenges. The energy-differential cross-section for inelastic scattering with energy loss $E$, denoted as $d\sigma/dE$, follows an inverse power-law relationship:[60]

$$\frac{d\sigma}{dE} \propto E^{-r} \tag{6}$$

where $r$ is a real and positive constant. This relationship implies that signal intensity decreases rapidly with increasing energy loss, making it significantly more difficult to detect and analyze ionization edges in the core-loss region compared to spectral features in the low-loss region. Moreover, plasmon scattering can complicate core-loss signals, particularly when multiple scattering becomes significant. This can increase the background intensity and distort the shape and energy of the core-loss spectrum. Additionally, core-loss peaks from shallower core levels, such as the M$_{2,3}$ edges, may overlap with the low-loss region, complicating peak assignment. To address these challenges and interpret both time-resolved and ultrafast low- and core-loss EELS measurements, theoretical simulations of the bulk plasmon and core-loss peaks are necessary. The next section will discuss how EELS spectra can be modeled using *ab initio* and numerical methods.

## III. TIME-RESOLVED AND ULTRAFAST EELS THEORY

While both *ab initio* methods and physics-based numerical models are used to simulate EEL spectra, their mathematical frameworks differ fundamentally. *Ab initio* calculations focus on predicting how the specimen responds to the perturbation introduced by the electron beam, typically by computing the material's dielectric response or loss function using quantum mechanical approaches such as density functional theory (DFT) and many-body perturbation theory. In contrast, numerical models based on classical electrodynamics, such as finite element method (FEM) or finite-difference time-domain method (FDTD), simulate how the fields induced within the specimen affect the electron beam. In essence, *ab initio* methods model the material's *intrinsic response*, while physics-based numerical models capture the *extrinsic interaction* between the electron and the local electromagnetic environment.

Both the *ab initio* and numerical approach have their own advantages and limitations. Numerical models can accurately approximate physical systems with complex geometries and boundary conditions, but they require the complex dielectric function to be input or estimated. Although they are computationally far less expensive for a macroscopic system, there is limited functionality for accurately simulating excited-state dynamics. In contrast, *ab initio* methods derive the dielectric response from first principles with minimal input parameters. However, they are less suited for handling complex geometries or interfaces due to computational expense. Various approximations can be applied in both frameworks to simulate photoexcited or time-resolved perturbations to the loss function.

In this section, we begin by introducing the fundamentals of EELS through the ground-state loss function and simplified electrodynamics, then build toward more advanced treatments of ultrafast and time-resolved EELS following photoexcitation or other perturbations. We exclusively describe the theoretical framework for low- and core-loss EELS simulations due to the current energy resolution limit of femtosecond electron pulses. Other studies report theoretical methods such as molecular dynamics and the frozen phonon approximation for simulating ultralow-loss EELS relevant to phonon and magnon excitations.[61–63]

### A. The Loss Function: Ground-State EELS Calculations

An EEL spectrum, or loss function, is described by the negative imaginary component of the inverse dielectric function, as in Eq. (1).[54,64,65] The loss function can be directly expressed using the real ($\epsilon_1$) and imaginary ($\epsilon_2$) components of the dielectric function ($\epsilon = \epsilon_1 + i\epsilon_2$) as:

$$\text{Im}\left[-\frac{1}{\epsilon(\mathbf{q},\omega)}\right] = \frac{\epsilon_2}{\epsilon_1^2 + \epsilon_2^2}. \tag{7}$$

Physically, $\epsilon_1$ represents the specimen's polarization when there is no absorption at that frequency, and $\epsilon_2$ represents single-particle excitations (i.e., absorption). The TEM's free electron beam also induces longitudinal oscillations of the specimen's electron charge density. As a result, the loss function measures both the longitudinal (out-of-plane) and transverse (in-plane) components of the dielectric function. This unique sensitivity to longitudinal excitations enables EELS to probe dark or optically forbidden electronic transitions, as it is not constrained by the transverse dipole selection rules that apply to most optical spectroscopies. The complex relation of the material's loss function and momentum transfer (**q**) are described in Sec. II, Eqs. (2)–(5).

As mentioned in Sec. I, EELS directly probes a plethora of excitations (e.g., vibrational modes, excitons, inter/intraband transitions, localized surface plasmons, core-level absorption). At low-loss energies (~5–50 eV), absorption features can become obscured by bulk plasmons, where the real part of the dielectric function is negative or small. These bulk plasmon peaks are generated when a specimen's free electron density is not polarizable and fails to screen the incoming electron beam's perturbation. As a result, the valence or free electrons within the specimen oscillate.

For excitations outside of the bulk plasmon's energetic range, and at ultralow-loss and core-loss energies, the loss function is only dependent on the dielectric function's imaginary component. As a result, an EEL spectrum is almost 1:1 comparable to infrared or X-ray absorption spectra (XAS). The loss function in these cases is represented by $\epsilon_2$ for core-loss excitations because $\epsilon_1 \to 1$ and $\epsilon_2 \ll 1$:

$$\text{Im}\left[-\frac{1}{\epsilon(\mathbf{q},\omega)}\right] = \frac{\epsilon_2}{\epsilon_1^2 + \epsilon_2^2} \approx \epsilon_2(\mathbf{q},\omega). \tag{8}$$

The Drude model in Eq. (1) can be used to directly predict the loss function and the bulk plasmon energy ($E_p$) through the relation to its oscillation frequency ($\omega_p$) and the free electron model:

$$E_p = \hbar\omega_p = \frac{h}{2\pi}\sqrt{\frac{ne^2}{m_e\epsilon_0}}. \tag{9}$$

Notably, the Drude and free electron models assume electrons move freely in the specimen, which is a relatively accurate approximation for metals. The electron density ($n$) largely dictates a specimen's $E_p$, while the effective mass ($m_e$) varies slightly depending on the specimen. Physical constants are Planck's constant ($h$), the elementary charge ($e$), and the permittivity of free space ($\varepsilon_0$).

The free electron model in Eq. (9) accurately captures the bulk plasmon energy in metals, where the absence of a bandgap permits conduction of the free electron gas.[66] In insulators, however, the free electron model breaks down because the loss function and $E_p$ in Eqs. (7)–(9) are complicated by dielectric screening, excitonic effects, and interband transitions. **Figure 2** depicts the complex dielectric function, loss function, and free electron model-calculated plasmon energy for both a metal (Al) and insulator (hBN).[67,68] The loss function was calculated *ab initio* using linearized time-dependent DFT (TDDFT).[69]

**Figure 2(b)** depicts the complex interpretation of the loss function for semiconducting and insulating specimens. Most notably, the Drude model fails to predict the loss function for insulating hBN, whereas an *ab initio* approach using TDDFT produces a far more accurate result. The Drude model calculation was performed using the electron density for a standard Al and hBN unit cell. Additionally, the *ab initio* approach enables a direct prediction of the loss function's momentum dependence, as approximated by Eqs. (2) and (5) depending on the loss region. We now utilize the remainder of Sec. III to describe the current methods for calculating a specimen's loss function beyond the Drude model and indicate which approach works best for a given application.

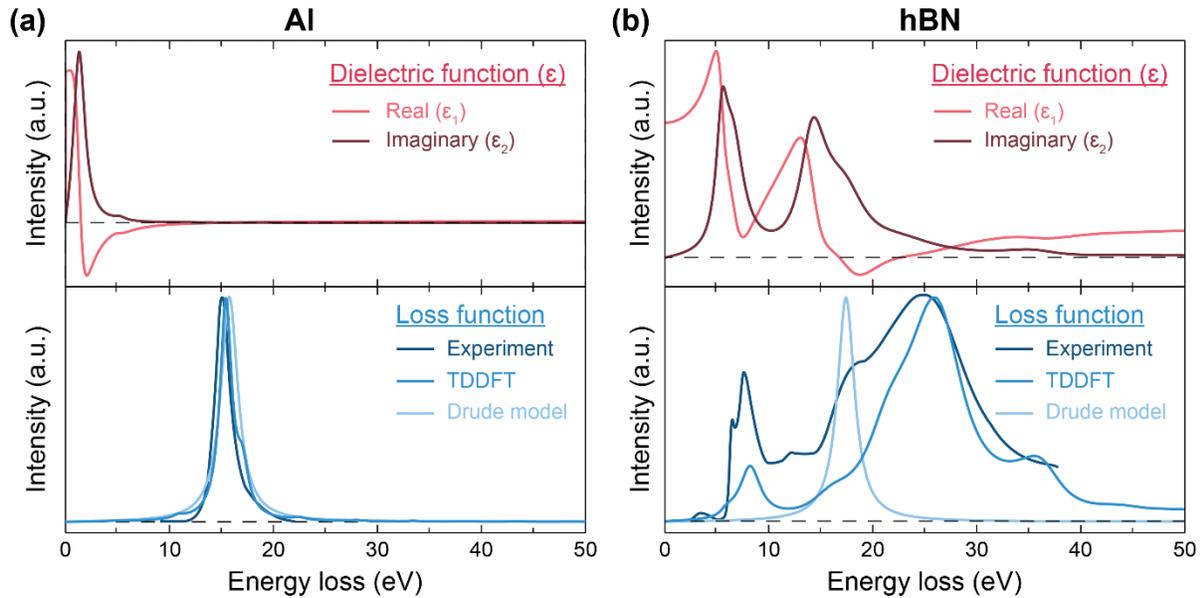

**Figure 2. Comparing loss functions for a metal and an insulator.** TDDFT-calculated dielectric function (top) is and loss function (bottom) for (a) aluminum (Al) and (b) hexagonal boron nitride (hBN). The loss functions for each material are derived from first-principles TDDFT, the Drude model, and experiment. Al experiment adapted with permission from Eswara Moorthy and Howe, J. Appl. Phys. **110**, 043515 (2011).[67] Copyright 2011 AIP Publishing LLC. hBN experiment adapted with permission from Nerl *et al.*, npj 2D Mater. Appl. **8**, 68 (2024).[68] Copyright 2024 Authors, licensed under a Creative Commons Attribution (CC BY) license.

### 1. *Ab Initio* EELS Calculations

*Ab initio* calculations, such as DFT, simulate electronic structure entirely from first principles. For a material, it predicts properties such as the valence band and select conduction band energies across the Brillouin zone, enabling downstream calculations of varying complexity (e.g., calculating an EELS loss function). By definition, *ab initio* methods rely solely on fundamental physical constants and atomic configurations without requiring measurements of empirical parameters.

**DFT.** In DFT, a material's or molecule's properties—such as atomic structure and bonding—are used to solve the Schrödinger-like equation. For example, a generic form of the non-interacting Schrödinger equation of a fictitious system is expressed by Eq. (10).[70] DFT's key innovation is that it avoids the intractable many-body wavefunction; instead, it approximates the system's orbital energies $\epsilon_i$ for all occupied orbitals $\psi_i(\mathbf{r})$, derived from the atoms' effective potential $V_{eff}(\mathbf{r})$. The electron density approximation dramatically reduces complexity and computational expense. This is achieved by solving Kohn–Sham equations,[71] which replace interacting electrons with particles moving in $V_{eff}(\mathbf{r})$ to a three-dimensional position in real space and each $i^{th}$ orbital:

$$\widehat{H}\psi_i(\mathbf{r}) = \left(-\frac{\hbar^2}{2m_e}\nabla^2 + V_{eff}(\mathbf{r})\right)\psi_i(\mathbf{r}) = \epsilon_i\psi_i(\mathbf{r}). \tag{10}$$

Here, the Hamiltonian $\widehat{H}$ is an operator applied to determine the system's total energy $\epsilon_i$. Pseudopotentials for each atom in a DFT calculation are often leveraged to approximate the effects of core electrons on the total wavefunction $\psi_i(\mathbf{r})$, further reducing computational cost. Using input pseudopotentials and crystalline properties, the ground-state energy and electron density $\rho_i(\mathbf{r})$ are determined. This determination is performed self-consistently through iterative calculations until convergence is achieved, where all occupied orbitals $\psi_i(\mathbf{r})$ are summed across all states as:

$$\rho_i(\mathbf{r}) = \sum_i^{\text{occupied}} |\psi_i(\mathbf{r})|^2. \tag{11}$$

**Low-Loss EELS.** The DFT framework can be expanded to TDDFT to simulate optical absorption and low-loss EEL spectra.[69] To do so, the density matrix $\rho_i(\mathbf{r})$, representing the specimen's dielectric response, needs to be inverted through a set of linear equations to determine its energy over time. The time-dependent Schrödinger equation can be evaluated by TDDFT, for example, by leveraging the Liouville von–Neumann equation with the Hamiltonian $\widehat{H}(t)$:

$$i\hbar\frac{d\rho(t)}{dt} = \left[\widehat{H}(t), \hat{\rho}(t)\right] \tag{12}$$

where the commutator $\left[\widehat{H}(t), \hat{\rho}(t)\right] = \widehat{H}(t)\hat{\rho}(t) - \hat{\rho}(t)\widehat{H}(t)$ describes how the state's populations (diagonal terms) and coherences (off-diagonal terms) evolve in time $t$. A Liouvillian superoperator $\check{L}$ succinctly expresses this evolution as $\check{L} \cdot \hat{\rho}(t) = \left[\widehat{H}(t), \hat{\rho}(t)\right]$. Using linear-response TDDFT, this is further reformulated by separating the ground-state ($\widehat{H}^0$ and $\hat{\rho}^0$) and first-order perturbed contributions.[69,72,73] For a perturbation at a specific momentum transfer direction **q**, the action of the Liouvillian superoperator becomes:

$$\check{L} \cdot \hat{\rho}'_\mathbf{q} = \left[\widehat{H}^0, \hat{\rho}'_\mathbf{q}\right] + \left[\hat{V}'_{HXC,\mathbf{q}}[\hat{\rho}'_\mathbf{q}], \hat{\rho}^0\right]. \tag{13}$$

Here, the first commutator describes the independent single-particle excitations in a non-interacting system, which is just the Kohn–Sham response using Eqs. (11) and (12). The second commutator incorporates many-body interactions via linearized Hartree-plus-exchange correlation (HXC) potential $\hat{V}'_{HXC,\mathbf{q}}[\hat{\rho}'_\mathbf{q}]$ that depends on the perturbed density $\hat{\rho}'_\mathbf{q}$.

To obtain the system's EEL spectrum and frequency-dependent response, the Liouville von–Neumann equation in Eq. (12) can be Fourier-transformed in time, yielding the linearized quantum Liouville equation:

$$(\omega - \check{L}) \cdot \hat{\rho}'_j(\omega, \mathbf{q}) = [\hat{r}_i, \hat{\rho}'_j(\omega, \mathbf{q})] \tag{14}$$

where $\hat{r}_i$ is the i$^{th}$ component of the dipole operator and $\hat{\rho}'_j(\omega, \mathbf{q}) = \hat{\rho}_j(\omega, \mathbf{q}) - \hat{\rho}^0$ is the response density matrix perturbed by an electric field polarized along the j$^{th}$ cartesian axis at a specific frequency $\omega$ and momentum $\mathbf{q}$.

Inverting the $(\omega - \check{L})$ Liouvillian matrix directly at each energy-loss frequency would be computationally expensive due to the matrix's massive size for any atomic system. To address this, a Liouville–Lanczos (LL) approach can be introduced.[69,74,75] This LL method employs a Lanczos bi-orthogonalization algorithm, which constructs a small matrix representation of the Liouvillian using one off-diagonal element. This transformation approximates the matrix inversion for all frequencies using a single, frequency-independent recursion, which is slightly less accurate due to the spectral interpolation. Yet, calculating the loss function is now realistic—achievable within minutes to hours on a standard high-performance computer for a typical unit cell. Essentially, the bulk of the numerical work (the Lanczos recursion) is performed only once, and the response at each frequency is obtained by a fast, low-dimensional matrix inversion.

A package within the Quantum ESPRESSO (QE) distribution called turboTDDFT has developed a LL solver to calculate the loss function and EEL spectra through turboEELS.[69,76,77] The turboEELS code has been utilized to compute temperature-dependence of the Si loss function,[78] momentum effects on EELS of TiN,[79] and the loss function of plasmonic $HfTa_4C_5$ carbides.[80] Other *ab initio* quantum calculation packages have similarly tackled this difficult treatment of the EELS loss function including VASP,[81] BerkeleyGW,[82,83] Yambo,[84] WIEN2k,[85,86] and exciting.[87]

**Core-Loss EELS.** Beyond *ab initio* DFT calculations of bulk plasmons, core-loss EEL spectra can also be simulated. The key additional requirement for core-loss simulations is the inclusion of the core hole, which calculates the exciton formed by transitioning a tightly bound core-level electron into a material's conduction band.[55] The highly localized electron-hole pair creates strong, many-body interactions that significantly modify both the electronic structure and the fine structure of the specimen. This is why a core-loss spectrum does not simply mirror the specimen's unoccupied DOS or conduction band.

The electron-hole interaction includes both *direct* and *exchange* interactions. The direct interaction, which describes the screened Coulomb attraction between the electron and hole, dramatically influences the exciton binding. The screening is from the material's dielectric response.[88] The exchange interaction, on the other hand, is repulsive and involves the bare

Coulomb interaction.[89] The exchange interaction is responsible for the splitting of spin-singlet and spin-triplet excitations.

Simulations of core-loss EELS typically begin with a ground-state DFT calculation to determine the underlying electronic structure, followed by a treatment of the core hole using either a supercell approach[90,91] or the Bethe–Salpeter equation (BSE).[88,89] In the supercell approach, a core hole is modeled within a large supercell (e.g., 4x4x4 unit cells) with a core-hole pseudopotential constrained on the central absorbing atom, while the surrounding lattice remains unperturbed. This method treats the core hole self-consistently and does not explicitly include the exciton's electron–hole binding interactions.

In contrast, the BSE directly solves for the exciton's direct and exchange interactions. While the BSE is mathematically more complex, it is made more computationally efficient by modeling a single unit cell and utilizing a Haydock recursion method, much like the LL solver used to calculate the low-loss spectrum as described above. The Haydock recursion avoids explicit diagonalization of the BSE Hamiltonian by generating a tridiagonal matrix through Lanczos iteration. This allows the computation of the imaginary component of the dielectric function $\epsilon_2(\mathbf{q}, \omega)$ in Eq. (8), or the absorption spectrum, at a dramatically reduced computational cost.

OCEAN is an example software package commonly used to simulate core-level spectra (e.g., EELS and XAS) and solves the BSE using either a Haydock recursion or generalized minimal residual method.[88] "Exciting" is another widely used BSE code for simulating core-level spectra.[87] Meanwhile, QE/XSpectra[90,92] and VASP[93] implement core-hole supercell methods, either through ΔSCF calculations or coupled with BSE post-processing. The FEFF code employs real-space multiple scattering theory,[94,95] whereas CTM4XAS utilizes charge transfer multiplet theory to model the multiplet and crystal field effects, providing efficient, non-DFT-based alternatives.[96] A list of example low- and core-loss theory packages, along with their underlying theoretical frameworks as of this writing, is provided in **Table I**.

2. Numerical Models of Low-Loss EELS using Maxwell's Equations

In contrast to *ab initio* methods that calculate the loss function and a complex dielectric response from quantum mechanics, physics-based numerical models solve Maxwell's equations using classical electrodynamics. Approaches such as FDTD and FEM models are utilized to simulate EEL spectra at interfaces and in nanoscale real-space dimensions. Because the entire three-dimensional system is directly modeled, these approaches are particularly effective at capturing interfacial quasiparticle excitations such as surface plasmons, excitons, and polaritons. It is more difficult for *ab initio* methods to simulate modes at nanoscale features as they typically simulate bulk systems using repeated unit cells. Previous studies have calculated localized surface plasmon modes of nanostructured Al, Ag, and Au or $MoS_2$/Au localized surface plasmons and interfacial excitons.[97–101]

Numerical models compute the energy loss probability, $\Gamma_{EELS}(\omega)$, by solving Maxwell's equations in the frequency domain and applying classical dielectric formalism.[97,102] The loss probability is calculated using the velocity ($v$), elementary charge ($e$), angular frequency ($\omega$), and direction ($z$) of the TEM's quickly moving electron beam, modeled as a line current density. The z-component of a specimen-induced electric field, $E_z^{ind}(z, \omega)$, largely dictates the energy loss probability $\Gamma_{EELS}(\omega)$ as:

$$\Gamma_{EELS}(\omega) = \frac{ve}{h\omega} \int dz \, \text{Re}\left[e^{-\frac{i\omega z}{v}} E_z^{ind}(z, \omega)\right]. \tag{15}$$

The real component of the induced field $E_z^{ind}(z, \omega)$ is essential, as it is the field that performs work on the moving electron probe, described by classical Lorentz forces in real space. The frequency domain solution of Maxwell's equations gives $E_z^{ind}(z, \omega)$ and thus $\Gamma_{EELS}(\omega)$. Specifically, the total electric field in the specimen $\mathbf{E}(\mathbf{r}, \omega)$ is given by the inhomogeneous wave equation:

$$\nabla \times [\nabla \times \mathbf{E}(\mathbf{r}, \omega)] = \frac{\omega^2}{c^2} \varepsilon_0 \varepsilon_\infty \mathbf{E}(\mathbf{r}, \omega) + i\mu_0 \omega \mathbf{J}_d(\mathbf{r}, \omega). \tag{16}$$

Here, the speed of light ($c$), angular frequency ($\omega$), permittivity of free space ($\varepsilon_0$), dielectric response at high frequencies ($\varepsilon_\infty$), and the permeability of free space ($\mu_0$) are considered. The nonlocal current ($\mathbf{J}_d(\mathbf{r}, \omega)$) has been described in detail.[97] Because the total electric field can be related to the specimen-induced electric field, while considering the external (incident) electric field of the electron beam as $\mathbf{E}(\mathbf{r}, \omega) = \mathbf{E}_0(\mathbf{r}, \omega) + \mathbf{E}^{ind}(\mathbf{r}, \omega)$, the energy loss probability can be calculated.

In practice, software packages such as the COMSOL Multiphysics RF Module can be used to calculate the energy loss probability in an arbitrary specimen geometry.[103] Other packages, such as Lumerical FDTD Solutions with the MATLAB toolbox for the simulation of metallic nanoparticles using a boundary element method (MNPBEM) approach, can also be utilized.[104–106]

## B. Simulating Transient and Photoexcited EELS

Advancements in *ab initio* simulations of ultrafast EELS are essential to understand charge carrier and lattice temperature dynamics at ultrafast timescales. An *ab initio* analysis of ultrafast core-loss EELS (discussed in Sec. IV) successfully modeled excited-state heating as measured by the C K edge of graphite (**Figs. 3(a)** and **3(b)**).[95] The temperature-dependent lattice response was simulated using molecular dynamics (MD) calculations that fed into FEFF calculations of the C K edge. Similar computational methods have been applied in excited-state X-ray spectroscopy, where carrier distributions and lattice heat were modeled in both steady-state and femtosecond regimes.[107–109] These approaches often introduce carriers into the DFT-calculated electronic structure, a method known as state blocking, which simulates photoexcited carrier effects on EEL spectra by constraining electrons and holes in respective bands outside equilibrium. On femtosecond timescales, carrier thermalization was captured in ZnTe using the Te $N_{4,5}$ edges after photoexcitation (**Figs. 3(c)** and **3(d)**).[108] This approach used in transient X-ray absorption spectroscopy can be applied to ultrafast core-loss EELS calculations that simulate carrier dynamics.

Density functional perturbation theory (DFPT) simulates a material's response to small perturbations such as atomic displacements, external fields, or applied strain. The material's excited-state atomic lattice and electronic structure are determined self-consistently by evaluating the perturbed total energy, interatomic forces, and electron–phonon coupling. DFPT calculations have been employed to investigate non-equilibrium dynamics, including phonon frequency shifts,[75,110,111] dielectric function modifications,[112,113] thermal expansion coefficients,[114,115] and lattice strain.[113]

**TABLE I. Available Software Distributions for Simulating EEL Spectra.**

| Software Distribution | Approach | Package / Module; Method; Solver (if applicable) | Loss Region |
|---|---|---|---|
| QE[69,76,77,92] | *Ab initio* | turboEELS and XSpectra; TDDFT and ΔSCF core-hole supercells; Liouville–Lanczos Solver; *constrained DFPT for excited-state calculations* | Low and Core |
| VASP[81,93] | *Ab initio* | Linear response TDDFT; ΔSCF core-hole supercells and optional BSE post-processing | Low and Core |
| WIEN2k[85,93] | *Ab initio* | OPTIC and TELNES3; TDDFT and optical response based on the linearized augmented plane-wave algorithm | Low and Core |
| Exciting[87] | *Ab initio* | All-electron linearized augmented-plane-wave code; BSE, TDDFT, and MD | Low and Core |
| FEFF[94] | *Ab initio* | Green's functions and real-space multiple scattering theory | Low and Core |
| BerkeleyGW[82] | *Ab initio* | GW + BSE; dielectric function calculations | Low |
| Yambo[84] | *Ab initio* | GW/BSE/TDDFT; real and reciprocal space solvers | Low |
| COMSOL Multiphysics[103] | Numerical | RF Module; Numerical electrodynamics FEM | Low |
| Lumerical[104-106] | Numerical | Numerical electrodynamics; FDTD and MODE solvers | Low |
| OCEAN[88] | *Ab initio* | DFT and BSE; Haydock solver; *constrained BSE for excited-state calculations* | Core |
| CTM4XAS[96] | Semi-empirical | Charge multiplet theory | Core |

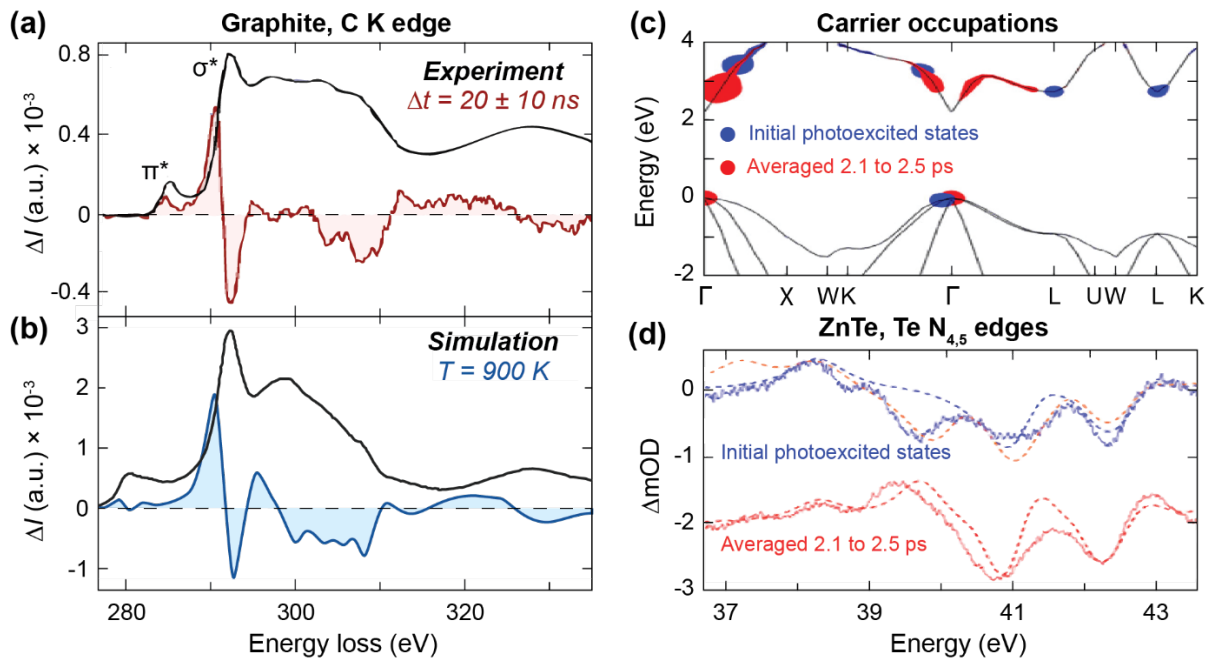

**Figure 3. Simulating heat and carrier effects in excited-state core-level spectra.** (a) Static ($<t_0$, black) and transient differential ($\Delta t = 20 \pm 10$ ns, red) C K-edge spectra at after 532 nm laser excitation of graphite. (b) Simulated static (black) and transient differential (blue) spectra from molecular dynamics (MD) simulations at room temperature and $\Delta$900 K, respectively. Adapted with permission from van der Veen *et al.*, Struct. Dyn. **2**, 024302 (2015).[95] Copyright 2015 Authors, licensed under a CC BY license. (c) Carrier occupations in the band structure of ZnTe are inferred by matching experiment to theory with the colors matching their corresponding spectrum in (d). (d) Experimental (solid) and theoretical (dashed) differential spectra for ZnTe at the initial (blue) and thermalized (red) photoexcited states. The best match with the initial state experiment is the simulation of ZnTe with carriers in midgap defect states (orange dashes). Reproduced with permission from Liu *et al.*, J. Phys. Chem. Lett. **14**, 2106–2111 (2023).[108] Copyright 2023 American Chemical Society.

Computational advances have extended DFPT to self-consistent calculations of excited-state carriers using constrained DFPT (cDFPT). Unlike conventional constrained DFT, which imposes physical constraints on the ground state, cDFPT introduces constraints to the perturbed system. For example, by constraining electrons and holes in a semiconductor's respective conduction and valence bands, the carrier-modified dielectric function can be simulated. These constrained carrier calculations are achieved by modeling two constrained chemical potentials in the material's unit cell—one potential each for electrons and holes.[116] When combined with tools such as turboEELS in QE, this approach provides a pathway to simulate transient carrier distributions measurable by ultrafast EELS.

Looking ahead, future developments will likely integrate more sophisticated theoretical methods with cDFPT to improve the accuracy and predictive power of excited-state simulations. Recent and promising extensions to open-source code distributions incorporate hybrid functionals, *ab initio* MD, and many-body perturbation theory (e.g., GW). These combined approaches will enable more accurate modeling of photoexcited dynamics associated with midgap states, phonon dispersions, and bandgap renormalization, especially for semiconductors. Such advancements will enhance the reliability of EELS simulations under non-equilibrium conditions and expand their applicability to more complex materials systems.

## C. Applying Denoising and Machine Learning to Time-Resolved and Ultrafast EELS

Machine and deep learning are numerical tools already being implemented to better interpret and resolve EEL spectra[78,117–120] and STEM imaging.[117,121–123] With these computational approaches, new microscopic and spectroscopic limits can be met. They will be particularly valuable for ultrafast EELS experiments, where signal quality is inherently constrained by the low beam currents associated with few-electron UEM pulses.[30] Among the most active areas of development are ZLP fitting and denoising of low signal-to-noise-ratio (SNR) spectra. Machine learning models ultimately expand the accessible energy-loss range to low-loss signals convoluted with the ZLP and high energy core-loss signals with fewer detected scattering events.

The use of supervised and unsupervised machine learning models to fit and subtract the ZLP from low-loss EEL spectra is pivotal for time-resolved and ultrafast EELS, where the Boersch effect's energy broadening of the ZLP bandwidth limits the technique's energy resolution. This approach has already enabled the accurate retrieval of low-energy excitations in steady-state EEL spectra that would otherwise be buried beneath the ZLP tail in non-monochromated microscopes.[124] Specifically, this approach reconstructs a parametrization of the ZLP intensity based on multi-layer feed-forward artificial neural networks to isolate the ZLP from scattering events. Excitations that now become readily imaged with STEM-EELS ZLP subtraction include semiconductors' bandgaps,[125,126] surface-induced energy-gain features,[127] and localized surface plasmon modes.[128]

Advanced denoising and deep learning frameworks based on physics-informed and data-driven models have also improved the interpretability of EEL spectra.[120,129,130] However, denoising-induced artifacts remain a significant challenge, and an understanding of the ground truth is essential. Recent work in denoising has enhanced the ability to resolve subtle inelastic vibrational modes of hBN with direct comparison to ground truth spectra with STEM-EELS (**Figs. 4(a)** and **4(b)**).[118,131] Notably, convolutional architectures such as U-Nets and Variational Networks have been trained on synthetic low-SNR datasets to learn noise characteristics while preserving spectral integrity. These models demonstrate superior performance over traditional filters by maintaining peak fidelity and avoiding the blurring of fine structures. Similarly, a self-supervised autoencoder was used to denoise differential EEL spectra for core-loss thermometry imaging in crystalline Si (**Figs. 4(c)–4(e)**).[78] These results ultimately enhance the accuracy of existing STEM-EELS capabilities while dramatically reducing the required acquisition time by improving the SNR.

In the context of time-resolved and ultrafast EELS, machine- and deep-learning as well as self-supervised denoising, will enable the acquisition of time-resolved spectral changes that would otherwise be obscured by shot noise. Each spectral feature is nearly obscured by the few-electron pulses necessary for temporal and spatial coherence in UEM. These new computational methods will allow faster acquisition and imaging of dynamic processes such as carrier transport, temperature-dependent plasmon energy shifts, intraband transitions, and core-level excitations.

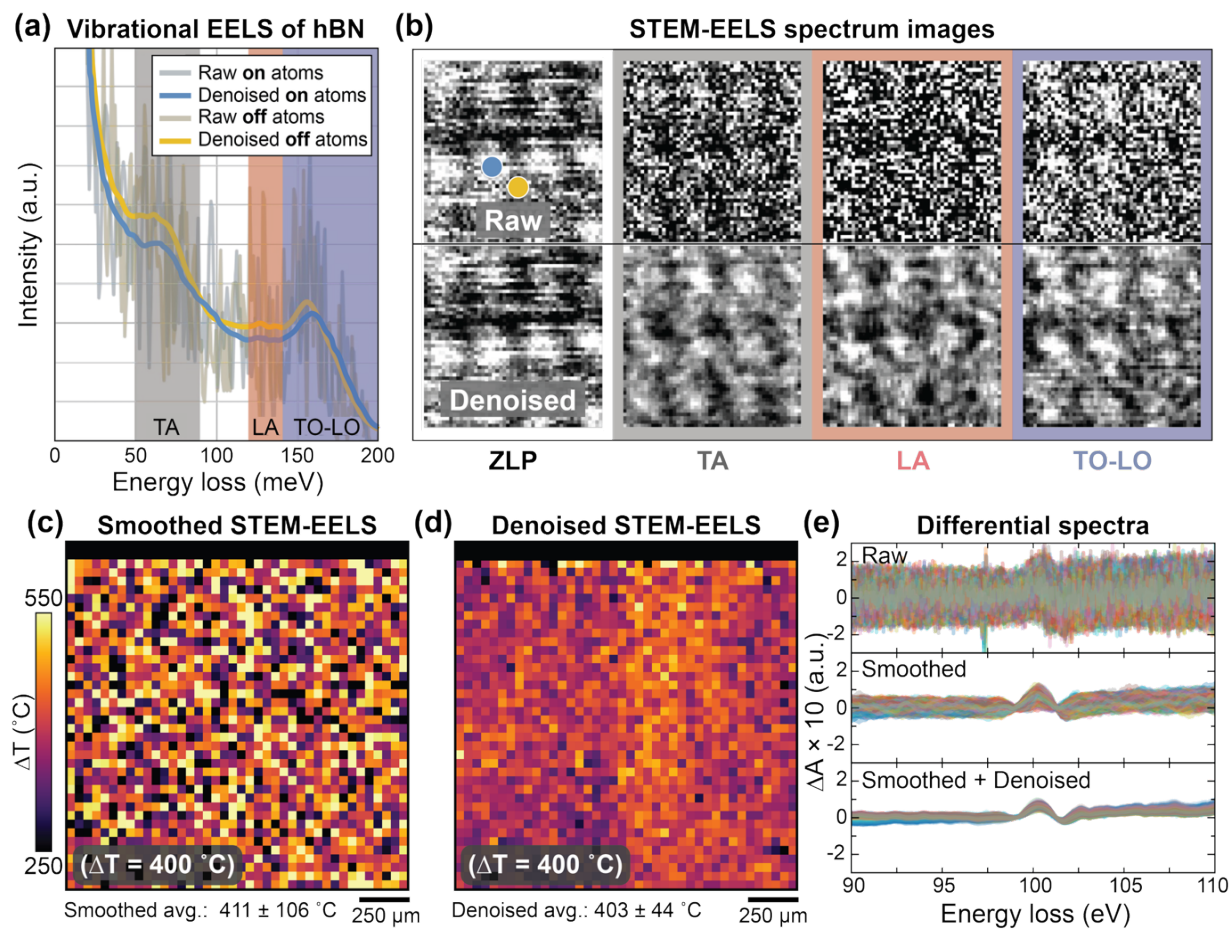

**Figure 4. Denoising EEL spectra to optimize EELS spectrum imaging.** (a) Raw (transparent) and denoised (opaque) EEL spectra on (yellow) and off (blue) an atomic column of hBN from pixel locations marked in (b). Energy windows: TA (50–90 meV), LA (120–140 meV), and TO–LO (140–200 meV). (b) STEM-EELS spectrum images of vibrational modes from raw (top) and denoised (bottom) data. Adapted with permission from Wang *et al.*, arXiv:2505.14032 [physics.ins–det; cond–mat.mrl–sci] (2025).[118] Copyright 2025 Authors, licensed under a CC BY license. (c) Core-loss thermometry image of Si at $\Delta T = 400$°C. Differential spectra were aligned, smoothed, and fitted for each pixel to extract the Si lattice temperature. The amplitude of the differential peak was converted to temperature using linear regression. (d) Core-loss thermometry image (c) with the differential spectrum at each pixel denoised. (e) Raw, smoothed, and denoised spectra from all pixels in the spectrum images. Reproduced with permission from Palmer *et al.*, ACS Phys. Chem. Au (2025). DOI: 10.1021/acsphyschemau.5c00044.[78] Copyright 2025 Authors. Published by American Chemical Society under a CC BY license.

## IV. ULTRAFAST EELS TECHNIQUES BEYOND PINEM

### A. Charge Dynamics Electron Microscopy

Charge dynamics electron microscopy (CDEM) exploits the interaction between an electron probe and terahertz (THz) near-fields with picosecond lifetimes. These THz fields are generated by moving charges within and around a specimen, where carriers around a specimen originate from a photoemitted electron plasma. As such, CDEM enables nanoscale imaging of charge generation, separation, and transport with femtosecond to picosecond time resolution.

**Figure 5(a)** illustrates how CDEM maps the motion of a photoemitted electron plasma evolving from a Cu rod by measuring how the plasma energetically impacts the electron probe.[132] The measurements reveal that the electron beam undergoes both acceleration and deceleration during its passage through or near the plasma's evolving electron cloud. Initially, acceleration dominates due to the motion of the emitted electrons along transverse directions to the electron beam. The electron beam not only gains energy but also shows spectral broadening, which is strongest within

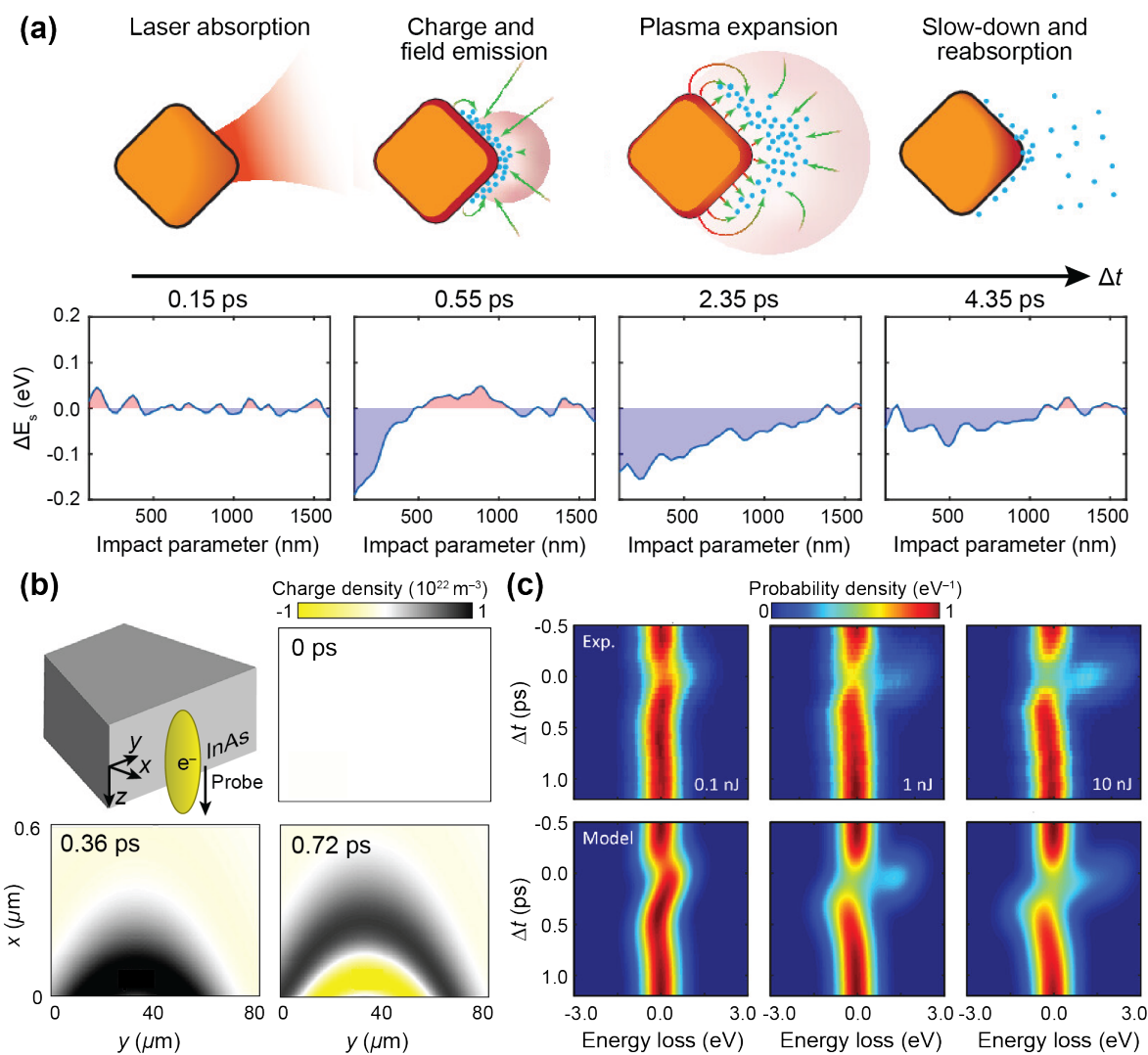

**Figure 5. Charge dynamics electron microscopy (CDEM).** (a) Evolution of the photoemitted electron cloud's plasma around a Cu rod (top) and the corresponding change in the ultrafast electron probe energy due to THz fields at fixed distances from the Cu rod (bottom). Adapted with permission from Madan *et al.*, ACS Nano **17**, 3657–3665 (2023).[132] Copyright 2023 Authors, licensed under a CC BY license. (b) Spatiotemporal evolution of the charge distribution in an InAs crystal. (c) Experimental (top) and simulated (bottom) ultrafast EEL spectra, measured near the InAs surface using CDEM, for varying pump pulse energies. Adapted with permission from Yannai *et al.*, ACS Nano **17**, 3645−3656 (2023).[133] Copyright 2023 American Chemical Society.

100 femtoseconds after laser excitation. As the plasma evolves, beginning to expand and reabsorb to the Cu surface, the electron probe decelerates. Through experimental data and physics-based numerical simulations, four distinct stages in the plasma's evolution are identified: laser irradiation, photoemission and THz field generation, plasma expansion, and surface reabsorption.

Beyond metallic specimens, CDEM also enables imaging of carrier dynamics in semiconductors.[133] The interaction is mediated by the THz field arising from the photo-Dember effect, which is where asymmetric electron–hole diffusion leads to dipolar field generation. By adjusting the electron beam position and time delay, the THz near-field is retrieved from the measured electron energy shifts. This approach not only spatially resolves the near-field but also enables the reconstruction of charge distributions. The differences in mobility between electrons and holes manifest as distinct spatiotemporal behaviors, as revealed by reconstructed charge density maps in InAs (**Fig. 5(b)**). Simulations using a hydrodynamic model of the photo-Dember effect predict transient currents and corresponding energy shifts of the probe electrons, showing strong agreement with the experimental results (**Fig. 5(c)**).

CDEM offers a powerful platform for exploring carrier dynamics in semiconductors, nanostructures, and optoelectronic devices. Its unique ability to capture transient THz fields associated with charge carrier motion provides critical insights for optimizing the performance of electronic components. By directly linking nanoscale material structure to functional electronic behavior, CDEM significantly advances our understanding of ultrafast charge carrier dynamics.

## B. Ultrafast Low-Loss EELS

As discussed in Sec. II, low-loss inelastic scattering probes inter/intraband transitions, as well as the excitation of quasiparticles such as phonons and plasmons. Although plasmon and near-field dynamics are widely studied in ultrafast low-loss EELS via PINEM,[45–50] this section specifically highlights photoexcited plasmon dynamics and their interplay with a material's electronic and structural properties.

Ultrafast low-loss EELS was first employed to map the chemical bonding dynamics in graphite, revealing how these dynamics modify the spectral features of the π, surface, and bulk σ + π plasmon excitations.[134,135] Graphite serves as an ideal model system for investigating the interplay between structural and electronic changes induced by laser irradiation. As a result, graphite has been extensively studied using other ultrafast and time-resolved EELS techniques, further discussed in Secs. IV. C and D. Experimental results show that graphite's bulk plasmon energy initially blueshifts, followed by a redshift at longer time delays (**Fig. 6(a)**). This behavior corresponds to interlayer contraction and subsequent expansion, as confirmed by ultrafast electron diffraction (UED) and DFT-based charge-density calculations. These structural changes were correlated with phase transitions from 2D ($sp^2$, graphene-like) to 3D ($sp^3$, diamond-like) electronic structures, highlighting the potential of ultrafast low-loss EELS for probing complex phenomena.

The photoinduced structural dynamics and resulting plasmonic response were further validated through *ab initio* DFT calculations using the MINDLab software.[136]

Ultrafast low-loss EELS was also employed to investigate plasmonic nanomaterials such as gold nanotriangles (AuNTs), enabling direct detection of both localized surface plasmons (LSPs) and bulk plasmons, which is difficult to achieve with purely optical methods.[137] The time-dependent evolution of the bulk plasmon and LSP peak energies in AuNTs (**Fig. 6(b)**) reveals opposite changes in the photoexcited electron density in the AuNTs' bulk and surface regions. This underscores the ability of ultrafast low-loss EELS to measure spatially dependent photoinduced effects in nanoparticles. Additionally, the relaxation times associated with electron–phonon and phonon–phonon interactions extracted from the EELS measurements were consistent with those observed in optical studies, further validating the technique's accuracy in probing ultrafast processes in nanoparticles. Analysis using a two-temperature model estimated the relaxation of electronic and lattice temperatures following laser irradiation, and these predictions were confirmed by the energy shifts and broadening of the plasmon peaks.

Ultrafast low-loss EELS has also been applied to other nanoscale systems such as carbon nanotubes (CNTs), gaining insight into their unique low-dimensional properties. Vanacore *et al.*

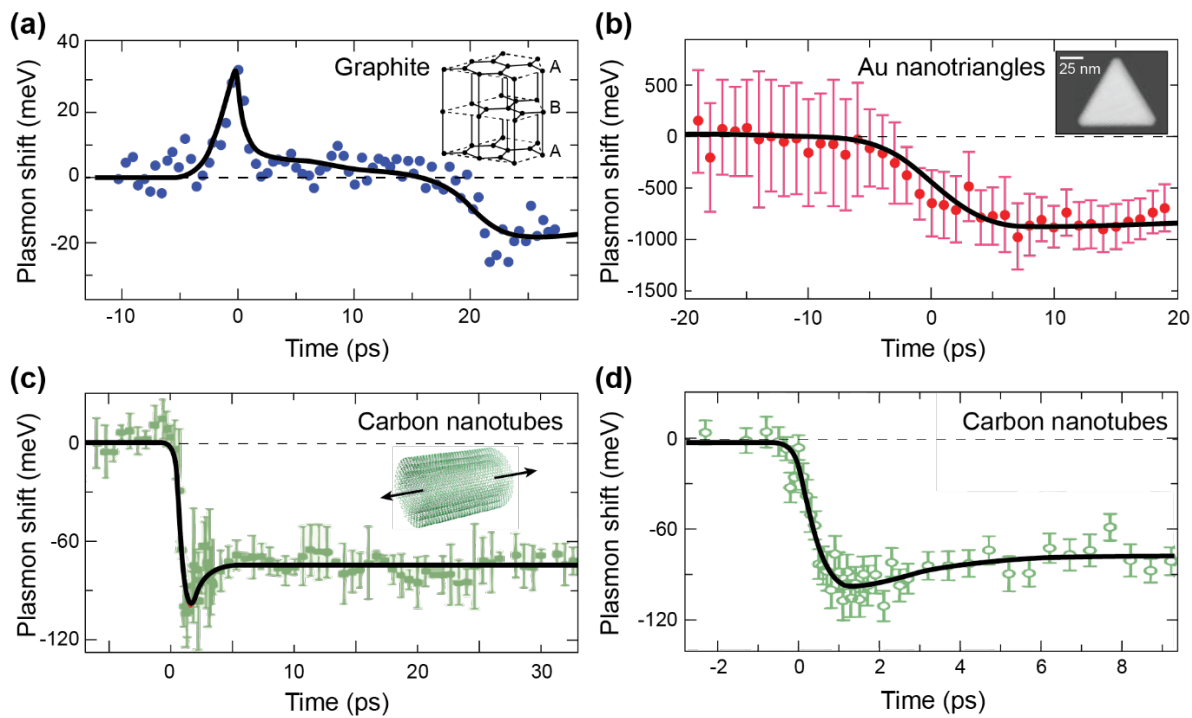

**Figure 6. Ultrafast low-loss EELS.** Time-dependent energy shifts of the bulk plasmon for (a) graphite, (b) gold (Au) nanotriangles, (c–d) and carbon nanotubes. Each inset depicts either the atomic structure or dark-field TEM image. (a) Adapted with permission from Carbone *et al.*, Science **325**, 181−184 (2009).[134] Copyright 2009 American Association for the Advancement of Science. (b) Adapted with permission from Kuwahara et al., Appl. Phys. Lett. **121**, 143503 (2022).[137] Copyright 2022 Authors, licensed under a CC BY license. (c) Adapted with permission from Vanacore *et al.*, ACS Nano **9**, 1721−1729 (2015).[138] Copyright 2015 American Chemical Society. (d) Adapted with permission from Zheng et al., Nanoscale Adv. **2**, 2808 (2020).[139] Copyright 2020 Authors, licensed under a CC BY license.

combined UED and ultrafast low-loss EELS to monitor lattice and charge distribution changes in CNTs under laser excitation.[138] UED data indicated that axial (along the tube axis) CNT deformation is dominated by electron-driven processes, as evidenced by a faster initial expansion, while radial deformation is slower and phonon-driven. The charge distribution dynamics were further examined in single-walled CNTs (SWCNTs) using ultrafast low-loss EELS. A redshift of the bulk plasmon (**Fig. 6(c)**) indicates a decreasing in-plane electron density, as predicted by the free-electron model. This suggests that lattice expansion occurs along the nanotube's axis, driven by occupation of antibonding $\pi^*$ state and a resulting decrease in intrawall electron density. *Ab initio* DFT calculations using ABINIT further reinforce this interpretation, showing significant axial C–C bond expansion upon electronic excitation. A separate study by Zheng *et al*. similarly investigated the ultrafast dynamics in SWCNTs (**Fig. 6(d)**).[139] In addition to UED and ultrafast low-loss EELS, optical transient absorption spectroscopy was also utilized to demonstrate how the redshift of the in-plane plasmons influences the ultrafast photoresponse of SWCNTs.

## C. Time- and Momentum-Resolved EELS

One of the key advantages of using electrons as a probe is their ability to provide momentum-resolved information (**Fig. 7(a)**). A recent study investigated the ultrafast dynamics of scattering processes and collective excitations in graphite, focusing on how photoexcitation with different pump wavelengths affects valley-specific charge carrier behavior and plasmon dynamics.[140] UED captured pump energy-dependent phonon population dynamics and lattice distortions, providing insight into graphite's structural changes (**Fig. 7(b)**). Complementing this, time- and momentum-resolved EELS (tr-q-EELS) tracked the evolution of in-plane $\pi$ and bulk $\sigma + \pi$ plasmons, which are sensitive to both electronic and structural changes (**Fig. 7(c)**). Together, these techniques offer a comprehensive view of a material's photoinduced response, allowing electron–phonon interactions to be directly mapped in reciprocal space.

Graphite's low-loss spectrum at $\Gamma_{000}$ blueshifts due to increased photocarrier density following photoexcitation (**Fig. 7(d)**). To resolve how this photocarrier density is localized within graphite's band structure, tr-q-EELS is measured along the $\Gamma \rightarrow M$ direction over 2–8 ps as shown in **Figs. 7(e)** and **7(f)**. The experimental results clearly demonstrate that plasmon responses depend on both momentum transfer and excitation energy. *Ab initio* DFT simulations using the Yambo code accurately reproduce the plasmon dynamics under visible excitation by incorporating effects of thermal lattice expansion and phonon population (**Fig. 7(f)**). However, for near-infrared excitation, the simulations fail to fully capture the observed plasmon dynamics (**Fig. 7(e)**), suggesting the involvement of additional mechanisms such as long-range charge interactions.

The study demonstrates the ability to control the scattering pathways and plasmonic responses in graphite by tuning the excitation energy. It also highlights the synergistic power of combining diffraction and spectroscopy for probing non-equilibrium states in quantum materials. These findings open avenues for the design of future nanoscale devices that leverage valleytronic and plasmonic functionalities, and they provide a methodological framework for studying related dynamics in other strongly corrected systems, including cuprates.

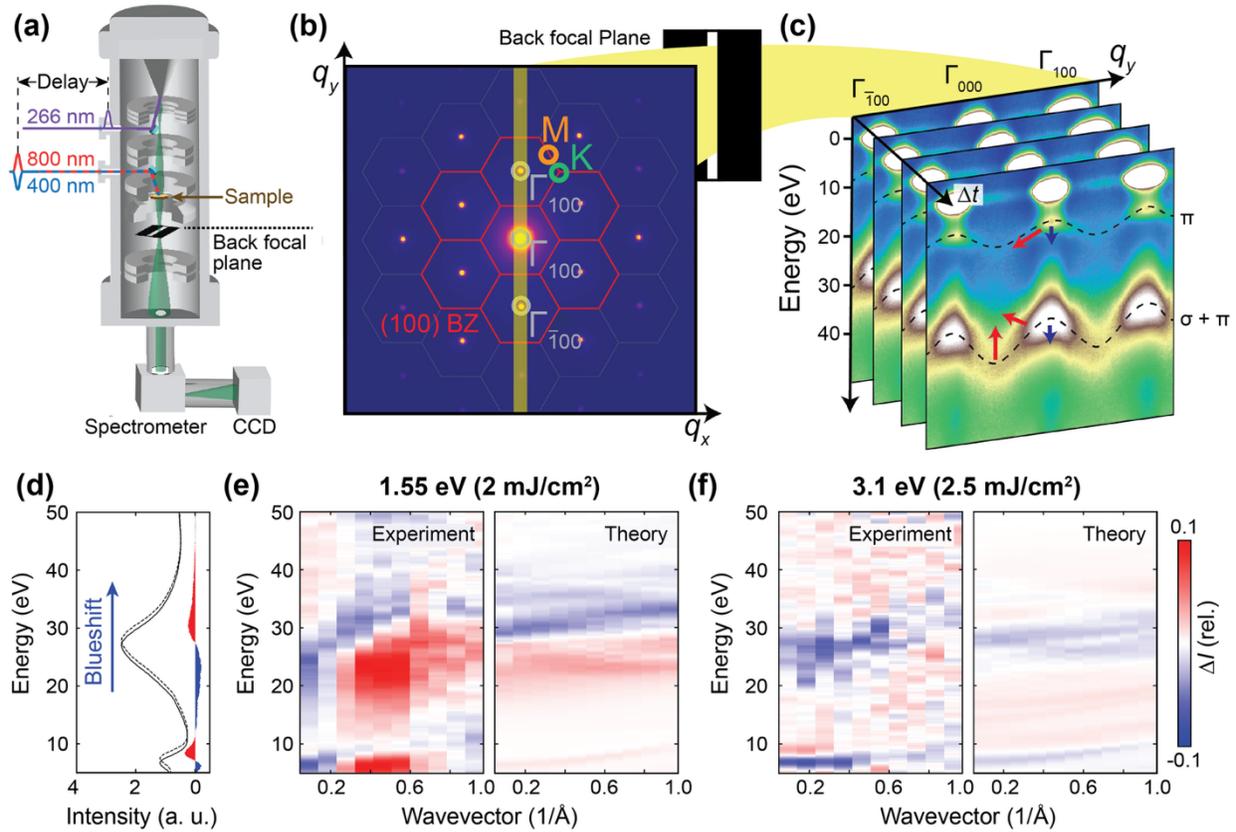

**Figure 7. Time- and momentum-resolved EELS (tr-q-EELS).** (a) Schematic illustration of the tr-q-EELS setup in a UEM where a q-slit at the back focal plane selectively collects scattered electrons along the Γ→M direction. (b) Electron diffraction pattern of graphite, highlighting the (100) Brillouin zone (BZ) in red; the yellow rectangle indicates the Γ→M direction. (c) Tr-q-EELS acquired along (100) BZ indicates the variable intensity of the π and σ + π plasmons resolved in $k$-space. (d) Graphite's EELS response at $Γ_{000}$. The differential EEL spectrum is obtained using the difference between blueshifted (dashed curve) and unperturbed (solid curve) spectra. (e,f) Measured (left) and simulated (right) tr-q-EELS maps along the Γ→M direction for (e) 1.55 eV and (f) 3.1 eV excitation. Measured spectra are averaged between 2 and 8 ps. Adapted with permission from Barantani *et al*., Sci. Adv. **11**, eadu1001 (2025).[140] Copyright 2025 Authors, licensed under a CC BY license.

## D. Ultrafast Core-Loss EELS

Ultrafast X-ray absorption spectroscopy probes element-specific carrier and structural dynamics through the X-ray probe's core-level excitation.[141,142] As discussed in Secs. II and III, ultrafast core-loss EELS can, in principle, access the same dynamics as X-ray absorption under the dipole approximation. However, the nanometer-scale spatial resolution of TEM-EELS allows for localized measurements at features such as junctions, dopants, and defects. Although previous ultrafast core-loss EELS studies have not yet fully exploited the microscope's ultimate spatial resolution, they have demonstrated unique capabilities beyond those of the low-loss EELS techniques described in Sec. IV (A)–(C).

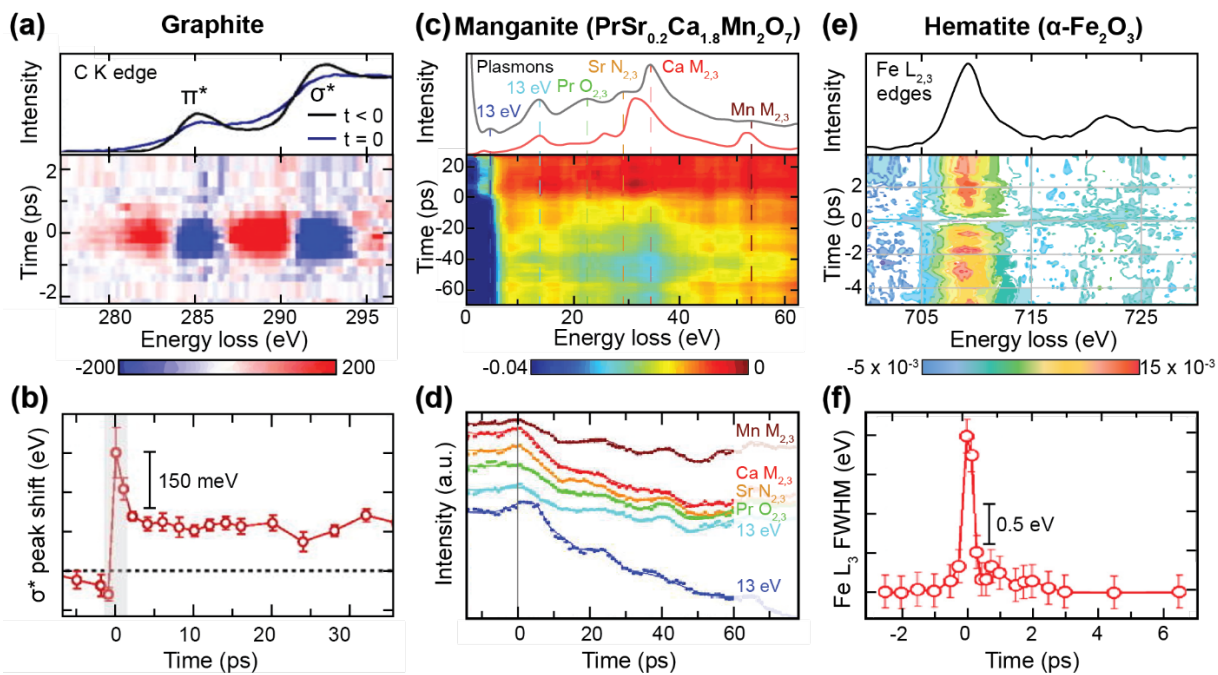

**Figure 8. Ultrafast core-loss EELS.** (a) Ultrafast core-loss EELS of graphite. The electron–photon interaction broadens the spectrum at $\Delta t = 0$ (top) and obscures core-loss dynamics near time zero (bottom). (b) Temporal evolution of the $\sigma^*$ peak. A redshift of the $\sigma^*$ peak is observed, and the gray-shaded area indicates the time window when PINEM occurs. Adapted with permission from van der Veen *et al.*, Struct. Dyn. **2**, 024302 (2015).[95] Copyright 2015 Authors, licensed under a CC BY license. (c) Static experimental (top, gray), simulated (top, red), and transient (bottom) EEL spectra of a $PrSr_{0.2}Ca_{1.8}Mn_2O_7$ film. (d) EELS intensity profile for the plasmon and core-level transitions as a function of time delay. Adapted with permission from Piazza *et al.*, Struct. Dyn. **1**, 014501 (2014).[86] Copyright 2014 Authors, licensed under a Creative Commons Attribution (CC BY) license. (e) Static (top) and transient (bottom) Fe $L_{2,3}$ edge of $\alpha$-$Fe_2O_3$. (f) Intrinsic dynamics of the full width at half maximum (FWHM) of the Fe $L_3$ edge. Adapted with permission from Su *et al.*, J. Am. Chem. Soc. **139**, 4916–4922 (2017).[143] Copyright 2017 American Chemical Society.

For example, **Figs. 8(a)** and **8(b)** present ultrafast core-loss EELS measurements of a graphite thin film.[95] In graphite, carbon atoms form strong intralayer $\sigma$-bonds with $sp^2$-hybridized orbitals and weaker interlayer $\pi$-bonds with $p_z$ orbitals. These bonding configurations result in distinct absorption peaks in the carbon K edge, as shown in the black spectrum of **Fig. 8(a)**. Upon photoexcitation, the transient feature near 290 eV in **Fig. 8(a)** arises from a redshift of the peak associated with transitions to empty states with $\sigma^*$ character (**Fig. 8(b)**). This redshift reflects the elongation of the $\sigma$-bonds in the basal plane of the photoexcited graphite, as supported by *ab initio* simulations combining MD and multiple scattering theory using FEFF. The prompt redshift of the $\sigma^*$ peak is attributed to ultrafast energy-gap shrinkage, driven by enhanced electron–phonon interactions following photoexcitation. Notably, a blueshift—typically expected due to thermal lattice contraction—is absent in the transient spectrum, which underscores the sensitivity of core-loss EELS to local bonding changes rather than global lattice responses.

Ultrafast core-loss EELS measurements on bi-layered manganite ($PrSr_{0.2}Ca_{1.8}Mn_2O_7$) demonstrate how correlating diffraction and imaging in the TEM yields a more comprehensive understanding of the electronic and structural dynamics influencing photoexcited core-loss EEL spectra.[86] Here, photothermally induced pressure waves are characterized using UED, while the resulting modulation of the lattice and electronic structure is monitored via ultrafast core-loss EELS. **Figure 8(c)** presents both static and transient EEL spectra of manganite in the low-loss region, with plasmon and core-loss features modeled using DFT calculations performed with the WIEN2k code. As shown in **Fig. 8(d)**, the Mn $M_{2,3}$ edges exhibit the highest sensitivity to the coherent structural distortions, and each elemental edge responds differently to these modulations. This highlights the element specificity of the technique.

Beyond photoinduced structural dynamics, ultrafast core-loss EELS has enabled direct observation of charge transfer processes. The charge-transfer dynamics in hematite ($α$-$Fe_2O_3$) are probed via the Fe $L_3$ edge in core-loss EELS, which is sensitive to changes in Fe oxidation state (**Fig. 8(e)**).[143] Upon photoexcitation, $Fe^{2+}$–$Fe^{4+}$ electron–hole pairs generated through Fe $3d$–$3d$ transitions cause a broadening of the Fe $L_3$ edge (**Fig. 8(f)**). Additionally, ultrafast energy-filtered TEM (EFTEM) is employed to image the charge carrier dynamics in single-crystal hematite, offering potential insights into coupled electronic and structural behavior in future studies.

Despite its potential, ultrafast core-loss EELS faces several challenges. The PINEM effect manifests in femtosecond core-loss EELS near time zero, shown as the spectral broadening in **Figs. 8(a)** and **8(e)**. This effect can be minimized using a convolution-based average method, which helps extract the intrinsic dynamics of the Fe $L_3$ edge, as demonstrated in **Fig. 8(f)**. Furthermore, core-loss peaks from shallow core levels—such as the peaks shown in **Fig. 8(c)**—can overlap with the bulk plasmon, complicating the analysis. As noted in Sec. II, core-level transitions have low scattering cross-sections, and the signal intensity decreases at higher energy loss following the power law. Therefore, careful experimental design is essential to account for both the core-loss scattering cross-section and plasmon overlap.

Ultrafast EELS, although still a rapidly developing technique, has made significant strides in resolving carrier and lattice dynamics in a variety of materials. Current efforts are now focusing on resolving dynamics spatially through spectrum imaging or in momentum space using tr-q-EELS. These ongoing efforts are summarized in **Table II**. Note that the experimental parameters for each report are listed for reference, with intentions to guide future time-resolved and ultrafast EELS efforts. The variables and acronyms within the table (and their meaning) include $\boldsymbol{F}$ (pump fluence), $\boldsymbol{\tau_{laser}}$ (laser pulse width), $\boldsymbol{f_{rep}}$ (laser repetition rate), $\boldsymbol{\beta}$ (collection angle), and $\boldsymbol{V_{acc}}$ (accelerating voltage). Additional experiment parameters for suitable ultrafast EELS conditions are summarized in the Appendix. We hope to inspire future works that emphasize the importance of reporting experimental parameters as we move toward exciting future directions for the field.

**TABLE II. Reported Imaging and Optics Conditions for Ultrafast EELS.**
*(NR = not reported)*

| Year | Specimen | Loss Region | $F$ (mJ/cm$^2$) | $\tau_{laser}$ | $f_{rep}$ | $\beta$ | $V_{acc}$ |
|---|---|---|---|---|---|---|---|
| 2009[134,135] | Graphite | Low | 5.3 | 220 fs | 100 kHz / 1 MHz | NR | 200 kV |
| 2015[138] | SWCNTs | Low | 20 | 220 fs | 500 kHz | ~7 mrad | NR |
| 2020[139] | SWCNTs | Low | ~44 | ~190 fs | 200 kHz | >100 mrad | NR |
| 2022[137] | AuNTs | Low | 0.008 | 150 fs | 80 MHz | NR | 80 kV |
| 2023[132] | Cu | Low | 189 | 50 fs | 100 kHz | NR | 200 kV |
| 2023[133] | InAs crystal | Low | 1.6 | 50 fs | 1 MHz | NR | 200 kV |
| 2025[140] | Graphite | Low | 2 and 2.5 | NR | 1 MHz | NR | 200 kV |
| 2014[86] | Manganite | Low and Core | NR | 80 fs | 1 MHz | NR | NR |
| 2015[95] | Graphite | Core (Low in Supporting Information) | 10 | ~10 ns | 6 kHz | 10 mrad | NR |
| | | | | ~250 fs | 500 kHz | | |
| 2017[143] | Hematite | Core | 12 | NR | 500 kHz | NR | NR |

## V. FUTURE DIRECTIONS

The continued expansion of ultrafast EELS capabilities is imminent considering the rapid development of technologies including laser-free electron beam pulsing, monochromation, *in situ* cells, and high-speed direct electron detection. Several recent works indicate the ongoing and new directions that will catapult the field toward new limits using ultrafast EELS in the TEM or STEM, even in operando and *in situ* environments. Other authors have outlined a roadmap for measuring quantum nanophotonics using free electrons and UEM techniques, including EELS.[144]

### A. Time-Resolved Monochromated EELS

Wien filters within the TEM have dramatically improved the energy resolution limits of EELS. Monochromation acts to minimize the energetic bandwidth of emitted free electrons, which is otherwise limited by the cathode's electron emission mechanism and surface condition. Typical Schottky emission microscopes have an energy bandwidth of <1 eV, and modern cold field-emission guns achieve a bandwidth of ~0.3 eV. Today, specialized TEMs now routinely achieve 3–5 meV resolutions with monochromation at low accelerating voltages, enabling measurements of excitations typically encompassed by the ZLP (e.g., vibrational modes, valence excitons, and surface plasmon polaritons).[145,146] Outside of the TEM, doubly monochromated EELS systems have reached 1 meV through toroidal lensing.[147–149]

Traditional UEMs operated with a photocathode are further plagued by large energy bandwidths due to energetic broadening from the Boersch effect. This broadening is a result of the Coulombic repulsion of pulsed, bunched electrons, which limits the ultimate photoelectron counts and coherence. Notably, the best energy resolution achieved by UEM has been 0.6 eV,[150] and typical UEM EELS measurements yield ~1 eV resolution. Current limiting factors are further discussed in the Appendix.

To minimize energetic broadening from the Boersch effect, recent developments in electron detectors have incorporated a time-to-digital converter (TDC) chip that measures the electron's time-of-arrival (ToA) on the detector down to 1.5 ns.[151] This achievement has been demonstrated in a monochromated STEM with a parabolic mirror for photoexcitation to achieve time-resolved spectrum imaging of thermal dissipation in a variety of solid-state materials (**Figs. 9(a)–9(f)**).[152] A $SiN_x$ optical phonon at approximately 100 meV EEL was clearly visible in the monochromated microscope. Following specimen photoexcitation, electron energy-gain (EEG) phonon modes were induced by an excited-state phonon population (**Fig. 9(a)**). Using a ratio of the EEL and EEG peaks, the photoexcited thermal dissipation in the $SiN_x$ window was resolved (**Fig. 9(d)**). The A exciton's redshift in $WS_2$ (**Figs. 9(b)** and **9(e)**) and bulk plasmon's redshift of Al (**Figs. 9(c)** and **9(f)**) were similarly tracked to resolve the microsecond cooling dynamics in each material. STEM-EELS spectrum imaging was also implemented to visualize heat dissipation at a defect in the Al foil.

Advanced operando measurements of thermal gradients will become resolvable at sub-nanosecond timescales with monochromated time-resolved EEL spectrum imaging of phonon and exciton dynamics. For example, steady-state non-equilibrium phonons were used to image an electrically bias-induced temperature gradient across an AlN–SiC interface (**Figs. 9(g)–9(i)**).[153] The EEL/EEG ratio again enabled detection of thermal gradients in the specimen. Furthermore, acoustic phonon imaging at individual crystal defects could ultimately resolve vibrational dynamics in response to a nanosecond heating front.[154] While these detection capabilities have not yet been successfully demonstrated at ultrafast femtosecond to picosecond timescales, the promise of achieving them with UEM in a standard monochromated STEM remains.

Another key advantage of time-resolved monochromated EELS is its potential to probe complex many-body interactions in strongly correlated materials. For instance, a recent study demonstrated the experimental observation of magnons using STEM-EELS, which were subsequently modeled with numerical simulations.[63] Although this work has not measured magnon dynamics over time, its extension into the time-resolved domain is highly anticipated. However, from a theoretical standpoint, significant challenges remain in modeling these interactions, particularly in materials with occupied d-orbitals,[155] magnetic ordering,[156] or intricate many-body phenomena observed in superconductors and charge- and spin-density-wave systems.[157] The difficulty stems from the need to accurately capture the dynamics of electronic, spin, and lattice degrees of freedom, especially when they exhibit strong correlations. Developing novel computational techniques and theoretical models, alongside a deeper understanding of time-resolved measurements, is crucial for advancing our knowledge of emergent quantum phases and high-temperature superconductivity.

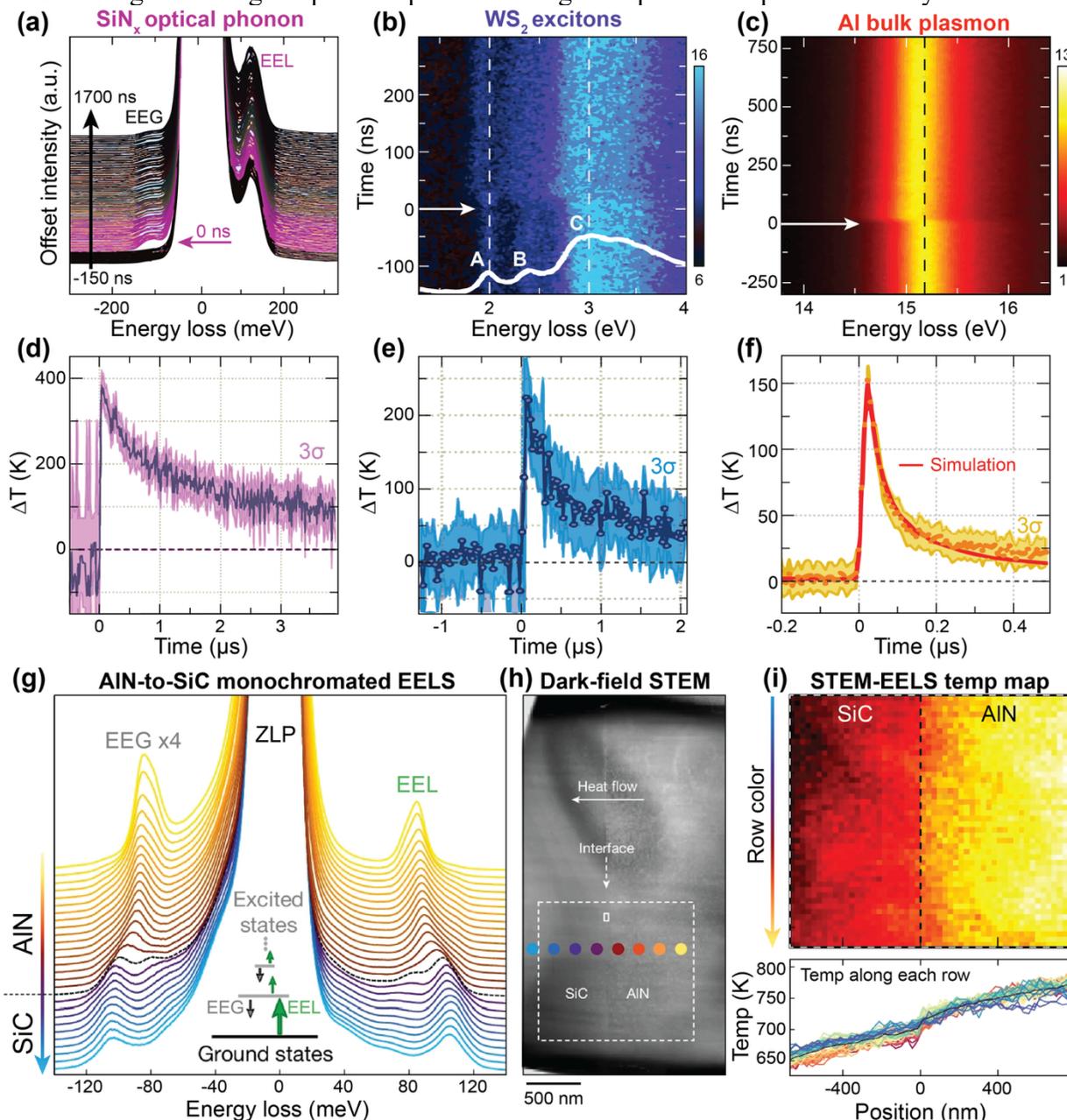

**Figure 9. Monochromated STEM-EELS maps of local temperature after optical or electrical excitation.** (a–c) Time-resolved EEL spectra of optically excited phonons, excitons, and plasmons in three materials; horizontal arrows mark time zero. Electron energy-gain (EEG) phonon peaks emerge after photoexcitation in (a). (d–f) Corresponding transient, thermal changes in each material derived from (a–c). Adapted with permission from Castioni *et al.*, Nano Lett. **25**, 1601–1608 (2025).[152] Copyright 2025 The Authors. Published by American Chemical Society. (g) EEL and EEG spectra acquired from hot AlN to cold SiC. The black dashed spectrum was acquired at the interface. Spectra were normalized to the EEL signal intensity, and EEG spectra were scaled 4x. (h) A dark-field STEM image of the AlN–SiC interface. (i) STEM-EELS temperature map and row-by-row temperature profiles from the dashed square in (h). Adapted with permission from Liu *et al.*, Nature, **642**, 941–946 (2025).[153] Copyright 2025 Springer Nature.

## B. Ultrafast *In Situ* and Operando EELS

Almost all UEM measurements to date have tracked photoexcitation-induced dynamics in a specimen under high vacuum. However, many photocatalytic, electrochemical, nanophotonic, and quantum materials systems operate under dynamic conditions with variable local environments. These conditions include exposure to reactive gases or liquids, electrical bias, and cryogenic or elevated temperature. In such systems, the specimen's environment directly influences the excited-state carrier, thermal, and structural dynamics.

Extending EELS to *in situ* and operando modes is therefore critical for capturing a material's true functional behavior. In principle, ultrafast EELS offers a powerful platform by combining nanometer spatial and femtosecond temporal resolutions, enabling direct spectral measurements of electronic and lattice excitations. However, several challenges must be addressed before these capabilities can be fully realized in practical environments. While time-resolved electron imaging and diffraction experiments under bias or gas exposure have been demonstrated,[158–163] ultrafast EELS in diverse operando conditions remains largely unexplored. In this section, we describe the opportunities and challenges for ultrafast EELS in gases and liquids, with variable temperature, and under applied bias.

### 1. Gas- and Liquid-Phase EELS

Extending EELS into fluidic gaseous and liquid environments is highly desirable for studying catalytic and photoelectrochemical systems, especially during active operation. Core-loss edges can directly probe local oxidation states and coordination environments of catalysts during gas adsorption or electrocatalysis, while low-loss features reveal ultrafast carrier densities and local heating, as described in Sec. IV. However, EELS of specimens in gases and liquids faces two significant obstacles: beam-induced radiolysis and fluid-induced inelastic scattering.[164,165]

Radiolysis refers to the decomposition of fluid molecules under electron irradiation, producing radicals and ions that may chemically alter the specimen's local environment. This effect is particularly severe in liquids due to their higher density but also occurs in gases at elevated pressures. Radiolysis scales with both beam dose and fluid pressure or thickness, since these factors increase the number of interactions with the electron beam and relate to each molecule's mean free path. Electron-beam radiolysis is an active area of EELS research,[166] with potential utility for controlled and irreversible *in situ* radiolysis studies.

The second limitation is strong inelastic scattering from fluids, which obscures a specimen's low-loss spectrum and broadens core-loss edges through multiple scattering.[167,168] This effect arises primarily from the relatively large fluid layer thickness (typically 100–1000 nm) compared to the specimen (<100 nm), as well as the intrinsic inelastic scattering of the gas or solvent. These strong inelastic peaks from the fluid, located at ~15–25 eV, obscure the material-specific signals from the specimen. Several reviews discuss the progress and challenges of gas-phase[169–171] and liquid-phase[171,172] TEM and EELS.

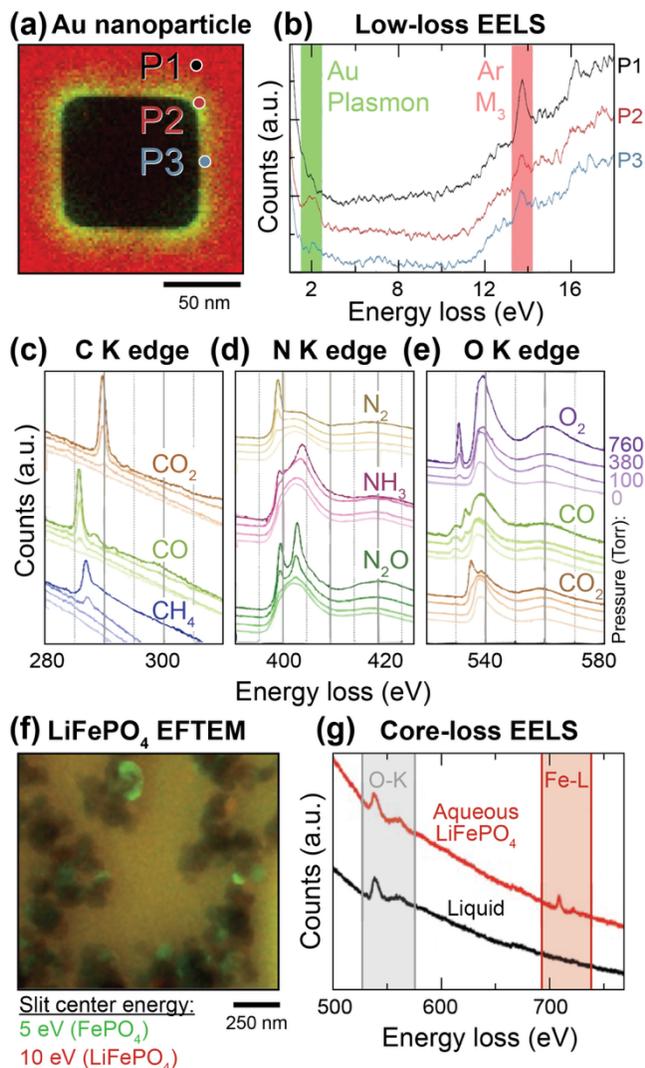

**Figure 10. Gas- and liquid-phase EELS.** (a) Overlaid EELS spectrum images of Au surface plasmon (green) and Ar $M_3$ (red) intensities for a Au nanocube in 1 atm Ar. (b) Low-loss EELS integrated from the spectrum image for each position ($P_x$). (c–e) Pressure-dependent radiolysis of different gases measured with EELS. (c) C K edge of $CO_2$, CO, and $CH_4$ gas. (d) N K edge of $N_2$, $NH_3$, and $N_2O$ gas. (e) O K edge of $O_2$, CO, and $CO_2$ gas. Spectra for each gas with colors from dark to light that correspond to different gas pressures (760, 380, 100, and 0 Torr). Adapted with permission from Koo *et al.*, Chem. Mater. **36**, 4078–4091 (2024).[166] Copyright 2024 American Chemical Society. (f, g) EFTEM and EELS of $LiFePO_4$ nanoparticles in a 180 nm aqueous solution. (f) Overlaid EFTEM images using a 5 eV energy-selecting slit centered around 5 eV = $FePO_4$ (green) and 10 eV = $LiFePO_4$ (red) based on their optical transitions. (g) Liquid-phase core-loss STEM-EELS on and off a nanoparticle. Adapted with permission from Holtz *et al.*, Microsc. Microanal. **19**, 1027–1035 (2013).[170] Copyright 2023 Microscopy Society of America.

**Gas-Phase EELS.** Despite challenges, *in situ* gas-phase EELS has made notable progress. In gas-phase environments, EELS is typically performed in either environmental TEM (ETEM) or closed-cell gas holders. However, achieving pressures above ~1 atm remains difficult, and EEL signals are often obscured by scattering when gas pressures/densities exceed that of their mean-free path. Recent improvements to gas flow cells, such as back-supported $SiN_x$ membranes, allow higher pressures (up to ~6 atm) while minimizing scattering.[166,173,174] As a result, gas-phase EELS has successfully imaged an Au nanocube's surface plasmon resonance in 1 atm of Ar using EEL spectrum imaging (**Figs. 10(a) and 10(b)**).[166] The inert Ar environment minimizes radiolysis and beam-induced contamination.

While radiolysis in gas-phase EELS is less severe than in liquids, it can still affect data interpretation for *in situ* measurements. As shown in **Figs. 10(c)–10(e)**, radiolysis can be directly probed through EELS, with its extent scaling with gas pressure.[166] Looking ahead, gas-phase EELS can be used to study dynamic processes while avoiding beam damage and contamination, as demonstrated in the studies of temperature-induced catalytic gas conversion,[175] gas adsorbates using time-resolved EELS,[176] and hydrogen oxidation reactions on a Pd nanocube.[177]

**Liquid-Phase EELS.** Liquid-phase EELS presents even greater challenges than gas-phase EELS due to the higher density of liquids, which causes stronger scattering and more severe beam-induced radiolysis. Standard liquid cells typically utilize ultrathin $SiN_x$ or polymeric window membranes, designed as closed cells or microfluidic channels optimized to minimize the liquid path length.[178–180] Graphene-encapsulated liquid pockets have also been widely adopted, as they further reduce both liquid and membrane scattering.[181–184]

To date, demonstrations of liquid-phase EELS have successfully resolved the electronic structure and composition of nanoparticles, with applications in energy conversion and storage. Several strategies have been implemented to mitigate the deleterious effects of the liquid environment. These strategies include evacuating liquid from the cell entirely or performing measurements in the liquid's optical gap (1–5 eV or >50 eV). Operating in the optical gap avoids the fluid's inelastic scattering peak and allows collection of both low- and core-loss EEL spectra from the specimen. For instance, low-loss EFTEM of $LiFePO_4$ in liquid revealed species-specific optical contrast in the optical gap, enabling differentiation between $LiFePO_4$ and $FePO_4$ based on their dielectric properties (**Fig. 10(f)**).[170] Additionally, the core-loss Fe $L_{2,3}$ edges were successfully resolved above the liquid's multiple scattering background, demonstrating the potential for element-specific chemical mapping under low-dose STEM-EELS conditions (**Fig. 10(g)**).

Other notable implementations of liquid-phase EELS include the observation of cerium salt oxidation via radiolysis in a graphene liquid cell,[176] self-hydrogenation and bubble formation on anatase $TiO_2$,[171] and STEM-EELS imaging of electrochemical Li ion insertion into a Si nanowire core.[159] Together, these studies illustrate the potential for chemical, structural, and electronic mapping in reactive liquid environments.

## 2. Variable Temperature Conditions

Exploring temperature-dependent behavior using EELS—either by applying controlled temperatures or by quantifying temperature at the nanoscale—is an increasingly important frontier in materials science. Temperature control enables access to a wide range of physical phenomena including phase transitions, thermal equilibration, and quantum effects. Many ultrafast phenomena in quantum materials[185,186] (e.g., coherent electronic dynamics, superconductivity, and exciton condensation) occur or are stabilized at cryogenic temperatures, where interactions with the phonon bath are suppressed. Conversely, elevated temperatures enable studies of phonon- or lattice-assisted transitions and phase instabilities. For ultrafast EELS, variable temperature environments open powerful avenues for measuring reproducible, thermally influenced dynamics in both equilibrium and nonequilibrium states.

**Cryogenic EELS.** Recent advances in cryogenic specimen holders capable of reaching liquid helium temperatures (5–50 K) have opened new directions in core-loss EELS.[187,188] Despite the ongoing challenges of mechanical instabilities caused by thermal drift, cryo-STEM-EELS has achieved atomic-resolution mapping in materials. One example is the study of a $FeSe/SrTiO_3$ heterojunction at 10 K, where the Fe $L_{2,3}$ edge blueshifts near the interface for eight unit cell-thick FeSe (**Fig. 11(a)**).[189] Below the superconducting transition temperature, the superconducting phase induces interfacial band-bending by retracting FeSe electrons transferred across the interface through phonon coupling (**Fig. 11(b)**). STEM-EELS spectrum imaging below and above the transition temperature further indicates the degradation of the superconducting phase.

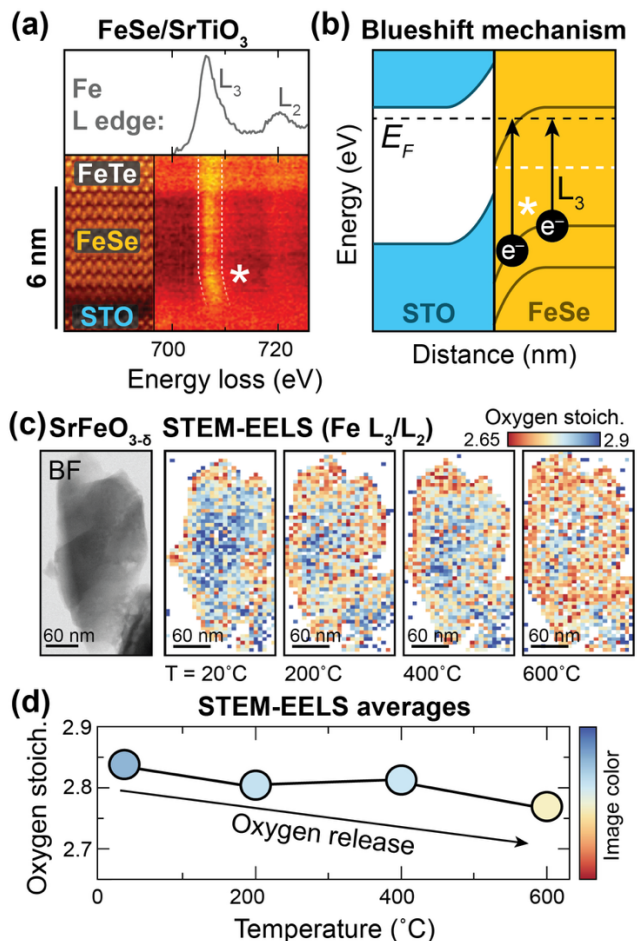

**Figure 11. EELS at cryogenic and elevated temperatures.** (a) Core-loss spectrum imaging of interfacial band-bending of the FeSe/SrTiO$_3$ interface at 10 K using liquid He. (Top) Reference Fe L$_{2,3}$-edge reference spectrum of bulk FeSe. (Bottom) Corresponding spectrum image. The Fe L$_{2,3}$ edge blueshifts at the SrTiO$_3$ (STO) interface when cooled to 10 K. The white dashed lines around Fe L$_3$ guide the eye. (b) The interfacial band bending that produces the L$_3$ edge blueshift. Adapted with permission from Zhao *et al.*, Sci. Adv. **4**, eaao2682 (2018).[189] Copyright 2018 Authors, licensed under a CC BY license. (c) *In situ* TEM images while heating SrFeO$_{3-\delta}$. (Left) Bright-field (BF) reference image of the SrFeO$_3$ particle. (Right) STEM-EELS maps showing changes in decreasing oxygen stoichiometry with increasing temperature. Pixels for <250 nm specimen thickness are excluded. (d) Average oxygen stoichiometry of SrFeO$_{3-\delta}$ from the temperature-dependent spectrum images. Adapted with permission from Harrison *et al.*, J. Mater. Chem. A, **13**, 32271 (2025).[162] Copyright 2025 The Royal Society of Chemistry, licensed under a CC BY license.

Cryogenic EELS is also critical for mitigating beam damage in beam-sensitive specimens. For example, the coexistence of LiH and Li metal was resolved in an electrochemical system using the Li K edge at cryogenic temperatures.[190] Such approaches hold strong promises for ultrafast EELS of soft matter and beam-sensitive specimens not measurable using room-temperature TEM.

**EELS at Elevated Temperatures.** At elevated temperatures, EELS can probe thermally induced changes in bonding, oxidation states, and band structure. These properties significantly impact charge transport, as measured by ultrafast EELS. As discussed in Secs. IV. B and V. A, phonons, excitons, and plasmons all display temperature-dependent shifts in the low-loss regime, forming the basis for EELS thermometry and plasmon energy expansion thermometry (PEET).[191,192] These features have also been tracked at ultrafast timescales following optical excitation.

More broadly, thermal cycling can reveal irreversible processes such as metal sublimation, oxygen vacancy formation, or phase decomposition, which may interfere with or evolve during ultrafast measurements. For instance, temperature-dependent STEM-EELS of SrFeO$_{3-\delta}$ mapped the decrease in oxygen stoichiometry by tracking the Fe L$_2$/L$_3$ ratio (**Figs. 11(c)** and **11(d)**).[162] This study underscores the value of elevated-temperature control for evaluating structural stability and oxidation dynamics over extended EELS scan acquisition times.

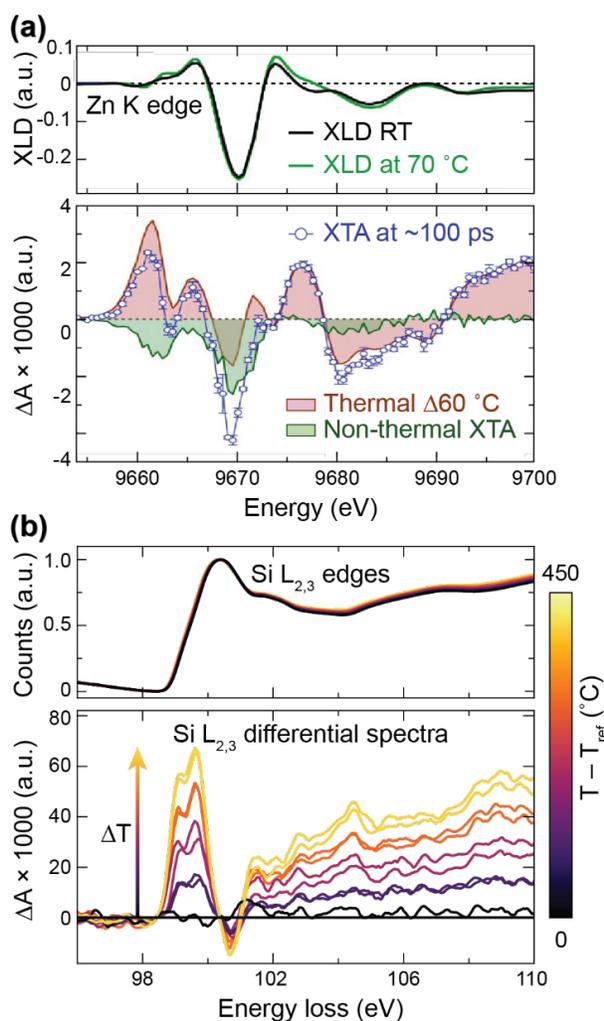

**Figure 12. Core-loss thermometry for element-specific lattice temperature measurements.** (a) X-ray absorption spectra of ZnO nanorods after femtosecond excitation. (Top) X-ray linear dichroism (XLD) at room temperature (RT; black) and 70°C (green). (Bottom) X-ray transient absorption (XTA) ~100 ps after 355 nm photoexcitation (blue), the simulated thermal response at Δ60°C (red), and the resulting calculated nonthermal carrier signal (green). Adapted with permission from Rossi *et al.*, Nano Lett. **21**, 9534–9542 (2021).[195] Copyright 2021 Authors, licensed under a CC BY license. (b) Si $L_{2,3}$ edge core-loss thermometry. (Top) As-measured reference spectra. (Bottom) Differential spectra, $\Delta A = (T - T_{ref})$, of the Si $L_{2,3}$ edge over heating/cooling cycles. Adapted with permission from Palmer *et al.*, ACS Phys. Chem. Au (2025). DOI: 10.1021/acsphyschemau.5c00044.[73] Copyright 2025 Authors. Published by American Chemical Society under a CC BY license.

**Core-Loss Thermometry.** Although bulk plasmon-based PEET has become a standard for nanoscale temperature measurements, its accuracy can be affected by factors such as thickness, momentum dispersion, strain, and contamination.[193] In contrast, core-loss EELS thermometry offers an element-specific approach with enhanced chemical sensitivity, particularly in semiconductors where the core-level edge shifts predictably with temperature as described by the Varshni equation.[78] Core-loss thermometry is insensitive to contamination and thickness effects by directly measuring an element-specific edge, and it is less sensitive to variable momenta and strain, as discussed in prior studies.

A linear redshift of a semiconductor's core-loss edge with temperature enables core-loss thermometry.[107,194,195] This redshift was observed using X-ray absorption spectroscopies and recently applied in time-resolved X-ray linear dichroism (XLD) measurements of ZnO nanowires, where temperature effects were cleanly separated from carrier dynamics (**Fig. 12(a)**).[195] Other work quantified the thermal effect on core-loss spectra using the Si $L_{2,3}$ edges in EEL spectra acquired at calibrated temperatures (**Fig. 12(b)**).[78] By quantifying sub-eV spectral shifts consistent with first-principles DFT-BSE simulations, these studies validate core-loss EELS as a robust, nanoscale thermometric technique. Core-loss thermometry holds strong potential for ultrafast studies of element-specific lattice temperatures, local heating, and energy dissipation even in stacked junctions.

The largest limitation to core-loss thermometry is the poor SNR typically observed for core-loss EELS. However, recent improvements in detecting core-loss edges at hard X-ray energies have made this technique far more accessible to additional edges and elements. Specifically, a new EELS detection scheme optimized the acquisition of core-loss edges out to 9 keV with enhanced SNR.[196]

## 3. UEM with Electrical Biasing

New frontiers in time-resolved and ultrafast electron imaging and diffraction are being unlocked by integrating electrical biasing with UEM.[158–163] This progress enables studies of field-induced structural and electronic phenomena at picosecond to microsecond timescales. The technique relies on applying either a radio-frequency (RF) or pulsed DC (direct current) field to the specimen followed by a synchronized electron pulse that probes transient responses such as lattice distortion and field-induced strain (**Fig. 13(a)**).[158]

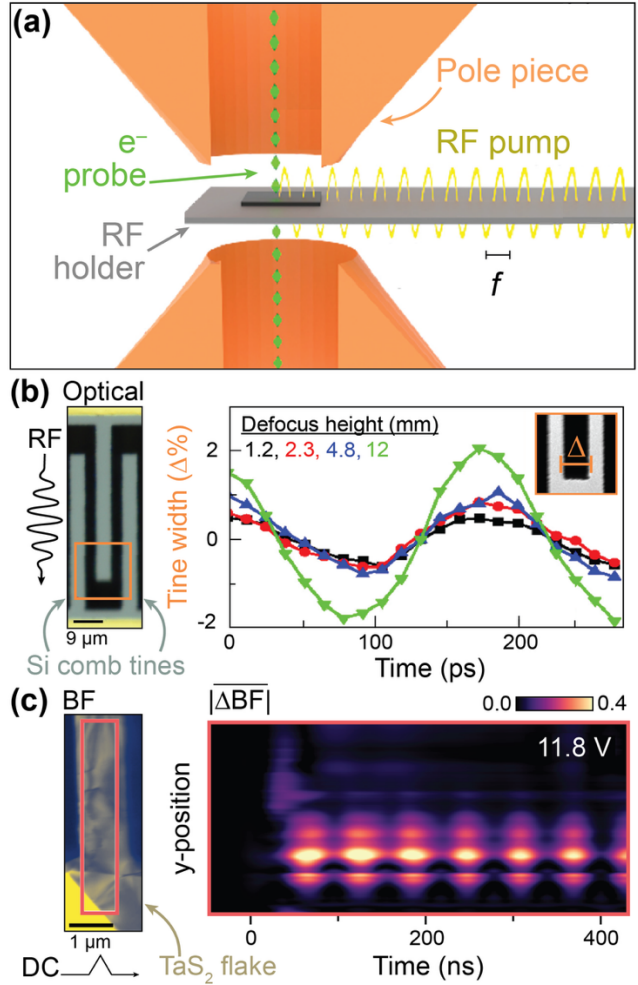

**Figure 13. UEM with electrical biasing.** (a) Schematic of UEM using an RF pump. (b, Left) Optical image of an interdigitated Si comb; RF pulse direction is indicated. (b, Right) Time-resolved tine width measured by UEM at four defocus heights using 0.5 W excitation; initial width is 4.5 μm. Adapted with permission from Reisbick *et al.*, Ultramicroscopy **235**, 113497 (2022).[158] Copyright 2022 Elsevier. (c) Mesoscopic, nanosecond strain dynamics in a TaS$_2$ flake. (Left) Initial bright-field (BF) TEM image with DC current direction and analysis region marked. (Right) Time-resolved mean absolute BF signal change ($|\overline{\Delta BF}|$) for varying (y) positions on the flake. The mean is computed along the x direction. Adapted with permission from Durham *et al.*, Phys. Rev. Lett. **132**, 226201 (2024).[159] Copyright 2024 American Physical Society.

The ultrafast RF-induced expansion and contraction of a Si comb was studied using a UEM with 10 ps resolution (**Fig. 13(b)**).[158] The RF field traversing the Si comb tines induced a local electric field that modulated the tine width on picosecond time and nanometer length scales. This effect is most clearly resolved with a defocused electron probe, which enhances the contrast mechanism by increasing the dispersion of the electron momenta during the beam-specimen interaction. In another study, a nanosecond pulsed DC bias and large electron probe were used to visualize the electrical melting of charge density waves in 1T–TaS$_2$ (**Fig. 13(c)**).[159] Under an applied 11.8 V DC bias, the integrated UEM image elucidates the flake's drumming on nanosecond timescales. The drumming results from an initial buckling due to heat-induced strain followed by thermal redistribution and oscillation at the acoustic resonance frequency.

Ultrafast EELS under electrical bias could enable real-time imaging of charge injection, field-induced band bending, and resistive switching, all of which are crucial mechanisms in memory architectures and semiconducting interfaces. Beyond purely structural dynamics such as strain, ultrafast EELS could provide insights into transient electronic and ionic transport phenomena resulting from applied bias. Several DC biasing TEM-EELS experiments have already been demonstrated in the steady state, including measurements on solid-state batteries[197] and semiconducting interfaces.[153] Adapting these methodologies to pulsed STEM-EELS holds immense promise for directly visualizing transport mechanisms in real time.

## C. New Hardware for Ultrafast EELS Spectrum Imaging

Advancing the capabilities of ultrafast EELS will rely heavily on continued innovation throughout the entire UEM, from electron sources and optics to detectors and spectral acquisition schemes. Each new generation of hardware has improved the accessibility and resolution of ultrafast EELS, enabling more reliable and versatile measurements of dynamic phenomena. Electron emitter designs for higher brightness, coherence, and energy resolution are also critical, as discussed in the Appendix Sec. B.

One persistent challenge in the global adoption of UEM is the cost and complexity of system installation. Beyond the base microscope, ultrafast optics, alignment, and synchronization demand specialized hardware and expertise. A promising recent solution is the development of laser-free pulsed electron sources, which eliminate the need for laser excitation altogether. These systems apply RF or DC pulses to blank or unblank the electron beam, generating electron pulses with picosecond to nanosecond widths (**Fig. 14(a)**).[198] In "sweeping mode", an RF pulse modulates beam position, sweeping the electron beam across an aperture and then realigning it along the electron beam axis via a second deflector (K2) post-aperture. While this mode achieves an improved temporal resolution, spatial coherence can suffer from imperfect realignment (**Fig. 14(b)**), and extremely high repetition rates from RF modulation can limit studies of long-lived dynamics, as discussed in Appendix Sec. A. Alternatively, in "chopping mode," a constant deflection is applied at K1, and a brief DC pulse redirects the beam to pass through the aperture, yielding ~250 ps resolution at modest repetition rates.

Improving probe brightness and SNR remains a main priority for UEM, especially for ultrafast EELS where inelastic scattering probabilities are inherently low. A recent approach addressed this by energy-filtering photoelectrons to select only the central 10 eV of a broad packet (~3000 e$^-$), significantly reducing pulse width and enhancing coherence (**Figs. 14(c)** and **14(d)**).[199] In measurements of VO$_2$ nanoparticles, this filtering strategy reduced pulse duration by over 50%, from $2.8 \pm 1.0$ ps to $700 \pm 200$ fs, improving temporal fidelity in EELS and imaging. Implementing beam monochromation before the specimen could further decrease the probe's pulse width without sacrificing total beam current.

Historically, stroboscopic pump–probe techniques have been necessary to resolve ultrafast dynamics due to detector speed limitations. However, direct electron detectors are transforming this paradigm. Time-to-digital converters (TDCs) now enable sub-nanosecond temporal resolution by capturing the precise ToA of each electron (**Fig. 14(e)**).[200] While current detectors have fewer pixels than conventional EELS detectors, they can achieve time resolution as low as 200 ps (**Fig. 14(f)**), providing a path toward UEM and ultrafast EELS in any STEM with specimen excitation.[201]

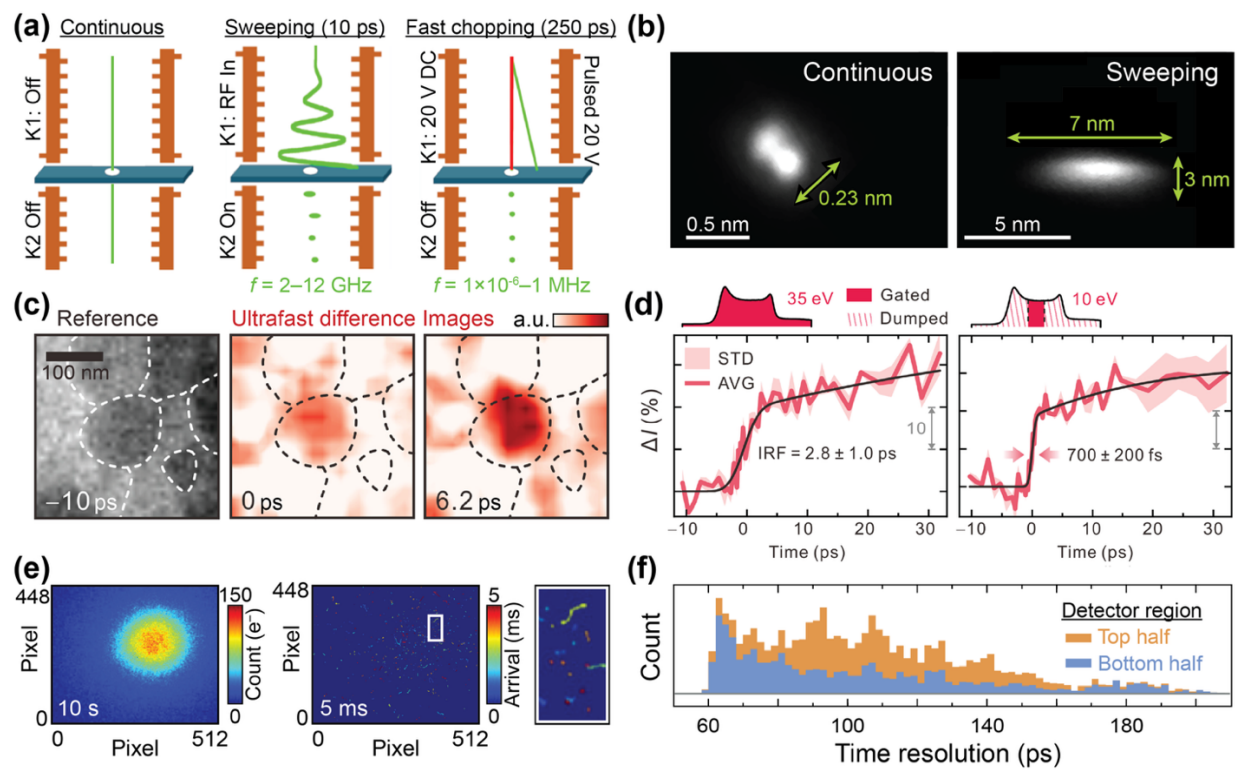

**Figure 14. Advanced hardware for ultrafast EELS.** (a) Electron gun with Euclid pulser in continuous (pulser off), high, and low frequency modes. (b) Minimum focused beam sizes in continuous (left) and sweeping (right) modes; spot sizes are the FWHM along the indicated direction. RF-induced beam expansion causes merging of the dual beams. Adapted with permission from Reisbick *et al.*, Ultramicroscopy **249**, 113733 (2023).[198] Copyright 2023 Elsevier. (c) Reference bright-field UEM image of $VO_2$ nanoparticles before time zero, and false-color difference images of the nanoparticles over time. The nanoparticles are outlined with dotted lines. (d) Energy-filtered gating narrows the electron probe pulse width; $\Delta I$ is extracted from the central particle's transient intensity profile and fit to an exponential (black). Adapted with permission from Kim *et al.*, Sci. Adv. **9**, eadd5373 (2023).[199] Copyright 2023 Authors, licensed under a CC BY license. (e) The Timepix4 detector: (Left) total counts per pixel over a 10 s acquisition time; (Right) the electron's time-of-arrival along a 5 ms time slice with the white rectangle highlighting the magnified region. Adapted with permission from Llopart *et al.*, JINST **17**, C01044 (2022).[200] Copyright 2022 Authors, licensed under a CC BY license. (f) The Timepix4 time-to-digital converter resolution across all pixels without any corrections. Adapted with permission from Heijhoff *et al.*, JINST **17**, P07006 (2022).[201] Copyright 2022 Authors, licensed under a CC BY license.

Looking forward, technological advances are making ultrafast EELS increasingly accessible and robust. New UEM modalities such as ultrafast 4D-STEM and STEM-EELS will soon be realized.[202,203] Further, optically gating the electron probe introduces new temporal limits, leading to the emerging field of attomicroscopy.[204–206] Attosecond temporal resolutions will enable the study of even faster carrier cooling and tunneling dynamics through ultrafast EELS.[108,109] As research groups and manufacturers collaborate to reduce costs and streamline integration, ultrafast EELS will transition from a niche capability to a widely adopted and indispensable tool for probing ultrafast dynamics in materials.

## VI. SUMMARY

Ultrafast EELS is rapidly evolving into a powerful technique for directly imaging charge, lattice, and heat dynamics across previously inaccessible time and length scales. With its ability to combine sub-nanometer spatial, femtosecond temporal, and variable momentum resolutions, ultrafast EELS uniquely enables the study of transient carrier dynamics and heat dissipation in complex materials and device architectures. We detailed the continued progress in theoretical methods for predicting and interpreting ground- and excited-state loss functions for low-loss plasmons and core-level spectra. New theoretical approaches leveraging cDFPT, TDDFT, and MD have laid the groundwork for interpreting complex EEL spectroscopic datasets. UEM EELS has elucidated charge carrier dynamics in real- and momentum-space through revolutionary methods including ultrafast low- and core-loss EELS, CDEM, and tr-q-EELS.

The next frontier for ultrafast EELS will be shaped by efforts to probe functional systems and devices in their working environments (e.g., under applied bias, at variable temperatures, and in gas or liquid media). Integrating *in situ* and operando cells with ultrafast EELS will enable transformative insights into catalytic cycles, thermal transport, quantum phase transitions, and interfacial charge transfer. Monochromated sources, laser-free ultrafast electron pulses, optimized energy-filtered detection, and high-speed TDC detectors offer the necessary technological advancements for enhancing technique accessibility and utility. Together, these advances position ultrafast EELS not only as a complementary tool to ultrafast electron diffraction and ultrafast optical and X-ray spectroscopies but as a central platform for visualizing electronic and thermal processes in emerging materials under working conditions. As the field continues to mature, ultrafast EELS will play a key role in uncovering structure–function relationships in next-generation quantum, energy, and nanoscale systems. Readers interested in implementing ultrafast EELS in their experimental workflow are invited to read curated suggestions within the Appendix.

## ACKNOWLEDGMENTS

The authors thank Professor Ye-Jin Kim, Nicholas Heller, and Audrey Washington for helpful discussions regarding future directions of ultrafast EELS and for their comments on the manuscript.

This research was supported as part of the Ensembles of Photosynthetic Nanoreactors, an Energy Frontier Research Center funded by the U.S. Department of Energy, Office of Science under Award

No. DE-SC0023431. Work performed at the Center for Nanoscale Materials, a U.S. Department of Energy Office of Science User Facility, was supported by the U.S. DOE, Office of Basic Energy Sciences, under Contract No. DE-AC02-06CH11357. The views expressed in this article do not necessarily represent the views of the U.S. Department of Energy, National Science Foundation, or the United States Government. The computations presented in **Fig. 2** were conducted in the Resnick High Performance Computing Center, a Resnick Sustainability Institute facility at the California Institute of Technology.

W.L. acknowledges support from the Korea Foundation for Advanced Studies. L.D.P. was supported by the National Science Foundation Graduate Research Fellowship under Grant No. DGE-1745301. This material is based upon work supported by the U.S. Department of Energy, Office of Science, Office of Workforce Development for Teachers and Scientists, Office of Science Graduate Student Research (SCGSR) program. The SCGSR program is administered by the Oak Ridge Institute for Science and Education for the DOE under contract number DE-SC0014664.

## AUTHOR DECLARATIONS

### Conflict of Interest

The authors have no conflicts to disclose.

### Author Contributions

† Wonseok Lee and Levi D. Palmer contributed equally to this work.
Please see CRediT statement.

## DATA AVAILABILITY

The data that support the findings of this study are available from the corresponding author upon reasonable request.

# APPENDIX: TECHNICAL CONSIDERATIONS OF ULTRAFAST EELS

UEM integrates diffraction, imaging, and spectroscopy to probe nanoscale dynamics. Ultrafast EELS poses unique experimental challenges that require careful optimization. In this section, we outline the key technical considerations for achieving high-quality ultrafast EELS measurements. These include specimen design, electron beam configuration, pump–probe laser alignment, and fundamental spatial resolution limits. Many of these parameters remain underexplored and warrant further systematic evaluation in future studies, as they are essential for reproducibility and accessibility of ultrafast EELS.

## A. Engineering an Ideal Specimen

Engineering an optimal specimen is essential for high-quality ultrafast EELS measurements. Key considerations include the specimen's heat dissipation, thickness, contamination, and support grid/film. These factors directly influence the measurement's spectral resolution, SNR, and accuracy by preventing artifacts.

**Thermal Management.** Efficient heat dissipation is critical, especially in high-repetition-rate experiments where residual heating between laser pump pulses can lead to irreversible changes such as melting or ablation. Optimizing the specimen's heat dissipation involves selecting or engineering through ion milling, growth, or etching materials with high thermal conductivity and/or optimizing the thermal conductance from the region of interest to a heat sink. The use of support films and grids that serve as heat sinks has been shown to effectively mitigate temperature buildup between laser pulses.[207–209]

**Specimen Thickness.** Specimen thickness directly affects EELS resolution. Thick specimens increase the likelihood of multiple scattering events, which broaden energy-loss features and obscure fine structure. In particular, low-loss spectra become dominated by bulk plasmon scattering, while core-loss edges are broadened and consumed by background. Ideally, specimen thickness should remain below its inelastic mean free path, which is typically on the order of tens of nanometers for 200–300 keV electrons. However, excessively thinning a specimen simultaneously reduces the signal intensity, so there is a careful balance between minimizing multiple scattering and maintaining a usable SNR.

**Electron Beam Contamination.** Hydrocarbon contamination remains a persistent issue in ultrafast EELS, where e-beam-induced deposition can degrade spectral quality and induce unwanted charging. Maintaining a clean specimen holder and microscope column is critical. Sample preparation strategies to prevent contamination spread include plasma cleaning, baking, and solvent washing. Washing or soaking specimen grids with various filtered, high-purity solvents can clean grids highly contaminated with loose hydrocarbons, and this method works particularly well for nanoparticles through centrifugation. Cold fingers can minimize hydrocarbon migration during imaging, but the liquid nitrogen evaporation rate may limit the UEM scan's duration. E-beam flooding or "beam showers" at low magnification and high current is used for atomic-resolution STEM to affix residual hydrocarbons into a more uniform layer. However, beam showering is less effective for UEM as the resulting affixation cannot last the duration of a standard UEM scan. Studies suggest that plasma cleaning is among the most effective approaches, though its efficacy depends on instrument-specific variables.[210]

**Support Grids and Films.** The specimen's support structure plays a pivotal role in both mechanical and electronic stability. Support films must minimize unwanted electron scattering, local charging, and background signal while ensuring specimen flatness and drift resistance. Support films with high atomic numbers or thickness should generally be avoided due to their increased scattering cross-sections, which can obscure weak spectral features. Grids with high thermal conduction (e.g., Cu, Al, and Au) are preferred to dissipate heat between pulses and minimize thermal drift. Similarly, the possibility of charge transfer between the support film, grid, and specimen must be considered, particularly in studies involving ultrafast photoexcitation or applied bias, where charge transfer at the junction influences local electronic environments. The support film composition and phase are also crucial because amorphous carbon films can decompose through laser ablation. Unless a TEM lamella is welded directly onto a lift-out grid, films such as graphene, hBN, or $SiN_x$ are often used as a specimen support.

## B. Optimizing Beam and Optics Conditions

### 1. Electron Beam Settings

The quality and interpretability of ultrafast EELS data are highly dependent on several experimental parameters that govern the generation, control, and detection of the electron beam. These include the characteristics of the photocathode, acceleration voltages, operation mode of the microscope, and key acquisition parameters such as collection angle and momentum transfer. Optimizing these factors is essential for enhancing spatiotemporal resolutions, SNR, and temporal precision, thereby maximizing the technique's ability to reveal photoexcited behaviors of materials.

**Photocathode Type and Geometry**. In ultrafast EELS using a UEM, the photocathode is a crucial component as it generates the electron pulses. The type and geometry of the photocathode determine key factors such as the electron beam brightness, energy spread of the emitted electrons, and temporal duration of the electron pulses—all of which impact the performance of ultrafast EELS measurements.[32] Due to Coulombic repulsive forces between electrons in an ultrafast pulse, there is an inherent tradeoff between emission coherence and number of electrons per pulse.[211] Larger, flat photocathodes, such as those made from $LaB_6$, Ta, and Au, can generate thousands of electrons in a single picosecond-width pulse, making them suitable for lower repetition rate experiments. In contrast, sharp tip photocathodes provide a much more coherent electron beam, enhancing both spatial and (typically) energy resolution, but they are often limited to producing only tens of electrons per pulse. As a result, experiments using sharp tips generally require much higher repetition rates to accumulate sufficient signal.[212–214] Methods to optimize the operation using nanoscale photocathodes in most UEMs are still underway. Other geometries, including the guard ring configuration[215] and off-axis photoemission geometry,[216] are also key considerations in optimizing electron pulse characteristics.

In addition to these established photocathode geometries, further improvements in beam brightness and temporal characteristics can be achieved through novel source concepts.[144] These include carbon-nanotube or $LaB_6$ needle photocathodes, which offer enhanced emission capabilities due to their high conductivity and unique geometric properties. Low-emittance planar photocathodes, designed to minimize transverse beam divergence, can also improve brightness and coherence. Additionally, advanced techniques such as RF and electric electron beam chopping or blanking are being explored to further enhance beam temporal resolution, enabling ultrafast EELS measurements with even greater precision and temporal control.

Beyond the photocathode's configuration and emission mechanism itself, another critical area of development is improving the gun and photocathode surface condition. Exposure to poor vacuum conditions leads to surface oxidation of the photocathode, creating a variable work function that broadens the ZLP. Maintaining a consistent ultra-high vacuum significantly improves the emitter's quantum efficiency and operational lifetime.[217] Nevertheless, some degree of surface oxidation is inevitable. Optimizing cathode fabrication, incorporating regular cleaning and heat treatments, and developing new cathode materials that sustain a stable work function and high coherence will further help reduce energy spread and enhance performance.[218–221]

**Accelerating Voltage.** Accelerating voltage is another component to be considered. Lower voltages can improve spectral resolution by the reduced energy spread of the electron beam. They also decrease the likelihood of knock-on damage, enabling measurements for beam-sensitive materials. However, the trade-off is a reduced SNR and decreased penetration depth, which may limit analysis of thicker specimens. Therefore, selecting an appropriate accelerating voltage involves balancing beam damage, spatial and spectral resolution, and specimen thickness to optimize data quality for specific applications.

**Imaging vs. Diffraction Mode.** UEM can be operated either imaging or diffraction mode. However, diffraction mode is generally preferred for EELS measurements. This preference arises because diffraction mode minimizes chromatic aberration, which can otherwise lead to energy-dependent variations in collection efficiency.[222] Moreover, diffraction mode provides more precise control over the collection angles—an essential factor for optimizing the SNR and signal-to-background ratio (SBR), as well as for suppressing contributions from multiple scattering. This geometry also simplifies the alignment of the spectrometer entrance aperture with specific momentum transfer, thereby improving the accuracy of q-EELS measurements.

**Collection Angle.** The collection angle ($\beta$), defining the angular range over which scattered electrons are collected by the spectrometer, is a critical parameter in EELS experiments. In diffraction or STEM mode, this angle is calculated as the radius of the spectrometer divided by the product of the camera length and a microscope-dependent geometric factor. The choice of collection angle has a significant impact on the quality of the EELS data. A larger $\beta$ improves signal intensity by capturing more scattered electrons, which is particularly important when detecting weak edges or trace elements. However, increasing $\beta$ also introduces more background and aberrations, degrading the SBR and spectral resolution. Conversely, using a smaller $\beta$ improves the spectral resolution and SBR, though it comes at the cost of reduced signal strength and lower SNR. **Figure 15** shows that excessive reduction of $\beta$ improves the SBR of the boron K edge but diminishes signal strength and a compromise is to set $\beta$ around 10 mrad.[223] To achieve a balance between these trade-offs, it is generally recommended to set the collection angle to 2–3 times the characteristic scattering angle ($\theta_E$) of the edge of interest. This characteristic angle is defined by the ratio of energy loss ($\Delta E$) to twice the primary beam energy ($E_0$):

$$\theta_E = \frac{\Delta E}{2E_0} \tag{A1}$$

For example, at 200 keV, $\beta$ = 10 mrad can capture ~90% of the Si $L_{2,3}$ edges (99 eV), but only ~28% of the Au $M_5$ edge (2200 eV). Therefore, $\beta$ needs to be optimized to the specific energy-loss edge to maximize the data quality.

**Momentum Transfer.** Momentum transfer is a critical parameter that significantly influences the interpretation of EEL spectra both in the low-loss and core-loss regions. Momentum transfer, denoted by **q** in Eq. (4), depends on both the scattering angle and the incident electron energy. Varying the momenta transfer also changes the inelastic scattering probability due to scattering selection rules. As discussed in Sec. II, EELS is sensitive to dipole-allowed transitions at low momentum transfer, providing information like optical spectroscopy. However, at higher momentum transfer, non-dipole transitions become accessible and controlling the momentum transfer is essential when optimizing the experimental conditions and accurately extracting material's properties such as the band structure, dielectric function, and bonding characteristics. Additionally, q-EELS can probe anisotropies in the electronic structure, further emphasizing the need for careful consideration of momentum transfer during both data acquisition and analysis.

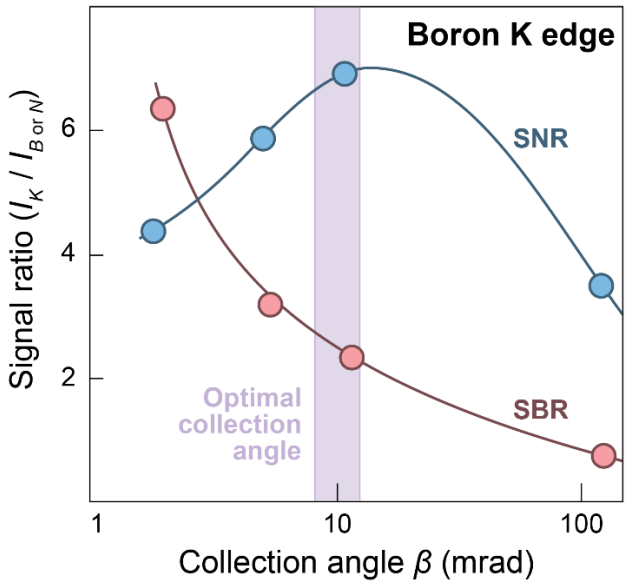

Figure 15. **Optimizing EELS collection angle.** Signal-to-background ratio (SBR, red) and signal-to-noise ratio (SNR, blue) calculated for the boron K edge of $B_2O_3$ as a function of collection angle ($\beta$). Each ratio is calculated using the intensity of the K edge ($I_K$) divided by either the background ($I_B$) or noise ($I_N$) The optimization of the two signal ratios results in a collection angle around 10 mrad for first-row elements (purple). Adapted with permission from Egerton, Springer eBook 3rd Ed., 111–229 (2011).[223] Copyright 2011 Springer Science+Business Media, LLC.

## 2. Laser and Optics Settings

UEM experiments present unique challenges due to their combination of advanced TEM imaging with ultrafast laser optics, requiring deep expertise in both fields. For pump-probe measurements, key laser parameters such as the pulse width, wavelength, fluence, and repetition rate must be carefully optimized for the specimen and the targeted excitation. Each of these laser optics parameters directly impacts the temporal resolution, signal strength, and quality of the measured dynamics. This section outlines practical considerations for optical configuration in UEM and ultrafast EELS.

**Optical Injection Mechanism.** The standard UEM architecture mirrors that of a conventional TEM but includes laser ports for photoexcitation of both the specimen and the electron emitter. However, the design is not quite that simple as femtosecond – nanosecond laser pulses must be focused precisely onto the specimen at typically 20–100 μm spot size. The original UEM design uses a laser table with one or more tiers and optical periscopes to bring the laser to the height of the TEM's optical ports (**Figs. 16(a)** and **16(b)**).[48] More compact approaches under development include stationary optics tubes or fiber-coupled holders (**Fig. 16(c)** and **16(d)**),[224,225] which improve user accessibility and widespread adoption but often restrict flexibility and alignment accuracy. For example, on-chip mirrors require realignment with any microscope stage movement, which is highly impractical for regular use. Other new innovative designs are under development from numerous manufacturers to use optical fibers for light injection either through an internal mirror, parabolic mirror, or the specimen holder.[226–229]

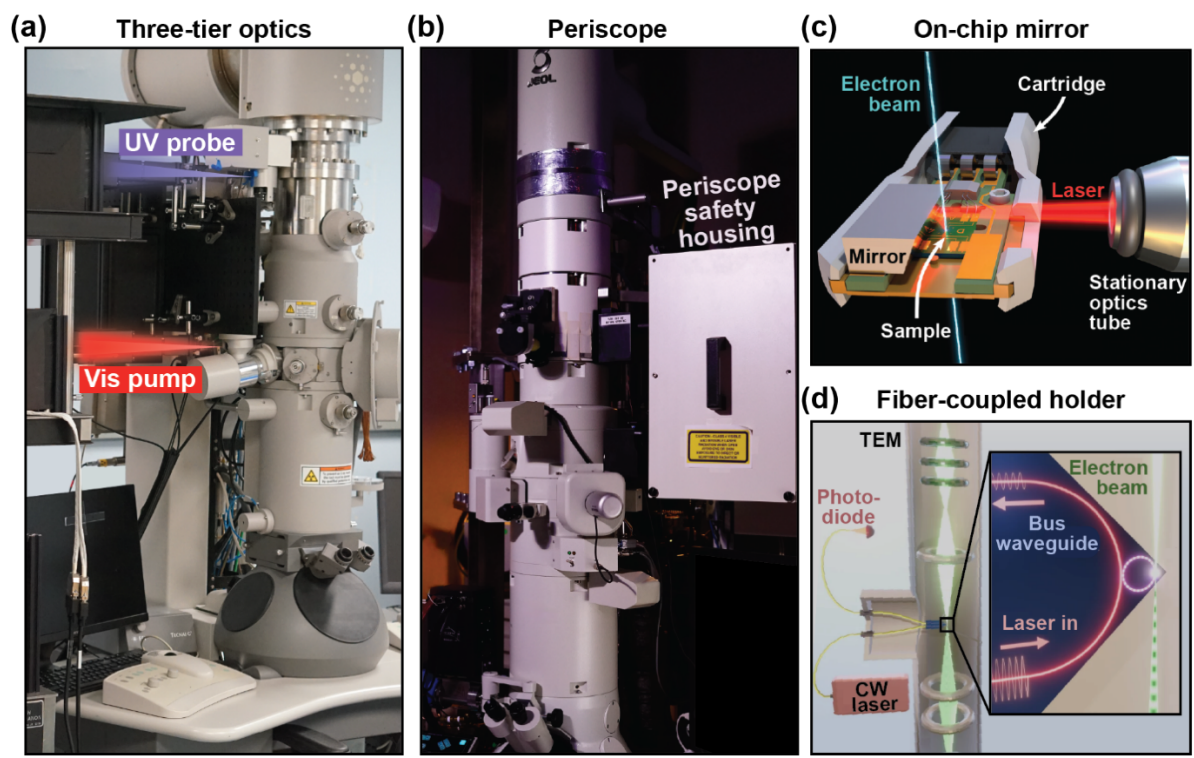

**Figure 16. Light injection schemes in TEM.** (a,b) Internal mirror-based systems. (a) Caltech UEM (modified Tecnai 200 kV Femto) with a three-tier optical table guiding pump and probe beams. (b) UEM at the Center for Nanoscale Materials at Argonne National Laboratory using a single optics table and dual periscopes directed into a safety housing.[48] (c) On-chip mirror approach with stationary optics; requires laser realignment for any microscope stage movement. Adapted with permission from Dyck *et al.*, Adv. Mater. Technol. **10**, 2401208 (2025).[224] Copyright 2025 Wiley-VCH GmbH. (d) Fiber-coupled holder delivering continuous-wave (CW) laser light directly to an optical resonator. Other systems direct the fiber's output to mirrors inside the TEM column. Adapted with permission from Henke *et al.*, Nature **600**, 653–658 (2021).[225] Copyright 2021 Authors, licensed under a CC BY license

Laser alignment typically proceeds in two steps. First, a coarse alignment directs the beam through a series of progressively smaller apertures visible on the TEM screen or flu-camera. Second, fine alignment is achieved by ablating a hole in an amorphous carbon film; if properly aligned, the hole matches the position of the electron probe, confirming spatial overlap and enabling spot size estimation.[230] The effects of the optical alignment and fluence are discussed further below.

**Laser Repetition Rate.** Laser repetition rate plays a central role in balancing signal strength with thermal and structural recovery times. While higher repetition rates yield faster signal accumulation, they can interfere with long-lived thermal or electronic relaxation, leading to time-averaged or non-representative dynamics. Most ultrafast EELS experiments operate below 1 MHz to avoid cumulative heating in typical energy materials. If variable, the laser's variable repetition rate mechanism utilizes a regenerative amplifier, the peak power may vary for different frequencies and impact fluence. Reported repetition rates for key studies are summarized in **Table II**.

**Pump Wavelength and Fluence at the Specimen.** Simply put, the choice of pump wavelength should match the material's absorption spectrum to efficiently excite the desired electronic or vibrational transitions. Shorter wavelengths offer higher energy excitation but can cause more damage or radiolysis. In semiconductors, tuning the pump wavelength can be an essential step so carriers access specific interband transitions or valley states. One must also consider the specimen's extinction coefficient, which can vary significantly even above the band gap and influence how deep in the band structure and efficiently the light is absorbed. The specimen's thickness is also a relevant consideration when it comes to pump fluence and damage.

Depending on the specimen's extinction coefficient and damage threshold at a given wavelength, a specific pump fluence can be determined for a given experiment (**Table II**), defined as the energy per unit area. Often if there are multiple specimens or nanoparticles to study, the pump fluence will be increased until laser damage is observed. At which point, the power would be decreased and the damage threshold documented. The same is often done for atomic-resolution TEM imaging, but the low beam current in UEM strongly minimizes electron beam effects.[231,232] Also, it is reported that some photoexcited carrier dynamics can change due to pump fluence.[233] Higher fluence increases the number of excited carriers, photothermal temperature, and the strength of the EELS signal, but excessive power can induce irreversible changes or nonlinear effects. The pump spot should be significantly larger than the electron beam size to ensure uniform excitation across the probed region and prevent spatial artifacts.

**Data Acquisition Scripting.** Lastly, the synchronization of all electronics for the UEM setup and its data acquisition typically utilize scripting in Digital Micrograph, Python, MATLAB, LabVIEW, or a homebuilt user interface. The UEM scripting process is described in depth elsewhere.[234] However, there are unique considerations that may benefit ultrafast EELS acquisition. Namely, incorporating a pump shutter may be useful as it allows for direct pump-on, pump-off referencing during the scan, rather than relying on timepoints before time zero as a reference.[235] It is also beneficial to carefully select timepoints based on the dynamics of interest. For example, timepoints can be acquired logarithmically to elucidate both fast and longer-lived dynamics without dramatically extending the total scan time. Other scripting settings are even able to acquire randomized timepoints to reduce artifacts created by laser damage or from the lab environment, either during the acquisition or post-processing.[236]

## C. Ultrafast EELS Spatial Resolution Limits

With continued advancement in UEM toward attosecond temporal, Ångström spatial, and meV energy scale resolutions,[204,237,238] it is critical to define the fundamental spatial resolution limits for ultrafast EELS. These limits are dictated primarily by three factors: (1) specimen drift during prolonged acquisition, (2) the excitation's intrinsic delocalization distance for a given energy loss and beam energy, and (3) effective electron source brightness. Accurately assessing these constraints prior to measurement is essential for optimizing scan strategies, minimizing artifacts, and improving data quality—especially as ultrafast STEM-EELS transitions toward widespread, routine application.

In general, the EELS probe size must be larger than both the specimen's thermal drift and the spatial extent of the excitation's wavefunction. For instance, if the specimen drifts by ~5 nm/hour over a 5-hour scan, a probe size of at least ~25 nm is advisable to avoid signal distortion. Similarly, for delocalized excitations such as phonons or plasmons, the probe size must also exceed the characteristic excitation delocalization to avoid false or unnecessary oversampling of the spatial resolution.

**Specimen Drift.** Specimen drift is influenced by thermal equilibration across the microscope system, including the lab environment, column, specimen holder, and sample itself. Careful experimental protocols such as proper thermal equilibration, post-acquisition drift correction (post-processing), and real-time automated spectrometer drift correction. These strategies are increasingly critical for prolonged STEM-EELS spectrum imaging. Specimen drift occurs across two general timescales.

At long timescales (minutes to hours) the specimen holder and TEM column thermally equilibrate with the surrounding laboratory. As such, the specimen can drift between nanometers and many microns over a multi-hour scan. This equilibration and drift directly correlate with the TEM's lab environment. As an approximate guideline, it is critical for the UEM lab to maintain a thermal drift of <1˚C/hour and ideally would always maintain a stability of ±1˚C, where even ±0.2˚C or less can be achieved.[239] Real-time drift correction in modern spectrometers further minimizes these drift effects. While the real-time correction is highly customizable depending on the drift rate, it can add significant time to the spectral acquisition, and the lab's temperature stability must still be a priority.

At short timescales (nanoseconds to microseconds) the specimen, support grid, and holder experience thermal equilibration induced by an optical or electrical pump. The pump-induced temperature creates strain in the support grid that causes drift over time. This drift is less significant and reproducible, as viewed by nanosecond UEM imaging, but it can be still on the order of ~1 nm over several nanoseconds depending on the specimen's geometry, laser fluence, and thermal expansion of the support film or grid.

**Excitation Delocalization Distance.** Each excitation captured in EELS has an associated wavefunction extent that limits the spatial resolution of the interaction. Low-energy excitations such as phonons and excitons are inherently delocalized, while core-loss edges are more localized, often within sub-Ångström scales. Other works have directly quantified and predicted EELS signal delocalization as compared to the dark-field STEM signal.[240,241]

We can calculate the wavefunction depth according to the angular distribution of the EELS scattering using the cutoff angle approximation. By approximating a Lorentzian angular distribution at half-width at a cutoff angle for 50% of signal transmission for the Bethe ridge, the delocalization distance ($d_{50}$) is calculated as:[241]

$$d_{50} \approx 0.71\lambda \left(\frac{E_0}{E}\right)^{3/4} \quad (A2)$$

In Eq. (A2), $d_{50}$ is calculated using the electron beam's relativistic wavelength ($\lambda$), incident energy ($E_0$), and energy loss energy ($E$). The delocalization distance with this approximation is depicted as the black dashed line in **Fig. 17**. We can alternatively calculate an EELS delocalization distance the root-mean-squared (RMS) impact parameter, $b_{RMS}$:

$$b_{RMS} \approx \left(\frac{h}{2\pi}\right)\left(\frac{v}{E}\right)\ln\left[\frac{4E_0}{E}\right] \quad (A3)$$

For Eq. (A3), principles from Fourier optics and the Heisenberg uncertainty principle under similar approximations as Eq. (A2) are incorporated to calculate the delocalization distance as $b_{RMS}$ while now considering the electron beam's velocity ($v$). Eq. (A3) is graphically represented in **Fig. 17** as the red dashed line, and both models agree well with experimental data.[241] Notably, for low-loss features below ~100 meV (e.g., phonons), the delocalization distance can exceed 100 nm, emphasizing that ultrafast EELS cannot spatially resolve these excitations at nanometer resolution unless they are tightly confined to local defects, grain boundaries, or interfaces.[242] This delocalization distance depicts the fundamental limit of spatial resolution for photoexcited dynamics during ultrafast EELS.

**Signal Detection Mechanism.** Signal detection is another crucial factor that strongly influences both the spatial resolution and detection limits of EELS. In particular, the choice of detectors significantly affects the number of collected counts and overall data quality. While CCD detectors are robust and widely used, they can become saturated by the intense ZLP due to its small spot size and often require additional dark reference acquisitions, effectively doubling the total scan time. Direct electron detectors, on the other hand, offer higher sensitivity, which enhances the SNR, especially for features with low scattering probabilities such as core-loss edges.[243] However, they are more susceptible to damage and generate large data volumes due to their fast readout rates. The use of dualEELS can further improve ultrafast EELS experiments by enabling optimized collection of both low- and high-loss signals. With modern detectors, it is now possible to rapidly switch the live time between these regions, allowing for accurate ZLP acquisition without damage while simultaneously optimizing inelastic signal collection.

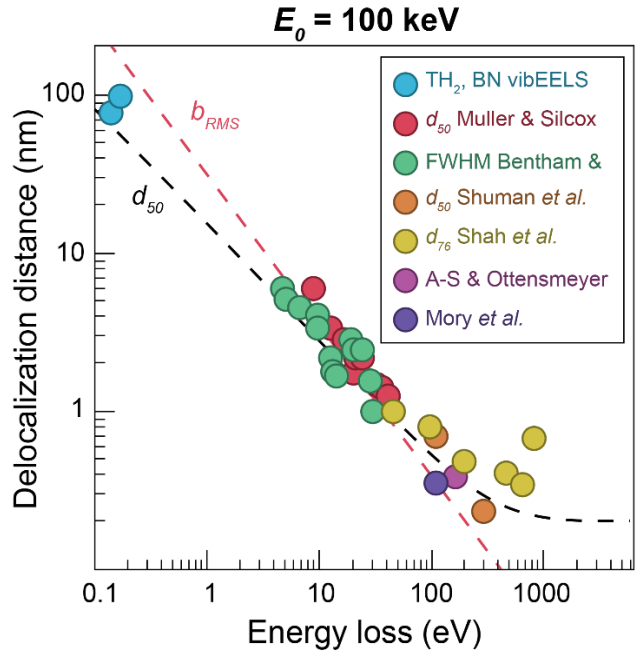

**Figure 17. Delocalization distance as a function of energy loss.** Delocalization distance for EELS excitations (wavefunction penetration depth) at 100 keV accelerating voltage as a function of energy loss. The dashed lines are based on theoretical predictions, as mentioned in the main text for both $d_{50}$ and $b_{RMS}$. The filled circles are colored according to experimental measurements reported by the authors in the legend. Adapted with permission from Egerton, Ultramicroscopy **180**, 115–124 (2017).[241] Copyright 2017 Elsevier B.V.

Overall, it is extremely critical to understand and be prepared to set up UEM experiments for ultrafast EELS. Successful measurements depend on careful attention to specimen design; beam, optical, and detection conditions; and the spatial resolution limits imposed by drift and excitation delocalization. Optimizing specimen thickness, minimizing contamination, and selecting appropriate support grids/films are essential for preserving spectral quality. Similarly, aligning the electron probe and optical pump to maintain probe coherences is key for maximizing sensitivity. Finally, understanding the spatial resolution limits from both specimen drift and energy-dependent delocalization helps guide probe size and scan settings. Together, these technical considerations form the foundation for reliable ultrafast EELS experiments.

# REFERENCES

[1] J. B. Baxter, C. Richter, and C. A. Schmuttenmaer, "Ultrafast Carrier Dynamics in Nanostructures for Solar Fuels," Annu. Rev. Phys. Chem. **65**, 423–447 (2014).

[2] J. Shi, Y. Li, Y. Li, D. Li, Y. Luo, H. Wu, and Q. Meng, "From Ultrafast to Ultraslow: Charge-Carrier Dynamics of Perovskite Solar Cells," Joule **2**(5), 879–901 (2018).

[3] A. Othonos, "Probing ultrafast carrier and phonon dynamics in semiconductors," J. Appl. Phys. **83**(4), 1789–1830 (1998).

[4] V. I. Klimov, "Optical Nonlinearities and Ultrafast Carrier Dynamics in Semiconductor Nanocrystals," J. Phys. Chem. B **104**(26), 6112–6123 (2000).

[5] L. Zhou, Q. Huang, and Y. Xia, "Plasmon-Induced Hot Electrons in Nanostructured Materials: Generation, Collection, and Application to Photochemistry," Chem. Rev. **124**(14), 8597–8619 (2024).

[6] S. V. Boriskina, H. Ghasemi, and G. Chen, "Plasmonic materials for energy: From physics to applications," Mater. Today **16**(10), 375–386 (2013).

[7] L. Mascaretti, and A. Naldoni, "Hot electron and thermal effects in plasmonic photocatalysis," J. Appl. Phys. **128**(4), 041101 (2020).

[8] H. Tang, C.-J. Chen, Z. Huang, J. Bright, G. Meng, R.-S. Liu, and N. Wu, "Plasmonic hot electrons for sensing, photodetection, and solar energy applications: A perspective," J. Chem. Phys. **152**(22), 220901 (2020).

[9] S. Cao, F. Tao, Y. Tang, Y. Li, and J. Yu, "Size- and shape-dependent catalytic performances of oxidation and reduction reactions on nanocatalysts," Chem. Soc. Rev. **45**(17), 4747–4765 (2016).

[10] K. An, and G. A. Somorjai, "Size and Shape Control of Metal Nanoparticles for Reaction Selectivity in Catalysis," ChemCatChem **4**(10), 1512–1524 (2012).

[11] M. A. El-Sayed, "Small Is Different: Shape-, Size-, and Composition-Dependent Properties of Some Colloidal Semiconductor Nanocrystals," Acc. Chem. Res. **37**(5), 326–333 (2004).

[12] O. V. Prezhdo, W. R. Duncan, and V. V. Prezhdo, "Dynamics of the Photoexcited Electron at the Chromophore–Semiconductor Interface," Acc. Chem. Res. **41**(2), 339–348 (2008).

[13] N. S. Ginsberg, and W. A. Tisdale, "Spatially Resolved Photogenerated Exciton and Charge Transport in Emerging Semiconductors," Annu. Rev. Phys. Chem. **71**, 1–30 (2020).

[14] A. Hon, "The Relation of Aperture and Power in the Microscope (continued).," J. R. Microsc. Soc. **2**(4), 460–473 (1882).

[15] P. E. Batson, N. Dellby, and O. L. Krivanek, "Sub-ångstrom resolution using aberration corrected electron optics," Nature **418**(6898), 617–620 (2002).

[16] P. D. Nellist, M. F. Chisholm, N. Dellby, O. L. Krivanek, M. F. Murfitt, Z. S. Szilagyi, A. R. Lupini, A. Borisevich, W. H. Sides, and S. J. Pennycook, "Direct Sub-Angstrom Imaging of a Crystal Lattice," Science **305**(5691), 1741 (2004).

[17] C. Kisielowski, B. Freitag, M. Bischoff, H. van Lin, S. Lazar, G. Knippels, P. Tiemeijer, M. van der Stam, S. von Harrach, M. Stekelenburg, M. Haider, S. Uhlemann, H. Müller, P. Hartel, B. Kabius, D. Miller, I. Petrov, E. Olson, T. Donchev, E. Kenik, A. Lupini, J. Bentley, S. Pennycook, I. Anderson, A. Minor, A. Schmid, T. Duden, V. Radmilovic, Q. Ramasse, M. Watanabe, R. Erni, E. Stach, P. Denes, and U. Dahmen, "Detection of Single Atoms and Buried Defects in Three Dimensions by Aberration-Corrected Electron Microscope with 0.5-Å Information Limit," Microsc. Microanal. **14**(5), 469–477 (2008).

[18] R. Erni, M. D. Rossell, C. Kisielowski, and U. Dahmen, "Atomic-Resolution Imaging with a Sub-50-pm Electron Probe," Phys. Rev. Lett. **102**(9), 096101 (2009).

[19] K. Kimoto, T. Asaka, T. Nagai, M. Saito, Y. Matsui, and K. Ishizuka, "Element-selective imaging of atomic columns in a crystal using STEM and EELS," Nature **450**(7170), 702–704 (2007).

[20] P. E. Batson, "Simultaneous STEM imaging and electron energy-loss spectroscopy with atomic-column sensitivity," Nature **366**(6457), 727–728 (1993).

[21] Y. Jiang, Z. Chen, Y. Han, P. Deb, H. Gao, S. Xie, P. Purohit, M. W. Tate, J. Park, S. M. Gruner, V. Elser, and D. A. Muller, "Electron ptychography of 2D materials to deep sub-ångström resolution," Nature **559**(7714), 343–349 (2018).

[22] Z. Chen, Y. Jiang, Y.-T. Shao, M.E. Holtz, M. Odstrčil, M. Guizar-Sicairos, I. Hanke, S. Ganschow, D. G. Schlom, and D. A. Muller, "Electron ptychography achieves atomic-resolution limits set by lattice vibrations," Science **372**(6544), 826–831 (2021).

[23] H. Yang, R. N. Rutte, L. Jones, M. Simson, R. Sagawa, H. Ryll, M. Huth, T. J. Pennycook, M. L. H. Green, H. Soltau, Y. Kondo, B. G. Davis, and P. D. Nellist, "Simultaneous atomic-resolution electron ptychography and Z-contrast imaging of light and heavy elements in complex nanostructures," Nat. Commun. **7**(1), 12532 (2016).

[24] C. Ophus, "Four-Dimensional Scanning Transmission Electron Microscopy (4D-STEM): From Scanning Nanodiffraction to Ptychography and Beyond," Microsc. Microanal. **25**(3), 563–582 (2019).

[25] A. H. Zewail, "Four-Dimensional Electron Microscopy," Science **328**(5975), 187–193 (2010).

[26] D. J. Flannigan, and A. H. Zewail, "4D Electron Microscopy: Principles and Applications," Acc. Chem. Res. **45**(10), 1828–1839 (2012).

[27] B. Barwick, and A. H. Zewail, "Photonics and Plasmonics in 4D Ultrafast Electron Microscopy," ACS Photonics **2**(10), 1391–1402 (2015).

[28] B. Barwick, H. S. Park, O.-H. Kwon, J. S. Baskin, and A. H. Zewail, "4D Imaging of Transient Structures and Morphologies in Ultrafast Electron Microscopy," Science **322**(5905), 1227–1231 (2008).

[29] Y.-J. Kim, W.-W. Park, H.-W. Nho, and O.-H. Kwon, "High-resolution correlative imaging in ultrafast electron microscopy," Adv. Phys. X **9**(1), 2316710 (2024).

[30] T. LaGrange, P. Cattaneo, B. Barwick, D. J. Flannigan, J. Weissenrieder, and F. Carbone, "Laser-driven ultrafast transmission electron microscopy," Nat Rev Methods Primers **5**(1), 61 (2025).

[31] N. J. Heller, A. J. Washington, and S. K. Cushing, *Electron Energy Loss Spectroscopy* (American Chemical Society, 2025).

[32] Y.-J. Kim, L. D. Palmer, W. Lee, N. J. Heller, and S. K. Cushing, "Using electron energy-loss spectroscopy to measure nanoscale electronic and vibrational dynamics in a TEM," J. Chem. Phys. **159**(5), 050901 (2023).

[33] T. Miyata, M. Fukuyama, A. Hibara, E. Okunishi, M. Mukai, and T. Mizoguchi, "Measurement of vibrational spectrum of liquid using monochromated scanning transmission electron microscopy–electron energy loss spectroscopy," Microscopy **63**(5), 377–382 (2014).

[34] F. S. Hage, D. M. Kepaptsoglou, Q. M. Ramasse, and L. J. Allen, "Phonon Spectroscopy at Atomic Resolution," Phys. Rev. Lett. **122**(1), 016103 (2019).

[35] C. S. Granerød, S. R. Bilden, T. Aarholt, Y.-F. Yao, C. C. Yang, D. C. Look, L. Vines, K. M. Johansen, and Ø. Prytz, "Direct observation of conduction band plasmons and the related Burstein-Moss shift in highly doped semiconductors: A STEM-EELS study of Ga-doped ZnO," Phys. Rev. B **98**(11), 115301 (2018).

[36] H.-R. Zhang, R. F. Egerton, and M. Malac, "Local thickness measurement through scattering contrast and electron energy-loss spectroscopy," Micron **43**(1), 8–15 (2012).

[37] L. Zhang, R. Erni, J. Verbeeck, and G. Van Tendeloo, "Retrieving the dielectric function of diamond from valence electron energy-loss spectroscopy," Phys. Rev. B **77**(19), 195119 (2008).

[38] A. Eljarrat, L. López-Conesa, J. López-Vidrier, S. Hernández, B. Garrido, C. Magén, F. Peiró, and S. Estradé, "Retrieving the electronic properties of silicon nanocrystals embedded in a dielectric matrix by low-loss EELS," Nanoscale **6**(24), 14971–14983 (2014).

[39] R. Erni, and N. D. Browning, "The impact of surface and retardation losses on valence electron energy-loss spectroscopy," Ultramicroscopy **108**(2), 84–99 (2008).

[40] V. J. Keast, "Ab initio calculations of plasmons and interband transitions in the low-loss electron energy-loss spectrum," J. Electron. Spectros. Relat. Phenomena **143**(2), 97–104 (2005).

[41] G. Haberfehlner, A. Orthacker, M. Albu, J. Li, and G. Kothleitner, "Nanoscale voxel spectroscopy by simultaneous EELS and EDS tomography," Nanoscale **6**(23), 14563–14569 (2014).

[42] H. Schmid, E. Okunishi, and W. Mader, "Defect structures in ZnO studied by high-resolution structural and spectroscopic imaging," Ultramicroscopy **127**, 76–84 (2013).

[43] D. A. Muller, "Simple model for relating EELS and XAS spectra of metals to changes in cohesive energy," Phys. Rev. B **58**(10), 5989–5995 (1998).

[44] M. D. Crescenzi, and G. Chiarello, "Extended energy loss fine structure measurement above shallow and deep core levels of 3d transition metals," J. Phys. C: Solid State Phys. **18**(18), 3595 (1985).

[45] B. Barwick, D. J. Flannigan, and A. H. Zewail, "Photon-induced near-field electron microscopy," Nature **462**(7275), 902–906 (2009).

[46] S. T. Park, M. Lin, and A. H. Zewail, "Photon-induced near-field electron microscopy (PINEM): theoretical and experimental," New J. Phys. **12**(12), 123028 (2010).

[47] I. Tanriover, Y. Li, T. E. Gage, I. Arslan, H. Liu, C. A. Mirkin, and K. Aydin, "Unveiling Spatial and Temporal Dynamics of Plasmon-Enhanced Localized Fields in Metallic Nanoframes through Ultrafast Electron Microscopy," ACS Nano **18**(41), 28258–28267 (2024).

[48] H. Liu, T. E. Gage, P. Singh, A. Jaiswal, R. D. Schaller, J. Tang, S. T. Park, S. K. Gray, and I. Arslan, "Visualization of Plasmonic Couplings Using Ultrafast Electron Microscopy," Nano Lett. **21**(13), 5842–5849 (2021).

[49] M. Liebtrau, M. Sivis, A. Feist, H. Lourenço-Martins, N. Pazos-Pérez, R. A. Alvarez-Puebla, F. J. G. de Abajo, A. Polman, and C. Ropers, "Spontaneous and stimulated electron–photon interactions in nanoscale plasmonic near fields," Light Sci. Appl **10**(1), 82 (2021).

[50] D. Zheng, S. Huang, C. Zhu, P. Xu, Z. Li, H. Wang, J. Li, H. Tian, H. Yang, and J. Li, "Nanoscale Visualization of a Photoinduced Plasmonic Near-Field in a Single Nanowire by Free Electrons," Nano Lett. **21**(24), 10238–10243 (2021).

[51] C. Kittel, and P. McEuen, *Introduction to Solid State Physics* (John Wiley & Sons, 2018).

[52] P. Drude, "Zur Elektronentheorie der Metalle," Ann. Phys. **306**(3), 566–613 (1900).

[53] P. Drude, "Zur Elektronentheorie der Metalle; II. Teil. Galvanomagnetische und thermomagnetische Effecte," Ann. Phys. **308**(11), 369–402 (1900).

[54] H. Raether, *Excitation of Plasmons and Interband Transitions by Electrons* (Springer, 2006).

[55] F. de Groot, and A. Kotani, *Core Level Spectroscopy of Solids* (CRC Press, 2008).

[56] J. M. Auerhammer, and P. Rez, "Dipole-forbidden excitations in electron-energy-loss spectroscopy," Phys. Rev. B **40**(4), 2024–2030 (1989).

[57] S. Löffler, I. Ennen, F. Tian, P. Schattschneider, and N. Jaouen, "Breakdown of the dipole approximation in core losses," Ultramicroscopy **111**(8), 1163–1167 (2011).

[58] V. J. Keast, A. J. Scott, R. Brydson, D. B. Williams, and J. Bruley, "Electron energy-loss near-edge structure – a tool for the investigation of electronic structure on the nanometre scale," J. Microsc. **203**(2), 135–175 (2001).

[59] M. Sarikaya, M. Qian, and E. A. Stern, "EXELFS revisited," Micron **27**(6), 449–466 (1996).

[60] P. J. Thomas, and P. A. Midgley, "An Introduction to Energy-Filtered Transmission Electron Microscopy," Top. Catal. **21**(4), 109–138 (2002).

[61] P. M. Zeiger, and J. Rusz, "Efficient and Versatile Model for Vibrational STEM-EELS," Phys. Rev. Lett. **124**(2), 025501 (2020).

[62] J. Á. Castellanos-Reyes, P. M. Zeiger, and J. Rusz, "Dynamical Theory of Angle-Resolved Electron Energy Loss and Gain Spectroscopies of Phonons and Magnons in Transmission Electron Microscopy Including Multiple Scattering Effects," Phys. Rev. Lett. **134**(3), 036402 (2025).

[63] D. Kepaptsoglou, J. Á. Castellanos-Reyes, A. Kerrigan, J. Alves do Nascimento, P. M. Zeiger, K. El hajraoui, J. C. Idrobo, B. G. Mendis, A. Bergman, V. K. Lazarov, J. Rusz, and Q. M. Ramasse, "Magnon spectroscopy in the electron microscope," Nature **644**(8075), 83–88 (2025).

[64] "electron energy-loss spectroscopy, EELS | Glossary | JEOL Ltd.," Electron Energy-Loss Spectroscopy, EELS | Glossary | JEOL Ltd., (n.d.). https://www.jeol.com/ (accessed 2025-06-30).

[65] J. Piprek, "Chapter 4 - Optical Waves," in *Semiconductor Optoelectronic Devices*, edited by J. Piprek, (Academic Press, Boston, 2003), pp. 83–120.

[66] H. Y. Li, S. M. Zhou, J. Li, Y. L. Chen, S. Y. Wang, Z. C. Shen, L. Y. Chen, H. Liu, and X. X. Zhang, "Analysis of the Drude model in metallic films," Appl. Opt. **40**(34), 6307–6311 (2001).

[67] S. K. Eswara Moorthy, and J. M. Howe, "Temperature dependence of the plasmon energy in liquid and solid phases of pure Al and of an Al-Si alloy using electron energy-loss spectroscopy," J. Appl. Phys. **110**(4), 043515 (2011).

[68] H. C. Nerl, J. P. Guerrero-Felipe, A. M. Valencia, K. F. Elyas, K. Höflich, C. T. Koch, and C. Cocchi, "Mapping the energy-momentum dispersion of hBN excitons and hybrid plasmons in hBN-WSe$_2$ heterostructures," npj 2D Mater. Appl. **8**(1), 68 (2024).

[69] I. Timrov, N. Vast, R. Gebauer, and S. Baroni, "turboEELS—A code for the simulation of the electron energy loss and inelastic X-ray scattering spectra using the Liouville–Lanczos approach to time-dependent density-functional perturbation theory," Comput. Phys. Commun. **196**, 460–469 (2015).

[70] R. M. Martin, *Electronic Structure: Basic Theory and Practical Methods* (Cambridge University Press, Cambridge, 2004).

[71] W. Kohn, "Nobel Lecture: Electronic structure of matter---wave functions and density functionals," Rev. Mod. Phys. **71**(5), 1253–1266 (1999).

[72] S. Baroni, and R. Gebauer, "The Liouville-Lanczos Approach to Time-Dependent Density-Functional (Perturbation) Theory," in *Fundamentals of Time-Dependent Density Functional Theory*, edited by M. A. L. Marques, N. T. Maitra, F. M. S. Nogueira, E. K. U. Gross, and A. Rubio, (Springer, Berlin, Heidelberg, 2012), pp. 375–390.

[73] O. B. Malcıoğlu, R. Gebauer, D. Rocca, and S. Baroni, "turboTDDFT – A code for the simulation of molecular spectra using the Liouville–Lanczos approach to time-dependent density-functional perturbation theory," Comput. Phys. Commun. **182**(8), 1744–1754 (2011).

[74] S. Baroni, P. Giannozzi, and A. Testa, "Green's-function approach to linear response in solids," Phys. Rev. Lett. **58**(18), 1861–1864 (1987).

[75] S. Baroni, S. de Gironcoli, A. Dal Corso, and P. Giannozzi, "Phonons and related crystal properties from density-functional perturbation theory," Rev. Mod. Phys. **73**(2), 515–562 (2001).

[76] P. Giannozzi, O. Andreussi, T. Brumme, O. Bunau, M. B. Nardelli, M. Calandra, R. Car, C. Cavazzoni, D. Ceresoli, M. Cococcioni, N. Colonna, I. Carnimeo, A. D. Corso, S. de Gironcoli, P. Delugas, R. A. DiStasio, A. Ferretti, A. Floris, G. Fratesi, G. Fugallo, R. Gebauer, U. Gerstmann, F. Giustino, T. Gorni, J. Jia, M. Kawamura, H.-Y. Ko, A. Kokalj, E. Küçükbenli, M. Lazzeri, M. Marsili, N. Marzari, F. Mauri, N. L. Nguyen, H.-V. Nguyen, A. Otero-de-la-Roza, L. Paulatto, S. Poncé, D. Rocca, R. Sabatini, B. Santra, M. Schlipf, A. P. Seitsonen, A. Smogunov, I. Timrov, T. Thonhauser, P. Umari, N. Vast, X. Wu, and S. Baroni, "Advanced capabilities for materials modelling with Quantum ESPRESSO," J. Phys.: Condens. Matter **29**(46), 465901 (2017).

[77] P. Giannozzi, S. Baroni, N. Bonini, M. Calandra, R. Car, C. Cavazzoni, D. Ceresoli, G. L. Chiarotti, M. Cococcioni, I. Dabo, A. D. Corso, S. de Gironcoli, S. Fabris, G. Fratesi, R. Gebauer, U. Gerstmann, C. Gougoussis, A. Kokalj, M. Lazzeri, L. Martin-Samos, N. Marzari, F. Mauri, R. Mazzarello, S. Paolini, A. Pasquarello, L. Paulatto, C. Sbraccia, S. Scandolo, G. Sclauzero, A. P. Seitsonen, A. Smogunov, P. Umari, and R. M. Wentzcovitch, "QUANTUM ESPRESSO: a modular and open-source software project for quantum simulations of materials," J. Phys.: Condens. Matter **21**(39), 395502 (2009).

[78] L. D. Palmer, W. Lee, D. B. Durham, J. Fajardo. Jr, Y. Liu, A. A. Talin, T. E. Gage, and S. K. Cushing, "Nanoscale and Element-Specific Lattice Temperature Measurements Using Core-Loss Electron Energy-Loss Spectroscopy," ACS Phys. Chem Au, (2025). DOI: 10.1021/acsphyschemau.5c00044.

[79] A. Catellani, and A. Calzolari, "Plasmonic properties of refractory titanium nitride," Phys. Rev. B **95**(11), 115145 (2017).

[80] A. Calzolari, C. Oses, C. Toher, M. Esters, X. Campilongo, S. P. Stepanoff, D. E. Wolfe, and S. Curtarolo, "Plasmonic high-entropy carbides," Nat. Commun. **13**(1), 5993 (2022).

[81] G. Onida, L. Reining, and A. Rubio, "Electronic excitations: density-functional versus many-body Green's-function approaches," Rev. Mod. Phys. **74**(2), 601–659 (2002).

[82] J. Deslippe, G. Samsonidze, D. A. Strubbe, M. Jain, M. L. Cohen, and S. G. Louie, "BerkeleyGW: A massively parallel computer package for the calculation of the quasiparticle and optical properties of materials and nanostructures," Comput. Phys. Commun. **183**(6), 1269–1289 (2012).

[83] E. A. Peterson, S. L. Watkins, C. Lane, and J.-X. Zhu, "Beyond-DFT *ab initio* Calculations for Accurate Prediction of Sub-GeV Dark Matter Experimental Reach," arXiv:2310.00147 (2023).

[84] D. Sangalli, A. Ferretti, H. Miranda, C. Attaccalite, I. Marri, E. Cannuccia, P. Melo, M. Marsili, F. Paleari, A. Marrazzo, G. Prandini, P. Bonfà, M. O. Atambo, F. Affinito, M. Palummo, A. Molina-Sánchez, C. Hogan, M. Grüning, D. Varsano, and A. Marini, "Many-body perturbation theory calculations using the yambo code," J. Phys.: Condens. Matter **31**(32), 325902 (2019).

[85] P. Blaha, K. Schwarz, F. Tran, R. Laskowski, G.K.H. Madsen, and L.D. Marks, "WIEN2k: An APW+lo program for calculating the properties of solids," J. Chem. Phys. **152**(7), 074101 (2020).

[86] L. Piazza, C. Ma, H. X. Yang, A. Mann, Y. Zhu, J. Q. Li, and F. Carbone, "Ultrafast structural and electronic dynamics of the metallic phase in a layered manganite," Struct. Dyn. **1**(1), 014501 (2014).

[87] A. Gulans, S. Kontur, C. Meisenbichler, D. Nabok, P. Pavone, S. Rigamonti, S. Sagmeister, U. Werner, and C. Draxl, "exciting: a full-potential all-electron package implementing density-functional theory and many-body perturbation theory," J. Phys.: Condens. Matter **26**(36), 363202 (2014).

[88] J. Vinson, "Advances in the OCEAN-3 spectroscopy package," Phys. Chem. Chem. Phys. **24**(21), 12787–12803 (2022).

[89] D. Y. Qiu, F. H. da Jornada, and S. G. Louie, "Solving the Bethe-Salpeter equation on a subspace: Approximations and consequences for low-dimensional materials," Phys. Rev. B **103**(4), 045117 (2021).

[90] O. Bunău, and M. Calandra, "Projector augmented wave calculation of x-ray absorption spectra at the $L_{2,3}$ edges," Phys. Rev. B **87**(20), 205105 (2013).

[91] M. Taillefumier, D. Cabaret, A.-M. Flank, and F. Mauri, "X-ray absorption near-edge structure calculations with the pseudopotentials: Application to the K edge in diamond and α-quartz," Phys. Rev. B **66**(19), 195107 (2002).

[92] D. Cabaret, and M. Calandra, "XSpectra: a density-functional-theory-based plane-wave pseudopotential code for XANES calculation," Int. Tables Crystallogr. **I**, 851–856 (2024).

[93] G. Donval, P. Moreau, J. Danet, S. Jouanneau-Si Larbi, P. Bayle-Guillemaud, and F. Boucher, "A hybrid method using the widely-used WIEN2k and VASP codes to calculate the complete set of XAS/EELS edges in a hundred-atoms system," Phys. Chem. Chem. Phys. **19**(2), 1320–1327 (2017).

[94] J. J. Kas, F. D. Vila, C. D. Pemmaraju, T. S. Tan, and J. J. Rehr, "Advanced calculations of X-ray spectroscopies with FEFF10 and Corvus," J. Synchrotron Rad. **28**(6), 1801–1810 (2021).

[95] R. M. van der Veen, T. J. Penfold, and A. H. Zewail, "Ultrafast core-loss spectroscopy in four-dimensional electron microscopy," Struct. Dyn. **2**(2), 024302 (2015).

[96] E. Stavitski, and F. M. F. de Groot, "The CTM4XAS program for EELS and XAS spectral shape analysis of transition metal L edges," Micron **41**(7), 687–694 (2010).

[97] A. Wiener, H. Duan, M. Bosman, A. P. Horsfield, J. B. Pendry, J. K. W. Yang, S. A. Maier, and A. I. Fernández-Domínguez, "Electron-Energy Loss Study of Nonlocal Effects in Connected Plasmonic Nanoprisms," ACS Nano **7**(7), 6287–6296 (2013).

[98] Y. Yang, R. G. Hobbs, P. D. Keathley, and K. K. Berggren, "Electron energy loss of ultraviolet plasmonic modes in aluminum nanodisks," Opt. Express, **28**(19), 27405–27414 (2020).

[99] J. Nelayah, M. Kociak, O. Stéphan, F. J. García de Abajo, M. Tencé, L. Henrard, D. Taverna, I. Pastoriza-Santos, L. M. Liz-Marzán, and C. Colliex, "Mapping surface plasmons on a single metallic nanoparticle," Nat. Phys. **3**(5), 348–353 (2007).

[100] Y. Cao, A. Manjavacas, N. Large, and P. Nordlander, "Electron Energy-Loss Spectroscopy Calculation in Finite-Difference Time-Domain Package," ACS Photonics **2**(3), 369–375 (2015).

[101] K. Reidy, P. E. Majchrzak, B. Haas, J. D. Thomsen, A. Konečná, E. Park, J. Klein, A. J. H. Jones, K. Volckaert, D. Biswas, M. D. Watson, C. Cacho, P. Narang, C. T. Koch, S. Ulstrup, F. M. Ross, and J. C. Idrobo, "Direct Visualization of Subnanometer Variations in the Excitonic Spectra of 2D/3D Semiconductor/Metal Heterostructures," Nano Lett. **23**(3), 1068–1076 (2023).

[102] F. J. García de Abajo, "Optical excitations in electron microscopy," Rev. Mod. Phys. **82**(1), 209–275 (2010).

[103] "RF Module User's Guide, version 5.3", COMSOL, Inc, www.comsol.com

[104] "Optical Simulation and Design Software | Ansys Optics," https://www.ansys.com/products/optics (accessed 2025-06-30).

[105] U. Hohenester, and A. Trügler, "MNPBEM – A Matlab toolbox for the simulation of plasmonic nanoparticles," Comput. Phys. Commun. **183**(2), 370–381 (2012).

[106] U. Hohenester, "Simulating electron energy loss spectroscopy with the MNPBEM toolbox," Comput. Phys. Commun. **185**(3), 1177–1187 (2014).

[107] L. D. Palmer, W. Lee, C.-L. Dong, R.-S. Liu, N. Wu, and S. K. Cushing, "Determining Quasi-Equilibrium Electron and Hole Distributions of Plasmonic Photocatalysts Using Photomodulated X-ray Absorption Spectroscopy," ACS Nano **18**(13), 9344–9353 (2024).

[108] H. Liu, J. M. Michelsen, J. L. Mendes, I. M. Klein, S. R. Bauers, J. M. Evans, A. Zakutayev, and S. K. Cushing, "Measuring Photoexcited Electron and Hole Dynamics in ZnTe and Modeling Excited State Core-Valence Effects in Transient Extreme Ultraviolet Reflection Spectroscopy," J. Phys. Chem. Lett. **14**(8), 2106–2111 (2023).

[109] S. K. Cushing, A. Lee, I. J. Porter, L. M. Carneiro, H.-T. Chang, M. Zürch, and S. R. Leone, "Differentiating Photoexcited Carrier and Phonon Dynamics in the Δ, L, and Γ Valleys of Si(100) with Transient Extreme Ultraviolet Spectroscopy," J. Phys. Chem. C **123**(6), 3343–3352 (2019).

[110] G. Petretto, S. Dwaraknath, H. P. C. Miranda, D. Winston, M. Giantomassi, M. J. van Setten, X. Gonze, K. A. Persson, G. Hautier, and G.-M. Rignanese, "High-throughput density-functional perturbation theory phonons for inorganic materials," Sci. Data **5**(1), 180065 (2018).

[111] A. Debernardi, S. Baroni, and E. Molinari, "Anharmonic Phonon Lifetimes in Semiconductors from Density-Functional Perturbation Theory," Phys. Rev. Lett. **75**(9), 1819–1822 (1995).

[112] A. Dal Corso, S. Baroni, and R. Resta, "Density-functional theory of the dielectric constant: Gradient-corrected calculation for silicon," Phys. Rev. B **49**(8), 5323–5328 (1994).

[113] X. Wu, D. Vanderbilt, and D. R. Hamann, "Systematic treatment of displacements, strains, and electric fields in density-functional perturbation theory," Phys. Rev. B **72**(3), 035105 (2005).

[114] L. He, F. Liu, G. Hautier, M. J. T. Oliveira, M. A. L. Marques, F. D. Vila, J. J. Rehr, G.-M. Rignanese, and A. Zhou, "Accuracy of generalized gradient approximation functionals for density-functional perturbation theory calculations," Phys. Rev. B **89**(6), 064305 (2014).

[115] T. Dengg, V. Razumovskiy, L. Romaner, G. Kresse, P. Puschnig, and J. Spitaler, "Thermal expansion coefficient of WRe alloys from first principles," Phys. Rev. B **96**(3), 035148 (2017).

[116] G. Marini, and M. Calandra, "Lattice dynamics of photoexcited insulators from constrained density-functional perturbation theory," Phys. Rev. B **104**(14), 144103 (2021).

[117] M. Botifoll, I. Pinto-Huguet, and J. Arbiol, "Machine learning in electron microscopy for advanced nanocharacterization: current developments, available tools and future outlook," Nanoscale Horiz. **7**(12), 1427–1477 (2022).

[118] Y. Wang, M. Tan, C. Fernandez-Granda, and P. A. Crozier, "Revealing Information from Weak Signal in Electron Energy-Loss Spectroscopy with a Deep Denoiser," arXiv:2505.14032 (2025).

[119] J. Lüder, "Machine learning approach to predict L-edge x-ray absorption spectra of light transition metal ion compounds," Phys. Rev. B **111**(8), 085110 (2025).

[120] C. M. Pate, J. L. Hart, and M. L. Taheri, "RapidEELS: machine learning for denoising and classification in rapid acquisition electron energy loss spectroscopy," Sci. Rep. **11**(1), 19515 (2021).

[121] G. Guinan, A. Salvador, M. A. Smeaton, A. Glaws, H. Egan, B. C. Wyatt, B. Anasori, K. R. Fiedler, M. J. Olszta, and S. R. Spurgeon, "Mind the gap: Bridging the divide between AI aspirations and the reality of autonomous microscopy," APL Mach. Learn. **3**(2), 020903 (2025).

[122] J. Schwartz, Z. W. Di, Y. Jiang, A. J. Fielitz, D.-H. Ha, S. D. Perera, I. El Baggari, R. D. Robinson, J. A. Fessler, C. Ophus, S. Rozeveld, and R. Hovden, "Imaging atomic-scale chemistry from fused multi-modal electron microscopy," npj Comput. Mater. **8**(1), 16 (2022).

[123] A. Ter-Petrosyan, M. Holden, J. A. Bilbrey, S. Akers, C. Doty, K. H. Yano, L. Wang, R. Paudel, E. Lang, K. Hattar, R. B. Comes, Y. Du, B. E. Matthews, and S. R. Spurgeon, "Revealing the Evolution of Order in Materials Microstructures Using Multi-Modal Computer Vision," arXiv:2411.09896 (2024).

[124] L. I. Roest, S. E. van Heijst, L. Maduro, J. Rojo, and S. Conesa-Boj, "Charting the low-loss region in electron energy loss spectroscopy with machine learning," Ultramicroscopy **222**, 113202 (2021).

[125] S. E. van Heijst, M. Bolhuis, A. Brokkelkamp, J. J. M. Sangers, and S. Conesa-Boj, "Heterostrain-Driven Bandgap Increase in Twisted WS$_2$: A Nanoscale Study," Adv. Funct. Mater. **34**(8), 2307893 (2024).

[126] A. Brokkelkamp, J. ter Hoeve, I. Postmes, S. E. van Heijst, L. Maduro, A. V. Davydov, S. Krylyuk, J. Rojo, and S. Conesa-Boj, "Spatially Resolved Band Gap and Dielectric Function in Two-Dimensional Materials from Electron Energy Loss Spectroscopy," J. Phys. Chem. A **126**(7), 1255–1262 (2022).

[127] H. La, A. Brokkelkamp, S. van der Lippe, J. ter Hoeve, J. Rojo, and S. Conesa-Boj, "Edge-induced excitations in Bi$_2$Te$_3$ from spatially-resolved electron energy-gain spectroscopy," Ultramicroscopy **254**, 113841 (2023).

[128] A. Brokkelkamp, S. E. van Heijst, and S. Conesa-Boj, "Edge-Localized Plasmonic Resonances in WS$_2$ Nanostructures from Electron Energy-Loss Spectroscopy," Small Sci. **5**(5), 2400558 (2025).

[129] Z. Ji, M. Hu, and H. L. Xin, "MnEdgeNet for accurate decomposition of mixed oxidation states for Mn XAS and EELS L$_{2,3}$ edges without reference and calibration," Sci. Rep. **13**(1), 14132 (2023).

[130] E. Hu, H. H. Choo, W. Zhang, A. Sumboja, I. T. Anggraningrum, A. Z. Syahrial, Q. Zhu, J. Xu, X. J. Loh, H. Pan, J. Chen, and Q. Yan, "Integrating Machine Learning and Characterization in Battery Research: Toward Cognitive Digital Twins with Physics and Knowledge," Adv. Funct. Mater. **35**(25), 2422601 (2025).

[131] Y. Wang, C. Fernandez-Granda, and P. A. Crozier, "Unsupervised Deep Video Denoiser: A Potential Key to Extracting Information from Monochromated EELS," Microsc. Microanal. **30**(Supplement_1), ozae044.1031 (2024).

[132] I. Madan, E. J. C. Dias, S. Gargiulo, F. Barantani, M. Yannai, G. Berruto, T. LaGrange, L. Piazza, T. T. A. Lummen, R. Dahan, I. Kaminer, G. M. Vanacore, F. J. García de Abajo, and F. Carbone, "Charge Dynamics Electron Microscopy: Nanoscale Imaging of Femtosecond Plasma Dynamics," ACS Nano **17**(4), 3657–3665 (2023).

[133] M. Yannai, R. Dahan, A. Gorlach, Y. Adiv, K. Wang, I. Madan, S. Gargiulo, F. Barantani, E. J. C. Dias, G. M. Vanacore, N. Rivera, F. Carbone, F. J. García de Abajo, and I. Kaminer, "Ultrafast Electron Microscopy of Nanoscale Charge Dynamics in Semiconductors," ACS Nano **17**(4), 3645–3656 (2023).

[134] F. Carbone, O.-H. Kwon, and A. H. Zewail, "Dynamics of Chemical Bonding Mapped by Energy-Resolved 4D Electron Microscopy," Science **325**(5937), 181–184 (2009).

[135] F. Carbone, B. Barwick, O.-H. Kwon, H. S. Park, J. Spencer Baskin, and A. H. Zewail, "EELS femtosecond resolved in 4D ultrafast electron microscopy," Chem. Phys. Lett. **468**(4), 107–111 (2009).

[136] F. Carbone, "The interplay between structure and orbitals in the chemical bonding of graphite," Chem. Phys. Lett. **496**(4), 291–295 (2010).

[137] M. Kuwahara, L. Mizuno, R. Yokoi, H. Morishita, T. Ishida, K. Saitoh, N. Tanaka, S. Kuwahara, and T. Agemura, "Transient electron energy-loss spectroscopy of optically stimulated gold nanoparticles using picosecond pulsed electron beam," Appl. Phys. Lett. **121**(14), 143503 (2022).

[138] G. M. Vanacore, R. M. van der Veen, and A. H. Zewail, "Origin of Axial and Radial Expansions in Carbon Nanotubes Revealed by Ultrafast Diffraction and Spectroscopy," ACS Nano **9**(2), 1721–1729 (2015).

[139] D. Zheng, C. Zhu, Z. Li, Z. Li, J. Li, S. Sun, Y. Zhang, F. Wang, H. Tian, H. Yang, and J. Li, "Ultrafast lattice and electronic dynamics in single-walled carbon nanotubes," Nanoscale Adv. **2**(7), 2808–2813 (2020).

[140] F. Barantani, R. Claude, F. Iyikanat, I. Madan, A. A. Sapozhnik, M. Puppin, B. Weaver, T. LaGrange, F. J. G. de Abajo, and F. Carbone, "Ultrafast momentum-resolved visualization of the interplay between phonon-mediated scattering and plasmons in graphite," Sci. Adv. **11**(14), eadu1001 (2025).

[141] P. M. Kraus, M. Zürch, S. K. Cushing, D. M. Neumark, and S. R. Leone, "The ultrafast X-ray spectroscopic revolution in chemical dynamics," Nat. Rev. Chem. **2**(6), 82–94 (2018).

[142] H. Liu, I. M. Klein, J. M. Michelsen, and S. K. Cushing, "Element-specific electronic and structural dynamics using transient XUV and soft X-ray spectroscopy," Chem **7**(10), 2569–2584 (2021).

[143] Z. Su, J. S. Baskin, W. Zhou, J. M. Thomas, and A. H. Zewail, "Ultrafast Elemental and Oxidation-State Mapping of Hematite by 4D Electron Microscopy," J. Am. Chem. Soc. **139**(13), 4916–4922 (2017).

[144] F. J. García de Abajo, A. Polman, C. I. Velasco, M. Kociak, L. H. G. Tizei, O. Stéphan, S. Meuret, T. Sannomiya, K. Akiba, Y. Auad, A. Feist, C. Ropers, P. Baum, J. H. Gaida, M. Sivis, H. Lourenço-Martins, L. Serafini, J. Verbeeck, A. Konečná, N. Talebi, B. M. Ferrari, C. J. R. Duncan, M. G. Bravi, I. Ostroman, G. M. Vanacore, E. Nussinson, R. Ruimy, Y. Adiv, A. Niedermayr, I. Kaminer, V. Di Giulio, O. Kfir, Z. Zhao, R. Shiloh, Y. Morimoto, M. Kozák, P. Hommelhoff, F. Barantani, F. Carbone, F. Chahshouri, W. Albrecht, S. Rey, T. Coenen, E. Kieft, H. L. Lalandec Robert, F. de Jong, and M. Solà-Garcia, "Roadmap for Quantum Nanophotonics with Free Electrons," ACS Photonics **12**(9), 4760–4817 (2025).

[145] J. A. Hachtel, J. R. Jokisaari, O. L. Krivanek, J. C. Idrobo, and R. F. Klie, "Isotope-Resolved Electron Energy Loss Spectroscopy in a Monochromated Scanning Transmission Electron Microscope," Microsc. Today **29**(1), 36–41 (2021).

[146] W. Zhao, Á. R. Echarri, A. Eljarrat, H. C. Nerl, T. Kiel, B. Haas, H. Halim, Y. Lu, K. Busch, and C. T. Koch, "Real-time surface plasmon polariton propagation in silver nanowires," arXiv:2411.19661 (2024).

[147] H. Ibach, "Electron energy loss spectroscopy with resolution below 1 meV," J. Electron. Spectros. Relat. Phenomena **64–65**, 819–823 (1993).

[148] P. Abbamonte, and J. Fink, "Collective Charge Excitations Studied by Electron Energy-Loss Spectroscopy," Annu. Rev. Condens. Matter Phys. **16**, 465–480 (2025).

[149] J. Chen, X. Guo, C. Boyd, S. Bettler, C. Kengle, D. Chaudhuri, F. Hoveyda, A. Husain, J. Schneeloch, G. Gu, P. Phillips, B. Uchoa, T.-C. Chiang, and P. Abbamonte, "Consistency between reflection momentum-resolved electron energy loss spectroscopy and optical spectroscopy measurements of the long-wavelength density response of $Bi_2Sr_2CaCu_2O_{8+x}$," Phys. Rev. B **109**(4), 045108 (2024).

[150] A. Feist, N. Bach, N. Rubiano da Silva, T. Danz, M. Möller, K. E. Priebe, T. Domröse, J. G. Gatzmann, S. Rost, J. Schauss, S. Strauch, R. Bormann, M. Sivis, S. Schäfer, and C. Ropers, "Ultrafast transmission electron microscopy using a laser-driven field emitter: Femtosecond resolution with a high coherence electron beam," Ultramicroscopy **176**, 63–73 (2017).

[151] D. Jannis, C. Hofer, C. Gao, X. Xie, A. Béché, T. J. Pennycook, and J. Verbeeck, "Event driven 4D STEM acquisition with a Timepix3 detector: Microsecond dwell time and faster scans for high precision and low dose applications," Ultramicroscopy **233**, 113423 (2022).

[152] F. Castioni, Y. Auad, J.-D. Blazit, X. Li, S. Y. Woo, K. Watanabe, T. Taniguchi, C.-H. Ho, O. Stéphan, M. Kociak, and L. H. G. Tizei, "Nanosecond Nanothermometry in an Electron Microscope," Nano Lett. **25**(4), 1601–1608 (2025).

[153] F. Liu, R. Mao, Z. Liu, J. Du, and P. Gao, "Probing phonon transport dynamics across an interface by electron microscopy," Nature **642**(8069), 941–946 (2025).

[154] X. Yan, C. Liu, C.A. Gadre, L. Gu, T. Aoki, T. C. Lovejoy, N. Dellby, O. L. Krivanek, D. G. Schlom, R. Wu, and X. Pan, "Single-defect phonons imaged by electron microscopy," Nature **589**(7840), 65–69 (2021).

[155] F. Nilsson, and F. Aryasetiawan, "Recent Progress in First-Principles Methods for Computing the Electronic Structure of Correlated Materials," Computation **6**(1), 26 (2018).

[156] M. H. Christensen, B. M. Andersen, and P. Kotetes, "Unravelling Incommensurate Magnetism and Its Emergence in Iron-Based Superconductors," Phys. Rev. X **8**(4), 041022 (2018).

[157] P. R. C. Kent, and G. Kotliar, "Toward a predictive theory of correlated materials," Science **361**(6400), 348–354 (2018).

[158] S. A. Reisbick, M.-G. Han, C. Liu, Y. Zhao, E. Montgomery, C. Jing, V. J. Gokhale, J. J. Gorman, J. W. Lau, and Y. Zhu, "Stroboscopic ultrafast imaging using RF strip-lines in a commercial transmission electron microscope," Ultramicroscopy **235**, 113497 (2022).

[159] D. B. Durham, T. E. Gage, C. P. Horn, X. Ma, H. Liu, I. Arslan, S. Guha, and C. Phatak, "Nanosecond Structural Dynamics during Electrical Melting of Charge Density Waves in 1T-TaS$_2$," Phys. Rev. Lett. **132**(22), 226201 (2024).

[160] S. A. Reisbick, A. Pofelski, M.-G. Han, C. Liu, E. Montgomery, C. Jing, K. Callaway, J. Cumings, J. W. Lau, and Y. Zhu, "Statistically elucidated responses from low-signal contrast mechanisms in ultrafast electron microscopy," Struct. Dyn. **12**(3), 034302 (2025).

[161] J. W. Lau, K. B. Schliep, M. B. Katz, V. J. Gokhale, J. J. Gorman, C. Jing, A. Liu, Y. Zhao, E. Montgomery, H. Choe, W. Rush, A. Kanareykin, X. Fu, and Y. Zhu, "Laser-free GHz stroboscopic transmission electron microscope: Components, system integration, and practical considerations for pump–probe measurements," Rev. Sci. Instrum. **91**(2), 021301 (2020).

[162] A. R. P. Harrison, F. Donat, J. M. A. Steele, J. C. Gebers, S. M. Fairclough, E. A. Willneff, A. J. Britton, C. L. Truscott, C. R. Müller, C. Ducati, C. P. Grey, and E. J. Marek, "In situ studies of oxygen transport mechanisms in Ag/SrFeO$_{3-\delta}$ materials for chemical looping catalysis," J. Mater. Chem. A **13**(38), 32271–32289 (2025).

[163] A. Sood, X. Shen, Y. Shi, S. Kumar, S.J. Park, M. Zajac, Y. Sun, L.-Q. Chen, S. Ramanathan, X. Wang, W. C. Chueh, and A. M. Lindenberg, "Universal phase dynamics in VO$_2$ switches revealed by ultrafast operando diffraction," Science **373**(6552), 352–355 (2021).

[164] M. A. Smeaton, P. Abellan, S. R. Spurgeon, R. R. Unocic, and K. L. Jungjohann, "Tutorial on In Situ and Operando (Scanning) Transmission Electron Microscopy for Analysis of Nanoscale Structure–Property Relationships," ACS Nano **18**(52), 35091–35103 (2024).

[165] D. Llorens Rauret, A. Garzón Manjón, and J. Arbiol, "Advances in *in situ* and *operando* TEM: From basic catalysis to industry-relevant reactions and future directions," Matter **8**(7), 102139 (2025).

[166] K. Koo, Y. Liu, Y. Cheng, Z. Cai, X. Hu, and V. P. Dravid, "Advances and Opportunities in Closed Gas-Cell Transmission Electron Microscopy," Chem. Mater. **36**(9), 4078–4091 (2024).

[167] M. Gu, L. R. Parent, B. L. Mehdi, R. R. Unocic, M. T. McDowell, R. L. Sacci, W. Xu, J. G. Connell, P. Xu, P. Abellan, X. Chen, Y. Zhang, D. E. Perea, J. E. Evans, L. J. Lauhon, J.-G. Zhang, J. Liu, N. D. Browning, Y. Cui, I. Arslan, and C.-M. Wang, "Demonstration of an Electrochemical Liquid Cell for Operando Transmission Electron Microscopy Observation of the

Lithiation/Delithiation Behavior of Si Nanowire Battery Anodes," Nano Lett. **13**(12), 6106–6112 (2013).

[168] P. A. Crozier, and S. Chenna, "*In situ* analysis of gas composition by electron energy-loss spectroscopy for environmental transmission electron microscopy," Ultramicroscopy **111**(3), 177–185 (2011).

[169] M. Tang, W. Yuan, Y. Ou, G. Li, R. You, S. Li, H. Yang, Z. Zhang, and Y. Wang, "Recent Progresses on Structural Reconstruction of Nanosized Metal Catalysts via Controlled-Atmosphere Transmission Electron Microscopy: A Review," ACS Catal. **10**(24), 14419–14450 (2020).

[170] M. E. Holtz, Y. Yu, J. Gao, H.D. Abruña, and D. A. Muller, "In Situ Electron Energy-Loss Spectroscopy in Liquids," Microsc. Microanal. **19**(4), 1027–1035 (2013).

[171] S. Ramasundaram, S. Jeevanandham, N. Vijay, S. Divya, P. Jerome, and T. H. Oh, "Unraveling the Dynamic Properties of New-Age Energy Materials Chemistry Using Advanced In Situ Transmission Electron Microscopy," Molecules **29**(18), 4411 (2024).

[172] N. Hodnik, G. Dehm, and K. J. J. Mayrhofer, "Importance and Challenges of Electrochemical in Situ Liquid Cell Electron Microscopy for Energy Conversion Research," Acc. Chem. Res. **49**(9), 2015–2022 (2016).

[173] K. Koo, Z. Li, Y. Liu, S. M. Ribet, X. Fu, Y. Jia, X. Chen, G. Shekhawat, P. J. M. Smeets, R. dos Reis, J. Park, J. M. Yuk, X. Hu, and V. P. Dravid, "Ultrathin silicon nitride microchip for in situ/operando microscopy with high spatial resolution and spectral visibility," Sci. Adv. **10**(3), eadj6417 (2024).

[174] K. Koo, P. J. M. Smeets, X. Hu, and V. P. Dravid, "Analytical In Situ Gas Transmission Electron Microscopy Enabled with Ultrathin Silicon Nitride Membranes," Microsc. Microanal. **29**(Supplement_1), 1597–1598 (2023).

[175] S. Chenna, and P. A. Crozier, "Operando Transmission Electron Microscopy: A Technique for Detection of Catalysis Using Electron Energy-Loss Spectroscopy in the Transmission Electron Microscope," ACS Catal. **2**(11), 2395–2402 (2012).

[176] Y. Wang, and P. A. Crozier, "Observation of Gas Adsorbates with Time-Resolved Vibrational EELS," Microsc. Microanal. **29**(Supplement_1), 631–632 (2023).

[177] Y. Liu, K. Koo, Z. Mao, X. Fu, X. Hu, and V. P. Dravid, "Unraveling the adsorption-limited hydrogen oxidation reaction at palladium surface via in situ electron microscopy," Proc. Natl. Acad. Sci. U. S. A. **121**(40), e2408277121 (2024).

[178] T.-W. Huang, S.-Y. Liu, Y.-J. Chuang, H.-Y. Hsieh, C.-Y. Tsai, Y.-T. Huang, U. Mirsaidov, P. Matsudaira, F.-G. Tseng, C.-S. Chang, and F.-R. Chen, "Self-aligned wet-cell for hydrated microbiology observation in TEM," Lab Chip **12**(2), 340–347 (2011).

[179] Y. Lu, W.-J. Yin, K.-L. Peng, K. Wang, Q. Hu, A. Selloni, F.-R. Chen, L.-M. Liu, and M.-L. Sui, "Self-hydrogenated shell promoting photocatalytic $H_2$ evolution on anatase $TiO_2$," Nat. Commun. **9**(1), 2752 (2018).

[180] H.-G. Liao, D. Zherebetskyy, H. Xin, C. Czarnik, P. Ercius, H. Elmlund, M. Pan, L.-W. Wang, and H. Zheng, "Facet development during platinum nanocube growth," Science **345**(6199), 916–919 (2014).

[181] J. Park, K. Koo, N. Noh, J. H. Chang, J. Y. Cheong, K. S. Dae, J. S. Park, S. Ji, I.-D. Kim, and J.M. Yuk, "Graphene Liquid Cell Electron Microscopy: Progress, Applications, and Perspectives," ACS Nano **15**(1), 288–308 (2021).

[182] D. J. Kelly, M. Zhou, N. Clark, M. J. Hamer, E. A. Lewis, A. M. Rakowski, S. J. Haigh, and R. V. Gorbachev, "Nanometer Resolution Elemental Mapping in Graphene-Based TEM Liquid Cells," Nano Lett. **18**(2), 1168–1174 (2018).

[183] S. M. Ghodsi, S. Anand, R. Shahbazian-Yassar, T. Shokuhfar, and C. M. Megaridis, "In Situ Study of Molecular Structure of Water and Ice Entrapped in Graphene Nanovessels," ACS Nano **13**(4), 4677–4685 (2019).

[184] M. F. Crook, I. A. Moreno-Hernandez, J. C. Ondry, J. Ciston, K. C. Bustillo, A. Vargas, and A. P. Alivisatos, "EELS Studies of Cerium Electrolyte Reveal Substantial Solute Concentration Effects in Graphene Liquid Cells," J. Am. Chem. Soc. **145**(12), 6648–6657 (2023).

[185] A. Zong, B. R. Nebgen, S.-C. Lin, J. A. Spies, and M. Zuerch, "Emerging ultrafast techniques for studying quantum materials," Nat. Rev. Mater. **8**(4), 224–240 (2023).

[186] P. Moradifar, Y. Liu, J. Shi, M. L. Siukola Thurston, H. Utzat, T. B. van Driel, A. M. Lindenberg, and J. A. Dionne, "Accelerating Quantum Materials Development with Advances in Transmission Electron Microscopy," Chem. Rev. **123**(23), 12757–12794 (2023).

[187] Y. Zhu, "Cryogenic Electron Microscopy on Strongly Correlated Quantum Materials," Acc. Chem. Res. **54**(18), 3518–3528 (2021).

[188] E. Rennich, S. H. Sung, N. Agarwal, M. Gates, R. Kerns, R. Hovden, and I. E. Baggari, "Ultracold cryogenic TEM with liquid helium and high stability," Proc. Natl. Acad. Sci. U. S. A. **122**(36), e2509736122 (2025).

[189] W. Zhao, M. Li, C.-Z. Chang, J. Jiang, L. Wu, C. Liu, J. S. Moodera, Y. Zhu, and M. H. W. Chan, "Direct imaging of electron transfer and its influence on superconducting pairing at FeSe/SrTiO$_3$ interface," Sci. Adv. **4**(3), eaao2682 (2018).

[190] R. A. Vilá, D. T. Boyle, A. Dai, W. Zhang, P. Sayavong, Y. Ye, Y. Yang, J. A. Dionne, and Y. Cui, "LiH formation and its impact on Li batteries revealed by cryogenic electron microscopy," Sci. Adv. **9**(12), eadf3609 (2023).

[191] M. Mecklenburg, B. Zutter, and B. C. Regan, "Thermometry of Silicon Nanoparticles," Phys. Rev. Appl. **9**(1), 014005 (2018).

[192] M. Mecklenburg, W. A. Hubbard, E. R. White, R. Dhall, S. B. Cronin, S. Aloni, and B. C. Regan, "Nanoscale temperature mapping in operating microelectronic devices," Science **347**(6222), 629–632 (2015).

[193] Y.-C. Yang, L. Serafini, N. Gauquelin, J. Verbeeck, and J. R. Jinschek, "Improving the accuracy of temperature measurement on TEM samples using plasmon energy expansion thermometry (PEET): Addressing sample thickness effects," Ultramicroscopy **270**, 114102 (2025).

[194] R. Nemausat, C. Gervais, C. Brouder, N. Trcera, A. Bordage, C. Coelho-Diogo, P. Florian, A. Rakhmatullin, I. Errea, L. Paulatto, M. Lazzeri, and D. Cabaret, "Temperature dependence of X-ray absorption and nuclear magnetic resonance spectra: probing quantum vibrations of light elements in oxides," Phys. Chem. Chem. Phys. **19**(8), 6246–6256 (2017).

[195] T. C. Rossi, C. P. Dykstra, T. N. Haddock, R. Wallick, J. H. Burke, C. M. Gentle, G. Doumy, A. M. March, and R. M. van der Veen, "Charge Carrier Screening in Photoexcited Epitaxial Semiconductor Nanorods Revealed by Transient X-ray Absorption Linear Dichroism," Nano Lett. **21**(22), 9534–9542 (2021).

[196] S. Lazar, P. Tiemeijer, C. S. Schnohr, M. Meledina, C. Patzig, T. Höche, P. Longo, and B. Freitag, "Enabling electron-energy-loss spectroscopy at very high energy losses: An opportunity to obtain x-ray absorption spectroscopy--like information using an electron microscope," Phys. Rev. Appl. **23**(5), 054095 (2025).

[197] Z. Wang, D. Santhanagopalan, W. Zhang, F. Wang, H. L. Xin, K. He, J. Li, N. Dudney, and Y. S. Meng, "In Situ STEM-EELS Observation of Nanoscale Interfacial Phenomena in All-Solid-State Batteries," Nano Lett. **16**(6), 3760–3767 (2016).

[198] S. A. Reisbick, A. Pofelski, M.-G. Han, C. Liu, E. Montgomery, C. Jing, H. Sawada, and Y. Zhu, "Characterization of transverse electron pulse trains using RF powered traveling wave metallic comb striplines," Ultramicroscopy **249**, 113733 (2023).

[199] Y.-J. Kim, H.-W. Nho, S. Ji, H. Lee, H. Ko, J. Weissenrieder, and O.-H. Kwon, "Femtosecond-resolved imaging of a single-particle phase transition in energy-filtered ultrafast electron microscopy," Sci. Adv. **9**(4), eadd5375 (2023).

[200] X. Llopart, J. Alozy, R. Ballabriga, M. Campbell, R. Casanova, V. Gromov, E. H. M. Heijne, T. Poikela, E. Santin, V. Sriskaran, L. Tlustos, and A. Vitkovskiy, "Timepix4, a large area pixel detector readout chip which can be tiled on 4 sides providing sub-200 ps timestamp binning," JINST **17**(01), C01044 (2022).

[201] K. Heijhoff, K. Akiba, R. Ballabriga, M. van Beuzekom, M. Campbell, A. P. Colijn, M. Fransen, R. Geertsema, V. Gromov, and X. Llopart Cudie, "Timing performance of the Timepix4 front-end," JINST **17**(07), P07006 (2022).

[202] P. Koutenský, N. L. Streshkova, K. Moriová, M. C. Chirita Mihaila, A. Knápek, D. Burda, and M. Kozák, "Ultrafast 4D Scanning Transmission Electron Microscopy for Imaging of Localized Optical Fields," ACS Photonics **12**(8), 4452–4459 (2025).

[203] T. Shimojima, A. Nakamura, and K. Ishizaka, "Development and applications of ultrafast transmission electron microscopy," Microscopy **72**(4), 287–298 (2023).

[204] D. Hui, H. Alqattan, M. Sennary, N. V. Golubev, and M. Th. Hassan, "Attosecond electron microscopy and diffraction," Sci. Adv. **10**(34), eadp5805 (2024).

[205] D. Hui, H. Alqattan, S. Zhang, V. Pervak, E. Chowdhury, and M. Th. Hassan, "Ultrafast optical switching and data encoding on synthesized light fields," Sci. Adv. **9**(8), eadf1015 (2023).

[206] D. Nabben, J. Kuttruff, L. Stolz, A. Ryabov, and P. Baum, "Attosecond electron microscopy of sub-cycle optical dynamics," Nature **619**(7968), 63–67 (2023).

[207] J. M. Voss, O. F. Harder, P. K. Olshin, M. Drabbels, and U. J. Lorenz, "Rapid melting and revitrification as an approach to microsecond time-resolved cryo-electron microscopy," Chem. Phys. Lett. **778**, 138812 (2021).

[208] M. Dąbrowski, S. Haldar, S. Khan, P. S. Keatley, D. Sagkovits, Z. Xue, C. Freeman, I. Verzhbitskiy, T. Griepe, U. Atxitia, G. Eda, H. Kurebayashi, E. J. G. Santos, and R. J. Hicken, "Ultrafast thermo-optical control of spins in a 2D van der Waals semiconductor," Nat. Commun. **16**(1), 2797 (2025).

[209] W. Hutchins, S. Zare, D. M. Hirt, J. A. Tomko, J. R. Matson, K. Diaz-Granados, M. Long, M. He, T. Pfeifer, J. Li, J. H. Edgar, J.-P. Maria, J. D. Caldwell, and P. E. Hopkins, "Ultrafast evanescent heat transfer across solid interfaces via hyperbolic phonon–polariton modes in hexagonal boron nitride," Nat. Mater. **24**(5), 698–706 (2025).

[210] M. Hugenschmidt, K. Adrion, A. Marx, E. Müller, and D. Gerthsen, "Electron-Beam-Induced Carbon Contamination in STEM-in-SEM: Quantification and Mitigation," Microsc. Microanal. **29**(1), 219–234 (2023).

[211] L. Zhang, J. P. Hoogenboom, B. Cook, and P. Kruit, "Photoemission sources and beam blankers for ultrafast electron microscopy," Struct. Dyn. **6**(5), 051501 (2019).

[212] P. Hommelhoff, Y. Sortais, A. Aghajani-Talesh, and M. A. Kasevich, "Field Emission Tip as a Nanometer Source of Free Electron Femtosecond Pulses," Phys. Rev. Lett. **96**(7), 077401 (2006).

[213] F. Houdellier, G. M. Caruso, S. Weber, M. Kociak, and A. Arbouet, "Development of a high brightness ultrafast Transmission Electron Microscope based on a laser-driven cold field emission source," Ultramicroscopy **186**, 128–138 (2018).

[214] A. Schröder, A. Wendeln, J. T. Weber, M. Mukai, Y. Kohno, and S. Schäfer, "Laser-driven cold field emission source for ultrafast transmission electron microscopy," Ultramicroscopy **275**, 114158 (2025).

[215] S. Ji, L. Piazza, G. Cao, S. T. Park, B. W. Reed, D. J. Masiel, and J. Weissenrieder, "Influence of cathode geometry on electron dynamics in an ultrafast electron microscope," Struct. Dyn. **4**(5), 054303 (2017).

[216] S. A. Willis, and D. J. Flannigan, "Influence of Photoemission Geometry on Timing and Efficiency in 4D Ultrafast Electron Microscopy," ChemPhysChem **26**(5), e202401032 (2025).

[217] J. Schaber, R. Xiang, and N. Gaponik, "Review of photocathodes for electron beam sources in particle accelerators," J. Mater. Chem. C **11**(9), 3162–3179 (2023).

[218] Y. Zhang, Y. Qian, J. Zhang, R. Fu, L. Liu, Y. Qiu, and B. Chang, "Integrated ultrahigh vacuum facility for photocathode preparation and in-situ characterization," in *Functional Material Applications: From Energy to Sensing*, (SPIE, 2023), pp. 77–83.

[219] C. Feng, Y. Zhang, J. Liu, Y. Qian, J. Zhang, J. Zhao, F. Shi, and X. Bai, "Optimized chemical cleaning procedure for enhancing photoemission from GaAs photocathode," Mater. Sci. Semicond. Process. **91**, 41–46 (2019).

[220] Y. Zhang, J. Zhang, C. Feng, H. Cheng, X. Zhang, and Y. Qian, "Improved performance of GaAs photocathodes using effective activation technique," in *Infrared Sensors, Devices, and Applications VII*, (SPIE, 2017), pp. 218–223.

[221] Y. Zhang, J. Niu, J. Zou, X. Chen, Y. Xu, B. Chang, and F. Shi, "Surface activation behavior of negative-electron-affinity exponential-doping GaAs photocathodes," Opt. Commun. **321**, 32–37 (2014).

[222] R. Brydson, *Electron Energy Loss Spectroscopy* (Garland Science, London, 2020).

[223] R. F. Egerton, *Electron Energy-Loss Spectroscopy in the Electron Microscope* (Springer US, Boston, MA, 2011).

[224] O. Dyck, O. Olunloyo, K. Xiao, B. Wolf, T. M. Moore, A. R. Lupini, and S. Jesse, "A Versatile Side Entry Laser System for Scanning Transmission Electron Microscopy," Adv. Mater. Technol. **10**(5), 2401208 (2025).

[225] J.-W. Henke, A. S. Raja, A. Feist, G. Huang, G. Arend, Y. Yang, F. J. Kappert, R. N. Wang, M. Möller, J. Pan, J. Liu, O. Kfir, C. Ropers, and T. J. Kippenberg, "Integrated photonics enables continuous-beam electron phase modulation," Nature **600**(7890), 653–658 (2021).

[226] "Optical Bulk Liquid Electrochemistry," Hummingbird Scientific, https://hummingbirdscientific.com/products/optical-bulk-liquid/ (accessed 2025-07-01).

[227] K. Karki, P. Kumar, A. Verret, N. Glachman, D. H. Alsem, D. Jariwala, N. Salmon, and E. Stach, "In situ/operando Study of Photoelectrochemistry Using Optical Liquid Cell Microscopy," Microsc. Microanal. **26**(S2), 2446–2447 (2020).

[228] "Luminary Micro Compact Specimen Photoexcitation System | Products | JEOL Ltd.," https://www.jeol.com/ (accessed 2025-07-01).

[229] "Mönch - CL Add-on for STEM - Attolight," https://www.attolight.com/products/light-collection-injection-for-stem (accessed 2025-07-01).

[230] J. Chen, and D. J. Flannigan, "A quantitative method for *in situ* pump-beam metrology in 4D ultrafast electron microscopy," Ultramicroscopy **234**, 113485 (2022).

[231] D. J. Flannigan, and E. J. VandenBussche, "Pulsed-beam transmission electron microscopy and radiation damage," Micron **172**, 103501 (2023).

[232] E. J. VandenBussche, and D. J. Flannigan, "Reducing Radiation Damage in Soft Matter with Femtosecond-Timed Single-Electron Packets," Nano Lett. **19**(9), 6687–6694 (2019).

[233] A. B. Swain, J. Kuttruff, J. Vorberger, and P. Baum, "Stronger femtosecond excitation causes slower electron-phonon coupling in silicon," Phys. Rev. Res. **7**(2), 023114 (2025).

[234] D. X. Du, S. A. Reisbick, and D. J. Flannigan, "UEMtomaton: A Source-Available Platform to Aid in Start-up of Ultrafast Electron Microscopy Labs," Ultramicroscopy **223**, 113235 (2021).

[235] D. J. Hilton, "Ultrafast Pump–Probe Spectroscopy," in *Optical Techniques for Solid-State Materials Characterization*, (CRC Press, 2012).

[236] P. G. Lynch, A. Das, S. Alam, C. C. Rich, and R. R. Frontiera, "Mastering Femtosecond Stimulated Raman Spectroscopy: A Practical Guide," ACS Phys. Chem. Au **4**(1), 1–18 (2024).

[237] Y. Morimoto, and P. Baum, "Single-Cycle Optical Control of Beam Electrons," Phys. Rev. Lett. **125**(19), 193202 (2020).

[238] K. E. Echternkamp, A. Feist, S. Schäfer, and C. Ropers, "Ramsey-type phase control of free-electron beams," Nat. Phys. **12**(11), 1000–1004 (2016).

[239] D. J. Flannigan, W. A. Curtis, E. J. VandenBussche, and Y. Zhang, "Low repetition-rate, high-resolution femtosecond transmission electron microscopy," J. Chem. Phys. **157**(18), 180903 (2022).

[240] D. A. Muller, and J. Silcox, "Delocalization in inelastic scattering," Ultramicroscopy **59**(1), 195–213 (1995).

[241] R. F. Egerton, "Scattering delocalization and radiation damage in STEM-EELS," Ultramicroscopy **180**, 115–124 (2017).

[242] C. A. Gadre, X. Yan, Q. Song, J. Li, L. Gu, H. Huyan, T. Aoki, S.-W. Lee, G. Chen, R. Wu, and X. Pan, "Nanoscale imaging of phonon dynamics by electron microscopy," Nature **606**(7913), 292–297 (2022).

[243] G. Haberfehlner, S. F. Hoefler, T. Rath, G. Trimmel, G. Kothleitner, and F. Hofer, "Benefits of direct electron detection and PCA for EELS investigation of organic photovoltaics materials," Micron **140**, 102981 (2021).